\documentclass[journal,final]{IEEEtran}
\usepackage[utf8]{inputenc}
\usepackage{amsmath}
\usepackage{multirow}
\usepackage{soul,color,xcolor}
\usepackage{cite}
\soulregister{\ref}7  
\soulregister{\cite}7 
\soulregister{\citep}7 
\soulregister{\citet}7 
\soulregister{\pageref}7 
\usepackage{graphicx}
\usepackage{float}
\usepackage{epstopdf}
\usepackage{subfloat}
\usepackage{caption}
\usepackage{subcaption}
\usepackage{makecell}
\usepackage[colorlinks,citecolor=blue]{hyperref}
\usepackage{stfloats}
\usepackage{txfonts}
\usepackage[ruled]{algorithm2e}
\usepackage{enumitem}
\usepackage{soul}
\usepackage[switch]{lineno}
\usepackage[T1]{fontenc}
\usepackage{dashrule}
\usepackage{multicol}

\usepackage{hyperref}
\usepackage{xcolor}
\usepackage{array}

\newtheorem{remark}{Remark}
\DeclareMathAlphabet{\mathpzc}{OT1}{pzc}{m}{it}
\newcolumntype{C}[1]{>{\centering\arraybackslash}m{#1}}

\begin{document}
\bibliographystyle{IEEEtran}
\title{Variational Bayesian Inference for Multiple Extended Targets or Unresolved Group Targets Tracking}
\author{Yuanhao Cheng,
	Yunhe Cao,~\IEEEmembership{Member,~IEEE,}
	Tat-Soon Yeo,~\IEEEmembership{Life~Fellow,~IEEE,}
	Yulin Zhang, Fu Jie
	\thanks{Yuanhao Cheng, Yunhe Cao, Yulin Zhang, and Fu Jie are with the National Key Laboratory of Radar Signal Processing, Xidian University, Xi'an 710071, China (email:chengyh@stu.xidian.edu.cn; caoyunhe@mail.xidian.edu.cn; zhangyulin@stu.xidian.edu.cn; fuj@stu.xidian.edu.cn).}
	\thanks{Tat-Soon Yeo is with the Department of Electrical and Computer Engineering, National University of Singapore, 119077, Singapore (email: eleyeots@nus.edu.sg).}
    \thanks{Corresponding author: Yunhe Cao. (email: caoyunhe@mail.xidian.edu.cn).}	
}

\maketitle

\begin{abstract}
In this work, we propose a method for tracking multiple extended targets or unresolvable group targets in a clutter environment. First, based on the Random Matrix Model (RMM), each target’s joint kinematic–extent state is modeled as a Gamma‑Gaussian‑Inverse‑Wishart (GGIW) distribution. Considering the uncertainty of measurement origin caused by the clutters, we adopt the idea of probabilistic data association and describe the joint association event as an unknown parameter in the joint prior distribution. Then, Variational Bayesian Inference (VBI) is used to approximate the intractable posterior distribution. To improve practicality, we propose two lightweight schemes to reduce computational complexity. The first is clustering‑based and effectively prunes joint association events. The second simplifies the variational posterior by using marginal association probabilities. Finally, we demonstrate effectiveness on simulations and real‑data experiments, and show that the method outperforms state‑of‑the‑art baselines in accuracy and adaptability. 
\end{abstract}


\section{INTRODUCTION}
\label{sec1}
\IEEEPARstart{I}{n} contrast to traditional point target tracking, when tracking extended targets or unresolvable group targets, it is typically assumed that the target will generate multiple measurements in each scan, and the tracking approach focuses on leveraging multiple measurements to jointly estimate the kinematic state and extended shape of the target \cite{r1,Granstrm2016ExtendedOT}.

Depending on the complexity of the silhouette, the extended shape of the target can be divided into simple axially symmetric shapes such as such as ellipses or rectangles \cite{2011Granström,2021Fowdur,2024Cao}, star-convex shapes \cite{2014Baum,2021Alqaderi}, or complex/irregular shapes \cite{2015Wahlström,2022Alqaderi,2019Aftab,2024Gao}. Different models can be selected to estimate the extended shape.

For targets modeled with regular elliptical shapes, the Random Matrix Model (RMM) \cite{2008Koch} is the most widely used model. The RMM represents the elliptical extent of the target as a Positive Semi-Definite Matrix (PSDM), and uses the inverse Wishart distribution to describe the uncertainty of the PSDM, which can ensure conjugacy in the Bayesian framework.

Since the RMM is easy to integrate into the Kalman-filter-based framework, it can be combined with traditional multi-target tracking methods such as the Joint Probabilistic Data Association (JPDA) or the Probabilistic Multi-Hypothesis Tracking (PMHT) to achieve Multiple Extended Target Tracking (METT) or Multiple Unresolvable Group Target Tracking (MUGTT) \cite{2012Wieneke,2015Schuster}. These methods incur a heavy computational burden due to complex data association and are limited to scenarios with fixed measurement rates. To mitigate association complexity, subsequent algorithms employ message passing \cite{2024LiYuanSheng}, but the resulting solutions are often suboptimal and still do not account for variable measurement rates.

Considering that the measurement rate of the target is uncertain, the gamma distribution is usually used to represent it. The joint state of the target can then be expressed as a Gamma Gaussian Inverse Wishart (GGIW) distribution based on the RMM. This distribution is usually embedded into the Random Finite Set (RFS)-based filters to deal with complex METT/MUGTT scenarios; representative methods include the GGIW-Probability Hypothesis Density (PHD) filter \cite{2012Granstrom}, the state-of-the-art GGIW-Poisson Multi-Bernoulli Mixture (PMBM) filter \cite{2020Granström_PMBM} and its approximate implementation \cite{2022Xia}. RFS-based methods achieve computational efficiency by avoiding data association. However, they are not trackers in the strict sense; they rely on auxiliary means (e.g., sets of trajectories \cite{2019Xia_T,2023XIAYx,2025WSX}) to generate target tracks, which greatly increases processing difficulty. Moreover, due to the lack of precise association, their accuracy cannot match that of data-association-based methods.

In addition, whether based on data association or on RFS, these methods still suffer from inaccurate shape estimation due to two major shortcomings of the RMM. One is that the standard RMM is unable to explicitly estimate the shape parameters of an ellipse (e.g., orientation, semi-axis lengths, etc.). Existing methods improve this in two ways: On the one hand, by introducing other auxiliary information, for example, \cite{2016Lan} uses an evolution matrix to describe changes in the extended shape, while \cite{2022Hoher} introduces a virtual measurement model to adaptively adjust the shape estimation. On the other hand, by modeling the shape parameters more finely, e.g., \cite{2019Yang} uses the Multiplicative Error Model (MEM) to decouple the shape parameters into orientation and semi-axis lengths. This decoupled form of the shape parameters is common in some studies \cite{2023Steuernagel,2023Li}, and the MEM can achieve accurate tracking of targets’ varying extent. However, the MEM has a risk of estimation collapse \cite{2021Tuncer}, which can lead to inaccuracies in shape estimation. This issue remains poorly addressed in many MEM-based METT/MUGTT applications, such as MEM combined with JPDA \cite{2018Yang_LJPDA,2020Yang} or with message passing \cite{2025GuoYunFei}.

Another shortcoming of the standard RMM is that it ignores the influence of measurement noise. As a result, the estimated shape is biased by the measurement noise covariance. An improved method \cite{2011Feldmann} introduces measurement noise at the expense of accuracy, but is suboptimal. Subsequent research has found that this shortcoming can be overcome using the Variational Bayesian Inference (VBI) \cite{2012Orguner}. The VBI aims to approximate the exact posterior using variational distributions by minimizing the Kullback-Leibler Divergence (KLD). In addition, the VBI performs well when dealing with over-parameterized models \cite{2019Ma} or nonlinear measurement equations \cite{2019Lan,2020ExtendedIET,2021Zhang,2021Tuncer}. 

For multi-target tracking in clutter with a fixed, known number of targets, or for single-target tracking in clutter. It is common to include the association variables as part of the parameters to be estimated by the VBI. For example, \cite{2024Gan} treats the association variables as dynamic parameters and combines them with VBI to track multiple point targets with a known target count. Building on \cite{2024Gan}, \cite{2025Gan} introduces a two-stage
variational inference routine to achieve higher robustness. \cite{2021Peng} performs approximate estimation of association variables and achieves single point target tracking with unknown measurement noise based on the VBI. \cite{2023Yang} derives a variational Bayesian approximation based on first-order statistical linearization for the association variables, which enables tracking and shape estimation of a single target in clutter. \cite{2023Su} uses the VBI and the message passing algorithm to achieve partially resolvable group target tracking, which is essentially the tracking of multiple point targets.

However, the above VBI-based methods either lack explicit modeling of the target shape, such as \cite{2024Gan,2025Gan} or do not consider joint data association for multi-target tracking, such as \cite{2021Peng,2023Yang,2023Su}. In \cite{2022Tuncer,2023Pei}, the VBI is used to implement target tracking with a multi-ellipsoidal shape model and to consider the association between the measurement and multiple sub-targets. This may seem to meet the need for METT/MUGTT, but because of the lack of prior knowledge of the number of measurements, this type of approach also cannot be used directly. In summary, VBI provides an effective solution to RMM’s two intrinsic shortcomings, yet there are currently no examples of the VBI being applied to METT/MUGTT.

Furthermore, from the data association perspective, existing methods that combine data association with the RMM, for example, \cite{2012Wieneke,2015Schuster,2024LiYuanSheng} not only fail to remedy RMM's inherent deficiencies but also require enumerating many feasible association hypotheses, incurring high computational costs and reduced robustness. Methods based on RFS and RMM, such as \cite{2012Granstrom,2020Granström_PMBM,2022Xia}, avoid data association and are more efficient, but they still suffer from RMM's limited capacity to estimate target shape and are unable to yield target trajectory. Therefore, none of the existing RMM-integrated methods are suitable for practical METT/MUGTT while simultaneously addressing RMM’s two key deficiencies.

In this work, we propose a novel method for METT/MUGTT with a fixed number of targets. The method is based on the RMM and defines the GGIW prior for the target's joint state, including measurement rate, kinematics, and extent. The GGIW prior is conjugate to the standard extended-target likelihood. This gives closed-form Bayesian updates, making the filter computationally efficient and easy to embed in tracking frameworks. It also provides principled uncertainty for centroid and extent and adapts to varying detection densities, improving robustness of gating and data association. To deal with the shortcomings of the standard RMM mentioned above, we first introduce an evolutionary model for the extended shape, and then derive the multi-target approximate posterior using VBI. This enables our method to take advantage of both RMM and VBI, thereby making it suitable for METT/MUGTT.

This work helps to inspire the extension of VBI applications to METT/MUGTT, and the main contributions are:
\begin{itemize}
\item We propose an RMM-based multi-target tracking method. The method can track multiple extended targets or unresolvable group targets in a cluttered environment with unknown measurement rates and association variables.
\item The proposed method incorporates a shape evolution model based on the RMM to obtain a more accurate estimate of target shape.
\item We combine the concept of JPDA and treat the multi-target joint data association as a static parameter; the well-known VBI is then employed to approximate the intractable joint posterior distribution to obtain estimates of the kinematic state, the extended shape, and the measurement rate of each target.
\item To reduce the computational complexity associated with enumerating all possible joint association events, we propose two lightweight schemes to improve the practicality of our method: one is based on gating and clustering, and the other is based on marginal association probabilities. These two schemes provide a balance between computational complexity and estimation accuracy while still retaining the essential advantages of the proposed method in applications.
\end{itemize}

This paper is organized as follows. In Section \ref{sec2}, the target is modeled using the RMM and we parameterize its joint state. We then introduce the concept of multi-target association events. In Sections \ref{sec3} and Section \ref{sec4}, we discuss our approach based on the Bayesian framework in terms of two parts, i.e., the time update and the measurement update. The VBI of the multi-target posterior distribution, as outlined in Section \ref{sec4}, represents a key area of interest. In Section \ref{sec5}, we present two feasible lightweight schemes to reduce the computational complexity of the proposed method, then the comparative experimental results with numerical simulations are given in Section \ref{sec6}. Finally, we draw conclusions in Section \ref{sec7}.

\begin{remark}
In general, the extended target and the unresolvable group target are subject to the same notion of tracking, and thus METT and MUGTT have the same meaning. Consequently, we do not differentiate between these two types of targets in detail, as our approach is applicable to both. In what follows, the term ``target'' will be used to refer to both types of targets for convenience. 
\end{remark}

\section{PROBLEM FORMULATION}
\label{sec2} 
\subsection{Modeling with the Random Matrix Model}
Consider an $n_d$-dimensional scenario, where the RMM models each target’s extent as an ellipse. At time $k$, the target’s kinematic state is denoted by $\boldsymbol{x}_k\in\mathbb{R}^{n_d\times n_s}$, containing position, velocity, etc., where $n_s$ is the dimension of the kinematic state in a one-dimensional physical space. Meanwhile, the target’s elliptical shape is represented by a PSDM.

The dynamic model of the target's kinematic state is:
\begin{equation}
\label{eq1}
\boldsymbol{x}_k=\mathcal{F}_k\left(\boldsymbol{x}_{k-1}\right)+\boldsymbol{\varepsilon}_k
\end{equation}

\noindent where the state transition function $\mathcal{F}(\cdot)$ governs the motion between consecutive time steps. $\boldsymbol{\varepsilon}_k$ denotes the process noise, typically modeled as Gaussian white noise $\boldsymbol{\varepsilon}_k\sim\mathcal{N}\left(0,\mathbf{G}_k\right)$ and $\mathbf{G}_k\in\mathbb{R}^{n_dn_s\times n_dn_s}$ is its covariance matrix. Under linear dynamics, this model simplifies to:
\begin{equation}
\label{eq2}
\boldsymbol{x}_k=\mathrm{\boldsymbol{\Phi}}_k\cdot\boldsymbol{x}_{k-1}+\boldsymbol{\varepsilon}_k
\end{equation}

\noindent where the state transition matrix satisfies  $\mathbf{\Phi}_k=\mathbf{F}_k\otimes \mathbf{I}_{n_d}$, $\mathbf{F}_k\in\mathbb{R}^{n_dn_s\times n_dn_s}$ is a dynamic matrix in a one-dimensional physical space, $\mathbf{I}_{n_d}$ denotes the $n_d$-th order identity matrix, and the symbol `$\otimes$' indicates the Kronecker product.

The dynamic model of the extension matrix $\boldsymbol{X}_k$ can be expressed as:
\begin{equation}
\label{eq3}
p\left(\boldsymbol{X}_k\mid \boldsymbol{X}_{k-1}\right)=\mathcal{W}\left(\boldsymbol{X}_k;\tau_k,\boldsymbol{E}_k\boldsymbol{X}_{k-1}\boldsymbol{E}_k^T\right)
\end{equation}

\noindent where $\mathcal{W}\left(\cdot; a,\mathbf{C}\right)$ denotes the Wishart density, the scalar $a$ is the degrees of freedom and $\mathbf{C}{\in\mathbb{R}}^{n_d\times n_d}$ is the scale matrix. Notice that in Eq. (\ref{eq3}) we adopt the shape‑evolution mechanism of Lan’s model \cite{2016Lan}. $\boldsymbol{E}_k{\in\mathbb{R}}^{n_d\times n_d}$ is an invertible evolution that describes the dependence of the extension on the shape characteristics of the target (e.g. orientation or size). The degrees of freedom $\tau_k$ control the stochasticity of the evolution. When $\boldsymbol{E}_k={\mathbf{I}_{n_d}}/{\sqrt{\tau_k}}$ and $\tau_k$ is time‑invariant, Eq. (\ref{eq3}) reduces to the standard RMM extension dynamics in \cite{2008Koch}.

Let the measurements received by the sensor at time $k$ be ${\mathcal{Y}_k=\left\{\boldsymbol{y}_k^j\right\}}_{j=1}^{m_k}$, where $m_k$ is the total number of measurements. We decompose $\mathcal{Y}_k$ into target‑generated measurements and clutter: $\mathcal{Y}_k={\widetilde{\mathcal{Y}}}_k\cup{\check{\mathcal{Y}}}_k$. Here, ${\widetilde{\mathcal{Y}}}_k=\left\{\boldsymbol{y}_k^{\widetilde{j}}\right\}_{\widetilde{j}=1}^{{\widetilde{m}}_k}$ denotes the measurements generated by the targets and ${\check{\mathcal{Y}}}_k=\left\{\boldsymbol{y}_k^{\check{j}}\right\}_{\check{j}=1}^{{\check{m}}_k}$ indicates the clutter. Clearly, $m_k={\widetilde{m}}_k+{\check{m}}_k$.

For the $\widetilde{j}$-th measurement $\boldsymbol{y}_k^{\widetilde{j}}$ in ${\widetilde{\mathcal{Y}}}_k$, we model:
\begin{equation}
\label{eq4}
\boldsymbol{y}_k^{\widetilde{j}}=\mathbf{H}_k\boldsymbol{x}_k+\boldsymbol{v}_k^{\widetilde{j}}
\end{equation}

\noindent where $\boldsymbol{x}_k$ is the kinematic state of the target that produced $\boldsymbol{y}_k^{\widetilde{j}}$. Target‑generated measurements are assumed independent and identically distributed.

The predicted distribution of $\boldsymbol{x}_k$ can be expressed as:
\begin{equation}
\label{eq5}
p\left(\boldsymbol{x}_k\mid{\widetilde{\mathcal{Y}}}^{k-1}\right)\approx\mathcal{N}\left({{\boldsymbol{x}}}_{k}; {\hat{\boldsymbol{m}}}_{k|k-1},{\hat{\boldsymbol{P}}}_{k|k-1}\right)
\end{equation}

\noindent where ${\hat{\boldsymbol{m}}}_{k|k-1}$ and ${\hat{\boldsymbol{P}}}_{k|k-1}$ are the mean and covariance of the predicted distribution $p\left(\boldsymbol{x}_k\mid{\widetilde{\mathcal{Y}}}^{k-1}\right)$, respectively. ${\widetilde{\mathcal{Y}}}^{k-1}=\left\{{\widetilde{\mathcal{Y}}}_1,{\widetilde{\mathcal{Y}}}_2,...,{\widetilde{\mathcal{Y}}}_k\right\}$ denotes all measurements of the target captured by the sensor before time $k$.

In Eq. (\ref{eq4}), $\mathbf{H}_k={\breve{\mathbf{H}}}_k\otimes \mathbf{I}_{n_d}$ with ${\breve{\mathbf{H}}}_k\in\mathbb{R}^{1\times n_s}$ is the measurement matrix in a one-dimensional physical space. The pseudo-measurement noise $\boldsymbol{v}_k^{\widetilde{j}}\sim\mathcal{N}\left(0,\boldsymbol{D}_k\boldsymbol{X}_k\boldsymbol{D}_k^\mathrm{T}\right)$ captures the deviation of $\boldsymbol{y}_k^{\widetilde{j}}$ from the target center. $\boldsymbol{D}_k{\in\mathbb{R}}^{n_d\times n_d}$ is an invertible matrix describing distortion in the observed extent and satisfies \cite{2016Lan}: 
\begin{equation}
\label{eq6}
\boldsymbol{D}_k=\left(s{\bar{\boldsymbol{X}}}_{k|k-1}+\mathbf{R}_k\right)^{-1/2}{\bar{\boldsymbol{X}}_{k|k-1}}^{-1/2}
\end{equation}

\noindent where $\textbf{R}_k$ is the covariance of the true measurement noise $\boldsymbol{\gamma}_k\sim\mathcal{N}\left(0,\textbf{R}_k\right)$ and ${\bar{\boldsymbol{X}}}_{k|k-1}$ is the predicted extension at time $k$. The scalar $s$ modulates the effect of ${\boldsymbol{X}}_k$ and can encode different target types\footnote{For a target with an elliptical shape, we have $s$ = 1/4\cite{2011Feldmann}.}. Note that $\boldsymbol{E}_k$ and $\boldsymbol{D}_k$ enhance the dynamic description of the extent, enabling more precise estimation of its size and orientation.

The predicted distribution of the extension ${\boldsymbol{X}}_k$ is assumed to be:
\begin{equation}
\label{eq7}
p\left(\boldsymbol{X}_k\mid{\widetilde{\mathcal{Y}}}^{k-1}\right)\approx\mathcal{IW}\left(\boldsymbol{X}_k;{\hat{\nu}}_{k|k-1},{\hat{\boldsymbol{V}}}_{k|k-1}\right)
\end{equation}

And we have:
\begin{equation}
\label{eq8}
{\bar{\boldsymbol{X}}}_{k|k-1}\triangleq\mathbb{E}\left(p\left(\boldsymbol{X}_k\mid{\widetilde{\mathcal{Y}}}^{k-1}\right)\right)={\hat{\boldsymbol{V}}}_{k|k-1}/\left({\hat{\nu}}_{k|k-1}-2n_d-2\right)
\end{equation}

\noindent where $\mathcal{IW}\left(\cdot;v,\boldsymbol{V}\right)$ denotes the inverse Wishart distribution.

The measurement model in Eq. (\ref{eq4}) specifies the spatial distribution of measurements, and the number of target‑generated measurements is usually modeled as Poisson with rate $\lambda_k$. To preserve conjugacy, the Poisson rate is given a gamma prior $\lambda_k\sim\mathcal{G}\!\left(\lambda_k;\alpha_k,\beta_k\right)$ with scalar shape $\alpha_k$ and scale $\beta_k$. Environmental clutter at time $k$ is modeled as a homogeneous Poisson Point Process (PPP): clutter is uniformly distributed over the surveillance area, and its cardinality is Poisson with time‑invariant rate $\lambda_c$.

\subsection{The State Parameters for Multiple Targets}
Consider $n_k, n_k>1$ targets. Let Let the joint state of all targets be $\boldsymbol{\Xi}_k^{1:n_k}=\left\{\mathbf{x}_k^{1:n_k},\mathbf{X}_k^{1:n_k},\mathbf{\Lambda}_k^{1:n_k}\right\}=\left\{\boldsymbol{x}_k^{\left(n\right)},\boldsymbol{X}_k^{\left(n\right)},\lambda_k^{\left(n\right)}\right\}_{n=1}^{n_k}$, where $\mathbf{x}_k^{1:n_k}=\left\{\boldsymbol{x}_k^{\left(n\right)}\right\}_{n=1}^{n_k}, \mathbf{X}_k^{1:n_k}=\left\{\boldsymbol{X}_k^{\left(n\right)}\right\}_{n=1}^{n_k},$ and $\mathbf{\Lambda}_k^{1:n_k}=\left\{\lambda_k^{\left(n\right)}\right\}_{n=1}^{n_k}$ collect the kinematic states, extension matrices, and measurement rates of all targets at time $k$, respectively. Define $\boldsymbol{\zeta}_k^{\left(n\right)}=\left\{\boldsymbol{m}_k^{\left(n\right)},\boldsymbol{P}_k^{\left(n\right)},v_k^{\left(n\right)},\boldsymbol{V}_k^{\left(n\right)},\alpha_k^{\left(n\right)},\beta_k^{\left(n\right)}\right\}$ represents all the state parameters of the $n$-th target. Then the joint state follows a GGIW distribution:
\begin{equation}
\label{eq9}
\begin{split}
&\boldsymbol{\xi}_k^{\left(n\right)}\sim\mathcal{GGIW}\left(\boldsymbol{\xi}_k^{\left(n\right)};\boldsymbol{\zeta}_k^{\left(n\right)}\right)\\
&=\mathcal{GGIW}\left(\boldsymbol{\xi}_k^{\left(n\right)};\boldsymbol{m}_k^{\left(n\right)},\boldsymbol{P}_k^{\left(n\right)},v_k^{\left(n\right)},\boldsymbol{V}_k^{\left(n\right)},\alpha_k^{\left(n\right)},\beta_k^{\left(n\right)}\right)\\
&=\mathcal{N}\left(\boldsymbol{x}_k^{\left(n\right)};\boldsymbol{m}_k^{\left(n\right)},\boldsymbol{P}_k^{\left(n\right)}\right)\cdot\mathcal{IW}\left(\boldsymbol{X}_k^{\left(n\right)};v_k^{\left(n\right)},\boldsymbol{V}_k^{\left(n\right)}\right)\cdot\mathcal{G}\left(\lambda_k^{\left(n\right)};\alpha_k^{\left(n\right)},\beta_k^{\left(n\right)}\right)
\end{split}
\end{equation}

\noindent where $\boldsymbol{\xi}_k^{\left(n\right)}=\left\{\boldsymbol{x}_k^{\left(n\right)},\boldsymbol{X}_k^{\left(n\right)},\lambda_k^{\left(n\right)}\right\}, n=1,2,\cdots,n_k$, is the joint state of the $n$-th target.

\begin{remark}
In Eq. (\ref{eq9}) we assume independence between the target's kinematic state and the extent, which is not strictly valid because target velocity often couples with extent orientation. This assumption, however, is needed to preserve conjugacy in the joint distribution. See Table \ref{tableNotations} for the complete list of notations.
\end{remark}

\begin{table}[h]
\renewcommand{\arraystretch}{1.5}
\caption{NOTATIONS}
\label{tableNotations}
\vspace{-5mm}
\centering
\begin{tabular}{p{0.43\textwidth}}	
\hrulefill
\setlist[itemize]{leftmargin=*}
\begin{itemize}
\setlength{\itemsep}{0pt}
\setlength{\parsep}{0pt}
\setlength{\parskip}{0pt}
\item Real number field $\mathbb{R}$; Positive integer field $\mathbb{Z}^+$ 
\item Set of real horizontal vector of length n is represented with $\mathbb{R}^n$.
\item Set of real matrices of size $m\times n$ is represented with $\mathbb{R}^{m\times n}$.
\item Set of symmetric positive definite and semi-definite matrices of size $n\times n$ is represented with $\mathbb{S}_{++}^n$ and $\mathbb{S}_+^n$, respectively.
\item $\mathcal{N}\left(\mathbf{x};\boldsymbol{\mu},\mathrm{\Sigma}\right)$ represents the multivariate Gaussian distribution with mean vector $\boldsymbol{\mu}\in\mathbb{R}^{n_\mathbf{x}}$ and covariance matrix $\mathrm{\Sigma}\in\mathbb{S}_{++}^{n_\mathbf{x}}$.
\item $\mathcal{W}\left(\mathbf{X};\upsilon,\mathbf{V}\right)$ denotes the Wishart distribution with degrees of freedom $\upsilon\in\mathbb{R}$ and semi-definite scale matrix $\mathbf{V}\in\mathbb{S}_+^{n_\mathbf{X}}$, and satisfied:
\begin{equation}
\nonumber
\mathcal{W}\left(\mathbf{X};\upsilon,\mathbf{V}\right)\propto\left|\mathbf{X}\right|^{\frac{\upsilon}{2}}\mathrm{etr}\left(-\frac{1}{2}\mathbf{X}^{-1}\mathbf{V}\right)
\end{equation}
where $\mathrm{etr}\left(\cdot\right)\triangleq {\mathtt
{e}}^{\mathrm{tr}\left(\cdot\right)}$.

\item $\mathcal{IW}\left(\mathbf{X};\dot{\upsilon},\mathbf{\dot{V}}\right)$ denotes the inverse Wishart distribution with degrees of freedom $\dot{\upsilon}\in\mathbb{R}$ and semi-definite scale matrix $\mathbf{\dot{V}}\in\mathbb{S}_+^{n_\mathbf{X}}$, and satisfied:
\begin{equation}
\nonumber
\mathcal{IW}\left(\mathbf{X};\dot{\upsilon},\mathbf{\dot{V}}\right)\propto\left|\mathbf{X}\right|^{-\frac{\dot{\upsilon}}{2}}\mathrm{etr}\left(-\frac{1}{2}\mathbf{X}^{-1}\mathbf{\dot{V}}\right)
\end{equation}

\item $\mathcal{G}\left(\lambda;a,b\right)$ denotes the gamma distribution with scalar shape parameter $a$ and scalar scale parameter $b$, and satisfied:
\begin{equation}
\nonumber
\mathcal{G}\left(\lambda;a,b\right)\propto \lambda ^{a-1} \mathtt{e}^{-b\lambda} 
\end{equation}
where $\Gamma\left ( \cdot \right )$ denotes the gamma function.

\item $\left|\mathbf{A}\right|$ denotes the determinant of the matrix $\mathbf{A}$, same as $\mathrm{det}[\mathbf{A}]$.
\item $a^\mathrm{T}$ means find the transpose of $a$.
\item $a\mathrm{!}$ means calculating the factorial of $a$.
\item $\mathbb{E}_P$ indicates the expectation operator, and subscript `$p$' emphasizes the underlying probability distribution(s).
\item diag[$a_1,a_2,\cdots,a_n$] returns the diagonal matrix whose diagonal elements are $a_1,a_2,\cdots,a_n$. 

\end{itemize} 
\hrulefill	
\end{tabular}
\end{table}

\subsection{Multi-Target Joint Association Events}
Data association for multi-target tracking means that the $m_k$ measurements from the set ${\mathcal{Y}_k=\left\{\boldsymbol{y}_k^j\right\}}_{j=1}^{m_k}$ at time $k$ are assigned to $n_k (n_k>1)$ targets. Since the ``target'' described in this paper is capable of generating multiple measurements at the same time, multiple measurements can be associated with the same target. In other words, data association is a many-to-one mapping problem. We denote a Joint Association Event (JAE) by $\boldsymbol{\theta}$, a $1\times m_k$ vector:
\begin{equation}
\label{eqadd1}
\boldsymbol{\theta}=\left[\vartheta_1,\vartheta_2,\cdots,\vartheta_{m_k}\right]_{1\times m_k}
\end{equation}

\noindent where the mapping $\vartheta$ assigns each measurement to a target index or to clutter, i.e.:
\begin{equation}
\label{eqadd2}
\begin{cases}
\vartheta_j=i,\ \ \ \text{if measurement $\boldsymbol{y}_k^j$ belongs to $i$-th target,}
\\
\vartheta_j=0,\ \ \  \text{if $\boldsymbol{y}_k^j$ is clutter.}
\end{cases}
\end{equation}

All JAEs at the time $k$ form the set ${\mathbf{\Theta}_k^{1:L_k^{\boldsymbol{\theta}}}=\left\{\boldsymbol{\theta}_k^l\right\}}_{l=1}^{L_k^{\boldsymbol{\theta}}}$, where $\boldsymbol{\theta}_k^l$ denotes the $l$-th JAE at the time $k$ and $L_k^\theta$ is the total number of JAEs. Suppose a given JAE assigns the measurements ${\widetilde{\mathcal{Y}}}_k^{l-\left(n\right)}=\left\{\boldsymbol{y}_k^{\widetilde{j}}\right\}_{\widetilde{j}=1}^{\phi_k^{l-\left(n\right)}}$ to the $n$-th target, where $\phi_k^{l-\left(n\right)}$ denotes the number of measurements assigned to the $n$-th target under $\boldsymbol{\theta}_k^l$. In particular, $\phi_k^{l-\left(0\right)}$ represents  the number of clutter measurements under $\boldsymbol{\theta}_k^l$. Then the likelihood of ${\widetilde{\mathcal{Y}}}_k^{\left(n\right)}$ can be expressed as:
\begin{equation}
\label{eqadd3}
\begin{split}
p\left({\widetilde{\mathcal{Y}}}_k^{\left(n\right)}\mid\boldsymbol{\theta}_k^l,\boldsymbol{\xi}_k^{\left(n\right)}\right)&=\prod_{\boldsymbol{y}_k^{\widetilde{j}}\in{\widetilde{\mathcal{Y}}}_k^{l-\left(n\right)}}{p\left(\boldsymbol{y}_k^{\widetilde{j}}\mid\boldsymbol{\xi}_k^{\left(n\right)}\right)} \\
&\mathrm{with} \ n=1,2,\ldots,n_k;\widetilde{j}=1,2,...,\phi_k^{l-\left(n\right)}
\end{split}
\end{equation}

\noindent where $p\left(\boldsymbol{y}_k^{\widetilde{j}}\mid\boldsymbol{\xi}_k^{\left(n\right)}\right)$ is the likelihood of  $\boldsymbol{y}_k^{\widetilde{j}}$. According to the measurement model in Eq. (\ref{eq4}), we have:
\begin{equation}
\label{eqadd4}
p\left(\boldsymbol{y}_k^{\widetilde{j}}\mid\boldsymbol{\xi}_k^{\left(n\right)}\right)=\mathcal{N}\left(\boldsymbol{y}_k^{\widetilde{j}}; \mathbf{H}_k^{\left(n\right)}\boldsymbol{x}_k^{\left(n\right)},\boldsymbol{D}_k^{\left(n\right)}\boldsymbol{X}_k^{\left(n\right)}\left(\boldsymbol{D}_k^{\left(n\right)}\right)^\mathrm{T}\right)
\end{equation}

For simplicity, all targets share the same measurement model; thus the superscript ‘$\left(n\right)$' on $\mathbf{H}_k^{\left(n\right)}$ can be dropped.

Given a JAE $\boldsymbol{\theta}_k^l$, the likelihood of the full measurement set $\mathcal{Y}_k$ with clutter is:
\begin{equation}
\label{eqadd5}
p\left(\mathcal{Y}_k\mid\boldsymbol{\theta}_k^l,\boldsymbol{\Xi}_k^{1:n_k}\right)=\left(\rho\right)^{\phi_k^{l-\left(0\right)}}\prod_{n=1}^{n_k}\prod_{\boldsymbol{y}_k^{\widetilde{j}}\in{\widetilde{\mathcal{Y}}}_k^{l-\left(n\right)}}{p\left(\boldsymbol{y}_k^{\widetilde{j}}\mid\boldsymbol{\xi}_k^{\left(n\right)}\right)}
\end{equation}

\noindent where $\rho$ denotes the clutter density. Since the Poisson parameter of the clutter $\lambda_c$ is time‑invariant, $\rho$ is also time-invariant and satisfies \cite{2005Kevin}:
\begin{equation}
\label{new_eqadd5}
\rho={\lambda_c}/{S_\mathrm{V}}
\end{equation}
if the volume of the scene $S_\mathrm{V}$ is fixed.

\begin{remark}
Given a JAE $\boldsymbol{\theta}_k^l$, the number of measurements assigned to each target or clutter in relation to that association event is directly known. Thus, a valid JAE can uniquely correspond to a measurement‑cardinality set, i.e, $\boldsymbol{\theta}_k^l\rightarrow\mathbf{\Pi}_k^l=\left\{\phi_k^{l-\left(n\right)}\right\}_{n=0}^{n_k}$. We illustrate the measurement cardinality set $\mathbf{\Pi}_k^l$ with a simple example.
\end{remark}
\begin{figure}[!htp]
\centering                                         
\includegraphics[width=2.2in]{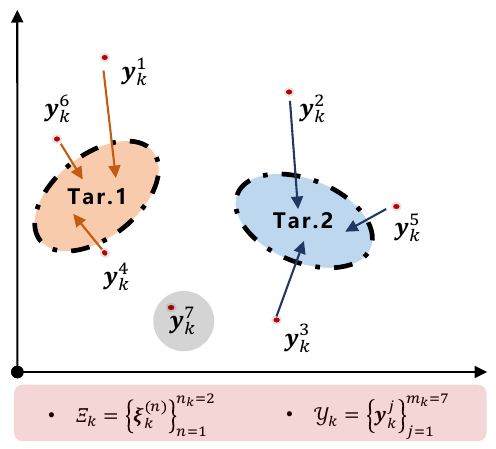}     
\caption{An example of a valid joint association event.} 
\label{fig1}                                     
\end{figure}

\textit{Example 1}: At time $k$, there are $n_k=2$ targets and $m_k=7$ unassigned and unknown-origin measurements. A reasonable JAE is illustrated in Fig. \ref{fig1}. In this association event, the measurements $\left\{\boldsymbol{y}_k^1,\boldsymbol{y}_k^4,\boldsymbol{y}_k^6\right\}$ are assigned to target 1 (Tar. 1), and the measurements $\left\{\boldsymbol{y}_k^2,\boldsymbol{y}_k^3,\boldsymbol{y}_k^5\right\}$ are assigned to target 2 (Tar. 2), while treating the $\boldsymbol{y}_k^7$ as clutter. Therefore, according to Eq. (\ref{eqadd1}) and ignore the time subscript ‘$k$’, this JAE can be expressed as $\boldsymbol{\theta}=\left[1,2, 2, 1, 2, 1, 0\right]$, and the corresponding set of measurement cardinality is $\boldsymbol{\theta}\rightarrow\boldsymbol{\Pi}=\left\{\phi^{\left(n\right)}\right\}_{n=0}^{n=2}=\left\{\phi^{\left(0\right)}=1,\phi^{\left(1\right)}=3,\phi^{\left(2\right)}=3\right\}$.

Note that the mapping from JAEs to cardinality sets is many‑to‑one. Given ${\boldsymbol{\Pi}_k}=\left\{\phi_k^{\left(n\right)}\right\}_{n=0}^{n_k}$, the number of JAEs it may correspond to is:
\begin{equation}
\label{eq10}
{n}_{\mathrm{JAE}}=\mathcal{C}_{m_k}^{\phi^{\left(1\right)}}\mathcal{C}_{m_k-\phi_k^{\left(1\right)}}^{\phi^{\left(2\right)}}\cdots \mathcal{C}_{m_k-\sum_{n=1}^{n_k-1}\phi_k^{\left(n\right)}}^{\phi^{\left(n_k\right)}}=\frac{m_k!}{\prod_{n=0}^{n_k}{\phi_k^{\left(n\right)}!}}
\end{equation}

\noindent where operation `$\mathcal{C}$' represents the calculation of the number of combinations.

The conditional probability of the $l$‑th JAE $\boldsymbol{\theta}_k^l$, $l=1,\ldots,L_k^\theta$, given the multi‑target state $\boldsymbol{\Xi}_k^{1:n_k}$ is:
\begin{equation}
\begin{split}
\label{eq11}
p\left(\boldsymbol{\theta}_k^l\mid\boldsymbol{\Xi}_k^{1:n_k}\right)&=p\left(\boldsymbol{\theta}_k^l,\boldsymbol{\Pi}_k^l\mid\boldsymbol{\Xi}_k^{1:n_k}\right)\\
&=p\left(\boldsymbol{\theta}_k^l\mid\boldsymbol{\Xi}_k^{1:n_k},\boldsymbol{\Pi}_k^l\right)p\left(\boldsymbol{\Pi}_k^l\mid\boldsymbol{\Xi}_k^{1:n_k}\right)
\end{split}
\end{equation}

\noindent where $\boldsymbol{\Pi}_k^l$ is the cardinality set corresponding to the JAE $\boldsymbol{\theta}_k^l$. The first step in Eq. (\ref{eq11}) can be derived by referring to \cite[Eq. (8)]{2018Yang_LJPDA}, and the last step of Eq. (\ref{eq11}) is obtained using the chain rule of conditional probability. For the first term in this step, since the number of measurements generated by the target is assumed to follow a Poisson distribution, there is:
\begin{equation}
\label{eq12}
p\left(\boldsymbol{\Pi}_k^l\mid\boldsymbol{\Xi}_k^{1:n_k}\right)=\prod_{n=0}^{n_k}\mathcal{P}\left(\phi_k^{l-\left(n\right)};\lambda_k^{\left(n\right)}\right)
\end{equation}

\noindent where $\mathcal{P}\left(\cdot;\lambda\right)$ represents the Poisson probability mass function, $\mathcal{P}\left(\phi,\lambda\right)=e^{-\lambda}\frac{\lambda^\phi}{\phi!}$. Eq. (\ref{eq12}) originates from \cite[Eq. (7)]{2018Yang_LJPDA}.For the second term $p\left(\boldsymbol{\theta}_k^l\middle|\boldsymbol{\Xi}_k^{1:n_k},\boldsymbol{\Pi}_k^l\right)$, since JAEs with the same $\boldsymbol{\Pi}$ are assumed uniformly distributed, we have:
\begin{equation}
\label{eq13}
p\left(\boldsymbol{\theta}_k^l\mid\boldsymbol{\Xi}_k^{1:n_k},\boldsymbol{\Pi}_k^l\right)=\frac{1}{{n}_{\mathrm{JAE}}}=\frac{\prod_{n=0}^{n_k}{\phi_k^{l-\left(n\right)}!}}{m_k!}
\end{equation}

Eq. (\ref{eq13}) is derived from \cite[Eq. (8)]{2018Yang_LJPDA}. And substituting Eq. (\ref{eq12}) and Eq. (\ref{eq13}) into Eq. (\ref{eq11}), it can be deduced that:
\begin{equation}
\label{eq14}
\begin{split}
p\left(\boldsymbol{\theta}_k^l\mid\boldsymbol{\Xi}_k^{1:n_k}\right)&=p\left(\boldsymbol{\theta}_k^l\mid\boldsymbol{\Xi}_k^{1:n_k},\boldsymbol{\Pi}_k^l\right)p\left(\boldsymbol{\Pi}_k^l\mid\boldsymbol{\Xi}_k^{1:n_k}\right) \\
&=\prod_{n=0}^{n_k}\mathcal{P}\left(\lambda_k^{\left(n\right)},\phi_k^{l-\left(n\right)}\right)\frac{\prod_{n=0}^{n_k}{\phi_k^{l-\left(n\right)}!}}{m_k!} \\
&=\frac{1}{m_k!}\prod_{n=0}^{n_k}{e^{-\lambda_k^{\left(n\right)}}\frac{\left(\lambda_k^{\left(n\right)}\right)^{\phi_k^{l-\left(n\right)}}}{\phi_k^{l-\left(n\right)}!}}\cdot\phi_k^{l-\left(n\right)}! \\
&=\frac{1}{m_k!}\prod_{n=0}^{n_k}e^{-\lambda_k^{\left(n\right)}}\left(\lambda_k^{\left(n\right)}\right)^{\phi_k^{l-\left(n\right)}}
\end{split}
\end{equation}

According to Bayes' rule, the posterior of the $l$-th JAE can be calculated as:
\begin{equation}
\label{eq15}
\begin{split}
&p\left(\boldsymbol{\theta}_k^l\mid\mathcal{Y}_k,\boldsymbol{\Xi}_k^{1:n_k}\right)\\
&\propto  p\left(\mathcal{Y}_k\mid\boldsymbol{\theta}_k^l,\boldsymbol{\Xi}_k^{1:n_k}\right)p\left(\boldsymbol{\theta}_k^l\mid\boldsymbol{\Xi}_k^{1:n_k}\right) \\
&\propto\left(\left(\rho\right)^{\phi_k^{l-\left(0\right)}}\prod_{n=1}^{n_k}\prod_{\boldsymbol{y}_k^{\widetilde{j}}\in{\widetilde{\mathcal{Y}}}_k^{l-\left(n\right)}}{p\left(\boldsymbol{y}_k^{\widetilde{j}}\mid\boldsymbol{\xi}_k^{\left(n\right)}\right)}\right)\left(\frac{1}{m_k!}\prod_{n=0}^{n_k}{e^{-\lambda_k^{\left(n\right)}}\left(\lambda_k^{\left(n\right)}\right)^{\phi_k^{l-\left(n\right)}}}\right) \\
&\propto\frac{\left(\rho\right)^{\phi_k^{l-\left(0\right)}}e^{-\lambda_c}\left(\lambda_c\right)^{\phi_k^{l-\left(0\right)}}}{m_k!}\prod_{n=1}^{n_k}\left(e^{-\lambda_k^{\left(n\right)}}\left(\lambda_k^{\left(n\right)}\right)^{\phi_k^{l-\left(n\right)}}\prod_{\boldsymbol{y}_k^{\widetilde{j}}\in{\widetilde{\mathcal{Y}}}_k^{l-\left(n\right)}}{p\left(\boldsymbol{y}_k^{\widetilde{j}}\mid\boldsymbol{\xi}_k^{\left(n\right)}\right)}\right)
\end{split}
\end{equation}

Since $m_k!$ and $e^{-\lambda_c}$ are independent of $\boldsymbol{\theta}_k^l$, Eq. (\ref{eq15}) simplifies to:
\begin{equation}
\begin{split}
&p\left(\boldsymbol{\theta}_k^l\mid\mathcal{Y}_k,\boldsymbol{\Xi}_k^{1:n_k}\right) \\
&\propto\left(\rho\right)^{\phi_k^{l-\left(0\right)}}\left(\lambda_c\right)^{\phi_k^{l-\left(0\right)}}\prod_{n=1}^{n_k}\left(e^{-\lambda_k^{\left(n\right)}}\left(\lambda_k^{\left(n\right)}\right)^{\phi_k^{l-\left(n\right)}}\prod_{\boldsymbol{y}_k^{\widetilde{j}}\in{\widetilde{\mathcal{Y}}}_k^{l-\left(n\right)}}{p\left(\boldsymbol{y}_k^{\widetilde{j}}\mid\boldsymbol{\xi}_k^{\left(n\right)}\right)}\right)
\end{split}
\end{equation}

We treat the JAE as an unknown parameter. Each individual JAE either occurs or does not occur, which is modeled by a Bernoulli distribution. Define the JAE indicator $w_k^l,l=1,2...,L_k^\theta$, as a Boolean variable. When $w_k^l=1$, the $l$-th JAE occurs; for all other JAEs, $w_k^{\dot{l}}=0$. The indicators satisfy $\sum_{l=1}^{L_k^\theta}w_k^l=1$. Naturally, the likelihood of the JAE set $\boldsymbol{\Theta}_k^{1:L_k^\theta}$ can be computed as:
\begin{equation}
\label{eq17}
\begin{split}
&p\left(\boldsymbol{\Theta}_k^{1:L_k^\theta}\mid\mathcal{Y}_k,\boldsymbol{\Xi}_k^{1:n_k}\right)\\
&\propto\prod_{l=1}^{L_k^\theta}\left[p\left(\boldsymbol{\theta}_k^l\mid\mathcal{Y}_k,\boldsymbol{\Xi}_k^{1:n_k}\right)\right]^{w_k^l}\\
&=\prod_{l=1}^{L_k^\theta}\left[\left(\rho\right)^{\phi_k^{l-\left(0\right)}}\left(\lambda_c\right)^{\phi_k^{l-\left(0\right)}}\prod_{n=1}^{n_k}\left(e^{-\lambda_k^{\left(n\right)}}\left(\lambda_k^{\left(n\right)}\right)^{\phi_k^{l-\left(n\right)}}\prod_{\boldsymbol{y}_k^{\widetilde{j}}\in{\widetilde{\mathcal{Y}}}_k^{l-\left(n\right)}}{p\left(\boldsymbol{y}_k^{\widetilde{j}}\mid\boldsymbol{\xi}_k^{\left(n\right)}\right)}\right)\right]^{w_k^l}
\end{split}
\end{equation}

Based on Eq. (\ref{eqadd3}), the likelihood of the measurement set $\mathcal{Y}_k$ can be written as:
\begin{equation}
\label{eqadd6}
\begin{split}
&p\left(\mathcal{Y}_k\mid\boldsymbol{\Theta}_k^{1:L_k^\theta},\boldsymbol{\Xi}_k^{1:n_k}\right)=\prod_{l=1}^{L_k^\theta}\left[\left(\rho\right)^{\phi_k^{l-\left(0\right)}}\prod_{n=1}^{n_k}\prod_{\boldsymbol{y}_k^{\widetilde{j}}\in{\widetilde{\mathcal{Y}}}_k^{l-\left(n\right)}}{p\left(\boldsymbol{y}_k^{\widetilde{j}}\mid\boldsymbol{\xi}_k^{\left(n\right)}\right)}\right]^{w_k^l} \\
&=\prod_{l=1}^{L_k^\theta}\left[\left(\rho\right)^{\phi_k^{l-\left(0\right)}}\prod_{n=1}^{n_k}\prod_{\boldsymbol{y}_k^{\widetilde{j}}\in{\widetilde{\mathcal{Y}}}_k^{l-\left(n\right)}} p\left(\boldsymbol{y}_k^{\widetilde{j}};\mathbf{H}_k^{\left(n\right)}\boldsymbol{x}_k^{\left(n\right)},\boldsymbol{D}_k^{\left(n\right)}\boldsymbol{X}_k^{\left(n\right)}\left(\boldsymbol{D}_k^{\left(n\right)}\right)^{\mathrm{T}}\right)\right]^{w_k^l}
\end{split}
\end{equation}

The term $\prod_{\boldsymbol{y}_k^{\widetilde{j}}\in{\widetilde{\mathcal{Y}}}_k^{l-\left(n\right)}} p\left(\boldsymbol{y}_k^{\widetilde{j}};\mathbf{H}_k^{\left(n\right)}\boldsymbol{x}_k^{\left(n\right)},\boldsymbol{D}_k^{\left(n\right)}\boldsymbol{X}_k^{\left(n\right)}\left(\boldsymbol{D}_k^{\left(n\right)}\right)^{\mathrm{T}}\right)$ in Eq. (\ref{eqadd6}) can be rewritten more compactly using the sample moments of ${\widetilde{\mathcal{Y}}}_k^{l-\left(n\right)}$, i.e., shown in Eq. (\ref{eq29}).
\begin{figure*}[t!]
\begin{equation}
\label{eq29}
\prod_{y_k^{\widetilde{j}}\in{\widetilde{\mathcal{Y}}}_k^{l-\left(n\right)}} p\left(\boldsymbol{y}_k^{\widetilde{j}};\mathbf{H}_k^{\left(n\right)}\boldsymbol{x}_k^{\left(n\right)},\boldsymbol{D}_k^{\left(n\right)}\boldsymbol{X}_k^{\left(n\right)}\left(\boldsymbol{D}_k^{\left(n\right)}\right)^{\mathrm{T}}\right)\propto\mathcal{N}\left({\bar{\boldsymbol{y}}}_k^{l-\left(n\right)};\mathbf{H}_k\boldsymbol{x}_k^{\left(n\right)},\frac{\boldsymbol{D}_k^{\left(n\right)}\boldsymbol{X}_k^{\left(n\right)}\left(\boldsymbol{D}_k^{\left(n\right)}\right)^{\mathrm{T}}}{\phi_k^{l-\left(n\right)}}\right)\mathcal{W}\left({\bar{\boldsymbol{Y}}}_k^{l-\left(n\right)}; \phi_k^{l-\left(n\right)}-1,\boldsymbol{D}_k^{\left(n\right)}\boldsymbol{X}_k^{\left(n\right)}\left(\boldsymbol{D}_k^{\left(n\right)}\right)^{\mathrm{T}}\right)
\end{equation}
\hrulefill
\end{figure*}

Eq. (\ref{eq29}) can be obtained from \cite[Eq. (16)]{2016Lan}. In Eq. (\ref{eq29}), ${\bar{\boldsymbol{y}}}_k^{l-(n)}$ and ${\bar{Y}}_k^{l-(n)}$ are the equivalent measurement and the equivalent spread of the measurements in ${\widetilde{\mathcal{Y}}}_k^{l-\left(n\right)}$, respectively. They can be calculated as:
 \begin{subequations}
 \begin{align}
 &{\bar{\boldsymbol{y}}}_k^{l-\left(n\right)}=\frac{1}{\phi_k^{l-\left(n\right)}}\sum_{\boldsymbol{y}_k^{\widetilde{j}}\in{\widetilde{\mathcal{Y}}}_k^{l-\left(n\right)}} \boldsymbol{y}_k^{\widetilde{j}} \\
 &{\bar{\boldsymbol{Y}}}_k^{l-\left(n\right)}=\sum_{\boldsymbol{y}_k^{\widetilde{j}}\in{\widetilde{\mathcal{Y}}}_k^{l-\left(n\right)}}{\left(\boldsymbol{y}_k^{\widetilde{j}}-{\bar{\boldsymbol{y}}}_k^{l-(n)}\right)\left(\boldsymbol{y}_k^{\widetilde{j}}-{\bar{\boldsymbol{y}}}_k^{l-(n)}\right)^{\mathrm{T}}}
 \end{align}
 \end{subequations}
 
These equivalent statistical moments ${\bar{\boldsymbol{y}}}_k^{l-\left(n\right)}$ and ${\bar{\boldsymbol{Y}}}_k^{l-\left(n\right)}$ are commonly used to approximate the measurement likelihood of the RMM \cite{2011Feldmann,2019Lan}. The loss of information from the approximation inevitably leads to an error. However, VBI can effectively reduce information loss to make more accurate estimates \cite{2012Orguner}. In what follows, we present the proposed method via time update and measurement update within a Bayesian iterative framework.

\section{TIME UPDATE}
\label{sec3}
Given the joint posterior at time $k-1$:
\begin{equation}
\label{eq18}
\begin{split}
&p\left(\boldsymbol{\Xi}_{k-1}^{1:n_{k-1}},\boldsymbol{\Theta}_{k-1}^{1:L_{k-1}^\theta}\mid\mathcal{Y}^{k-1}\right) \\
&=p\left(\boldsymbol{\Xi}_{k-1}^{1:n_{k-1}}\mid\mathcal{Y}^{k-1}\right)p\left(\boldsymbol{\Theta}_{k-1}^{1:L_{k-1}^\theta}\mid\mathcal{Y}^{k-1},\boldsymbol{\Xi}_{k-1}^{1:n_{k-1}}\right)
\end{split}
\end{equation}

According to the target's parametric model in Eq. (\ref{eq9}), the posterior $p\left(\boldsymbol{\Xi}_{k-1}^{1:n_{k-1}}\middle|\mathcal{Y}^{k-1}\right)$ factorizes as:
\begin{equation}
\label{eq_19}
\begin{split}
&p\left(\boldsymbol{\Xi}_{k-1}^{1:n_{k-1}}\mid\mathcal{Y}^{k-1}\right) \\
&=p\left(\mathbf{x}_{k-1}^{1:n_{k-1}}\mid\mathcal{Y}^{k-1}\right)p\left(\mathbf{X}_{k-1}^{1:n_{k-1}}\mid\mathcal{Y}^{k-1}\right)p\left(\mathbf{\Lambda}_{k-1}^{1:n_{k-1}}\mid\mathcal{Y}^{k-1}\right) \\
&=\prod_{n=1}^{n_{k-1}}\left[\begin{matrix}\mathcal{N}\left(\boldsymbol{x}_{k-1}^{\left(n\right)};{\hat{\boldsymbol{m}}}_{k-1|k-1}^{\left(n\right)},{\hat{\boldsymbol{P}}}_{k-1|k-1}^{\left(n\right)}\right)\cdot\mathcal{IW}\left(\boldsymbol{X}_{k-1}^{\left(n\right)};{\hat{v}}_{k-1|k-1}^{\left(n\right)},{\hat{\boldsymbol{V}}}_{k-1|k-1}^{\left(n\right)}\right)\\\cdot\mathcal{G}\left(\lambda_{k-1}^{\left(n\right)};{\hat{\alpha}}_{k-1|k-1}^{\left(n\right)},{\hat{\beta}}_{k-1|k-1}^{\left(n\right)}\right)\\\end{matrix}\right]
\end{split}
\end{equation}

The $n$-th component in Eq. (\ref{eq_19}) is the posterior of the $n$-th target at time $k-1$.

For simplicity, the target kinematics is assumed to follow linear dynamics. The predicted distribution $p\left(\boldsymbol{x}_k^{\left(n\right)}\mid \mathcal{Y}^{k-1}\right)$ of the $n$-th target's kinematic state $\boldsymbol{x}_k^{\left(n\right)},n=1,2,...,n_{k-1}$$n=1,2,\ldots,n_{k-1}$, is given by the standard Kalman time update:
\begin{equation}
p\left(\boldsymbol{x}_k^{\left(n\right)}\mid\mathcal{Y}^{k-1}\right)\sim\mathcal{N}\left(\boldsymbol{x}_k^{\left(n\right)};{\hat{\boldsymbol{m}}}_{k|k-1}^{\left(n\right)},{\hat{\boldsymbol{P}}}_{k|k-1}^{\left(n\right)}\right) \label{eq20}
\end{equation}

\noindent with 
\begin{align}
& {\hat{\boldsymbol{m}}}_{k|k-1}^{\left(n\right)}=\boldsymbol{\Phi}_k\cdot{\hat{\boldsymbol{m}}}_{k-1|k-1}^{\left(n\right)} \tag{\ref{eq20}{a}}\\
&{\hat{\boldsymbol{P}}}_{k|k-1}^{\left(n\right)}=\boldsymbol{\Phi}_k{\hat{\boldsymbol{P}}}_{k-1|k-1}^{\left(n\right)}\boldsymbol{\Phi}_k^{\mathrm{T}}+\mathbf{G}_k \tag{\ref{eq20}{b}} &    
\end{align}

\noindent where ${\hat{\boldsymbol{m}}}_{k|k-1}^{\left(n\right)}$ and ${\hat{\boldsymbol{P}}}_{k|k-1}^{\left(n\right)}$ are the mean and covariance of the predicted distribution, respectively.

For the extension matrix $\boldsymbol{X}_k^{\left(n\right)},n=1,2,...,n_{k-1}$, we adopt the time update from \cite{2016Lan}, i.e., the predicted distribution $p\left(\boldsymbol{X}_k^{\left(n\right)}\mid\mathcal{Y}^{k-1}\right)$ obeys:
\begin{equation}
p\left(\boldsymbol{X}_k^{\left(n\right)}\mid\mathcal{Y}^{k-1}\right)\sim\mathcal{IW}\left(\boldsymbol{X}_k^{\left(n\right)};{\hat{v}}_{k|k-1}^{\left(n\right)},{\hat{\boldsymbol{V}}}_{k|k-1}^{\left(n\right)}\right) \label{eq21}
\end{equation}

\noindent with
\begin{align}
&{\hat{v}}_{k|k-1}^{\left(n\right)}=\frac{2\tau_k^{\left(n\right)}\left(\gamma_{k-1}^{\left(n\right)}+1\right)\left(\gamma_{k-1}^{\left(n\right)}-1\right)\left(\gamma_{k-1}^{\left(n\right)}-2\right)}{\left(\gamma_{k-1}^{\left(n\right)}\right)^2\left(\gamma_{k-1}^{\left(n\right)}+\tau_k^{\left(n\right)}\right)}+2n_d+4 \tag{\ref{eq21}{a}} \\
&{\hat{\boldsymbol{V}}}_{k|k-1}^{\left(n\right)}=\frac{\tau_k^{\left(n\right)}}{\gamma_{k-1}^{\left(n\right)}}\left({\hat{v}}_{k|k-1}^{\left(n\right)}-2n_d-2\right)\boldsymbol{E}_k^{\left(n\right)}{\hat{\boldsymbol{V}}}_{k-1|k-1}^{\left(n\right)}\left(\boldsymbol{E}_k^{\left(n\right)}\right)^\mathrm{T} \tag{\ref{eq21}{b}} \\
& \gamma_{k-1}^{\left(n\right)}={\hat{v}}_{k-1|k-1}^{\left(n\right)}-2n_d-2 \tag{\ref{eq21}{c}}&
\end{align}

For the measurement rate $\lambda_{k}^{\left(n\right)},n=1,2,...,n_{k-1}$, we employ the forgetting factor $\iota_{k-1}^{\left(n\right)}\in\mathbb{R},\ \iota_{k-1}^{\left(n\right)} > 1$, to model temporal changes, i.e.,
\begin{equation}
p\left(\lambda_k^{\left(n\right)}\mid \mathcal{Y}^{k-1}\right)\sim\mathcal{G}\left(\lambda_k^{\left(n\right)};{\hat{\alpha}}_{k|k-1}^{\left(n\right)},{\hat{\beta}}_{k|k-1}^{\left(n\right)}\right) \label{eq22}
\end{equation}
\noindent with
\begin{align}
&{\hat{\alpha}}_{k|k-1}^{\left(n\right)}={\hat{\alpha}}_{k-1|k-1}^{\left(n\right)}/{\iota_{k-1}^{\left(n\right)}} \tag{\ref{eq22}{a}}\\
&{\hat{\beta}}_{k|k-1}^{\left(n\right)}={\hat{\beta}}_{k-1|k-1}^{\left(n\right)}/{\iota_{k-1}^{\left(n\right)}} \tag{\ref{eq22}{b}}
\end{align}

In general, we assume the same forgetting pattern across targets, so the superscript ‘$n$’ on $\iota_{k-1}^{\left(n\right)}$ can be dropped.

Eq. (\ref{eq20}) and Eq. (\ref{eq21}) come from \cite[Table $\text{\uppercase\expandafter{\romannumeral1}}$]{2016Lan}. Eq. (\ref{eq22}) adopts the conventional treatment of the measurement rate in the time update, which can be traced back to \cite[Eq. (31)]{2013Lundquist}. For the posterior density of the JAE set $p\left(\boldsymbol{\Theta}_{k-1}^{1:L_{k-1}^\theta}\mid\mathcal{Y}^{k-1},\boldsymbol{\Xi}_{k-1}^{1:n_{k-1}}\right)$ at time $k-1$, replacing the state parameters with their predicted values yields the joint predicted distribution:
\begin{equation}
\label{eq23}
\begin{split}
&p\left(\mathrm{\Xi}_k^{1:n_k},\mathrm{\Theta}_k^{1:L_k^\theta}\mid\mathcal{Y}^{k-1}\right)\\
&=p\left(\mathrm{\Xi}_k^{1:n_k}\mid\mathcal{Y}^{k-1}\right)p\left(\mathrm{\Theta}_k^{1:L_k^\theta}\mid\mathcal{Y}^{k-1},\mathrm{\Xi}_k^{1:n_k}\right) \\
&=\prod_{n=1}^{n_{k-1}}\left[\begin{matrix}\mathcal{N}\left(\boldsymbol{x}_{k}^{\left(n\right)};{\hat{\boldsymbol{m}}}_{k|k-1}^{\left(n\right)},{\hat{\boldsymbol{P}}}_{k|k-1}^{\left(n\right)}\right)\cdot\mathcal{IW}\left(\boldsymbol{X}_{k}^{\left(n\right)};{\hat{v}}_{k|k-1}^{\left(n\right)},{\hat{\boldsymbol{V}}}_{k|k-1}^{\left(n\right)}\right)\\\cdot\mathcal{G}\left(\lambda_{k}^{\left(n\right)};{\hat{\alpha}}_{k|k-1}^{\left(n\right)},{\hat{\beta}}_{k|k-1}^{\left(n\right)}\right)\\\end{matrix}\right] \\
&\qquad\times p\left(\boldsymbol{\Theta}_k^{1:L_k^\theta}\mid\mathcal{Y}^{k-1},\boldsymbol{\Xi}_{k}^{1:n_k}\right)
\end{split}
\end{equation}

\section{MEASUREMENT UPDATE}
\label{sec4}
Since the target state parameters follow a GGIW distribution, the joint posterior $p\left(\boldsymbol{\Xi}_k^{1:n_k},\boldsymbol{\Theta}_k^{1:L_k^\theta}\mid\mathcal{Y}^k\right)$ involves a total of $6\cdot n_k$ unknown state parameters. The solution of the exact posterior density is extremely complex, especially with the JAE also unknown. VBI approximates the true posterior with factorized variational posteriors and iteratively updates each factor via Coordinate‑Ascent Variational Inference (CAVI). By doing so, it yields more accurate likelihood estimation, mitigating the impact of ignoring measurement noise in the RMM.

Specifically, the joint posterior $p\left(\boldsymbol{\Xi}_k^{1:n_k},\boldsymbol{\Theta}_k^{1:L_k^\theta}\mid\mathcal{Y}^k\right)$ can be decomposed into a product of several variational posterior densities according to \cite[Eq. (10)]{2021Tuncer}:
\begin{equation}
\label{eq24}
\begin{split}
p\left(\boldsymbol{\Xi}_k^{1:n_k},\boldsymbol{\Theta}_k^{1:L_k^\theta}\mid\mathcal{Y}^k\right)
&=p\left(\mathbf{x}_k^{1:n_k},\mathbf{X}_k^{1:n_k},\mathbf{\Lambda}_k^{1:n_k},\boldsymbol{\Theta}_k^{1:L_k^\theta}\mid\mathcal{Y}^k\right) \\
&\approx q_\mathbf{x}\left(\mathbf{x}_k^{1:n_k}\right)q_\mathbf{X}\left(\mathbf{X}_k^{1:n_k}\right)q_\mathbf{\Lambda}\left(\mathbf{\Lambda}_k^{1:n_k}\right)q_{\mathbf{\Theta}}\left(\mathbf{\Theta}_k^{1:L_k^\theta}\right)
\end{split}
\end{equation}

\noindent  where, for $\ast\in\{\mathbf{x},\mathbf{X},\mathbf{\Lambda}\}$, $q_\ast\left(\ast_k^{1:n_k}\right)=\prod_{n=1}^{n_k}q_{\ast_k^{\left(n\right)}}\left(\ast_k^{\left(n\right)}\right)$. For the JAE factor, $q_{\mathbf{\Theta}}\left(\mathbf{\Theta}_k^{1:L_k^\theta}\right)$ follows its own product form over events as in Eq. (\ref{eq28}). As an example, if ‘$\ast$’ denotes the kinematic state $\mathbf{x}$, then we have $q_\mathbf{x}\left(\mathbf{x}_k^{1:n_k}\right)=\prod_{n=1}^{n_k}{q_{\boldsymbol{x}_k^{\left(n\right)}}\left(\boldsymbol{x}_k^{\left(n\right)}\right)}$.

CAVI minimizes the KL divergence between the variational posterior and the true posterior by alternating updates of each factor. This avoids calculating the exact posterior and the KL divergence between two distributions can be defined as:
\begin{equation}
\label{eq25}
\mathcal{KL}\left(q\left(y\right)||p\left(y\right)\right)=\int{q\left(y\right)\mathrm{ln} \frac{q\left(y\right)}{p\left(y\right)}dy}
\end{equation}

Based on the above rules for variational inference, \cite{2021Tuncer} has the following relation:
\begin{equation}
\label{eq26}
\mathrm{ln}\ q_\psi\left(\psi_k\right)\propto\mathbb{E}_{\psi}\left[\mathrm{ln}\ p\left(\mathbf{x}_k^{1:n_k},\mathbf{X}_k^{1:n_k},\mathbf{\Lambda}_k^{1:n_k},\boldsymbol{\Theta}_k^{1:L_k^\theta},\mathcal{Y}_k\mid\mathcal{Y}^{k-1}\right)\right]+\mathtt{C}_{\psi}
\end{equation}

\noindent where $\psi\in\left\{\mathbf{x}_k^{1:n_k},\mathbf{X}_k^{1:n_k},\mathbf{\Lambda}_k^{1:n_k},\boldsymbol{\Theta}_k^{1:L_k^\theta}\right\}$ and the symbol `$\psi$' represents the set consisting of all unknown parameters except the parameter $\psi$, e.g., $\mathbb{E}_{\boldsymbol{x}_k^{\left(i\right)}}$ denotes the parameter ``$\mathbf{X}_k^{1:n_k},\mathbf{\Lambda}_k^{1:n_k},\boldsymbol{\Theta}_k^{1:L_k^\theta}$'' as well as $\boldsymbol{x}_k^{\left(j\right)}$ in $\mathbf{x}_k^{1:n_k}$, with $i\neq j$. And the term independent of $\psi$ can be absorbed into the constant $\mathtt{C}_{\psi}$.

The joint density $p\left(\mathbf{x}_k^{1:n_k},\mathbf{X}_k^{1:n_k},\mathbf{\Lambda}_k^{1:n_k},\boldsymbol{\Theta}_k^{1:L_k^\theta},\mathcal{Y}_k\mid\mathcal{Y}^{k-1}\right)$ in Eq. (\ref{eq26}) contains all the information of the posterior density and can be expanded using the chain rule of conditional probability:
\begin{equation}
\label{eq27}
\begin{split}
&p\left(\mathbf{x}_k^{1:n_k},\mathbf{X}_k^{1:n_k},\mathbf{\Lambda}_k^{1:n_k},\boldsymbol{\Theta}_k^{1:L_k^\theta},\mathcal{Y}_k\mid\mathcal{Y}^{k-1}\right) \\
&=p\left(\mathcal{Y}_k\mid\boldsymbol{\Xi}_k^{1:n_k},\boldsymbol{\Theta}_k^{1:L_k^\theta}\right)p\left(\boldsymbol{\Theta}_k^{1:L_k^\theta}\mid\boldsymbol{\Xi}_k^{1:n_k},\mathcal{Y}^{k-1}\right)p\left(\boldsymbol{\Xi}_k^{1:n_k}\mid\mathcal{Y}^{k-1}\right)
\end{split}
\end{equation}

\noindent where $p\left(\mathcal{Y}_k\mid\boldsymbol{\Xi}_k^{1:n_k},\boldsymbol{\Theta}_k^{1:L_k^\theta}\right)$, $p\left(\boldsymbol{\Theta}_k^{1:L_k^\theta}\mid\boldsymbol{\Xi}_k^{1:n_k},\mathcal{Y}^{k-1}\right)$$p\left(\boldsymbol{\Xi}_k^{1:n_k}\mid\mathcal{Y}^{k-1}\right)$ are given in Eq. (\ref{eqadd6}) and Eq. (\ref{eq23}), respectively.

Since the VBI is a process with multiple iterations, here we give the computation of each variational posterior in a single variational Bayesian inference below:

$\bullet$\quad For the JAE set $\boldsymbol{\Theta}_k^{1:L_k^\theta}$ at time $k$, its variational posterior density $q_{\boldsymbol{\Theta}}\left(\boldsymbol{\Theta}_k^{1:L_k^\theta}\right)$ can be calculated as Eq. (\ref{eq28}). 

In Eq. (\ref{eq28}), ${\bar{\mathcal{U}}}_{\lambda_k^{\left(n\right)}}\left(\phi_k^{l-\left(n\right)}\right)=\left(\mathbb{E}_{\lambda_k^{\left(n\right)}}\left(\lambda_k^{\left(n\right)}\right)\right)^{\phi_k^{l-\left(n\right)}}$, and $\mathbb{E}_{\lambda_k^{\left(n\right)}}\left(\lambda_k^{\left(n\right)}\right)$ is the expectation of the $n$-th target measurement rate, which can be calculated as Eq. (\ref{eq32}). In addition, the term $\prod_{\boldsymbol{y}_k^{\widetilde{j}}\in{\widetilde{\mathcal{Y}}}_k^{\left(n\right)}}\mathcal{N}\left(\boldsymbol{y}_k^{\widetilde{j}};\mathbf{H}_k{\hat{\boldsymbol{m}}}_{k|k-1}^{\left(n\right)}, \boldsymbol{S}_k^{\left(n\right)}\right)$ is the general likelihood of the measurements generated by the $n$-th target under the $l$-th JAE. $\boldsymbol{S}_k^{\left(n\right)}$ is the innovation covariance of the $n$-th target which can be calculated according to Eq. (\ref{eq37}). The proof of Eq. (\ref{eq28}) can be found in \cite[Appendix A]{Cheng2025arvix}.
\begin{figure*}[t!]
\begin{equation}
\label{eq28}
q_{\boldsymbol{\Theta}}\left(\boldsymbol{\Theta}_k^{1:L_k^\theta}\right)\propto\prod_{l=1}^{L_k^\theta}\left[\left(\rho\right)^{\phi_k^{l-\left(0\right)}}\left(\lambda_c\right)^{\phi_k^{l-\left(0\right)}}\prod_{n=1}^{n_k}{\bar{\mathcal{U}}}_{\lambda_k^{\left(n\right)}}\left(\phi_k^{l-\left(n\right)}\right)\prod_{\boldsymbol{y}_k^{\widetilde{j}}\in{\widetilde{\mathcal{Y}}}_k^{l-\left(n\right)}}{\mathcal{N}\left(\boldsymbol{y}_k^{\widetilde{j}};\mathbf{H}_k\hat{\boldsymbol{m}}_{k|k-1}^{\left(n\right)},\boldsymbol{S}_k^{\left(n\right)}\right)}\right]^{w_k^l}
\end{equation}
\hrulefill
\end{figure*}

\begin{remark}
The variational posterior $q_\mathrm{\boldsymbol{\Theta}}\left(\mathrm{\boldsymbol{\Theta}}_k^{1:L_k^\theta}\right)$ in Eq. (\ref{eq28}) is rigorously derived by the probabilistic data association, i.e., for each JAE $\boldsymbol{\theta}_k^l,l=1,2..., L_k^\theta$ in the set $\boldsymbol{\Theta}_k^{1:L_k^\theta}$, the variational posterior can be expressed as the product of the prior and the likelihood. However, since the measurement rate ${\lambda}_k^{\left(n\right)},n=1,2,...,n_k$ of the target is unknown, we use an expectation $\mathbb{E}_{\lambda_k^{\left(n\right)}}\left(\lambda_k^{\left(n\right)}\right)$ instead. This expectation of $\lambda_k^{\left(n\right)}$ can be derived by solving for the variational posterior $q_\mathbf{\Lambda}\left(\mathbf{\Lambda}_k^{1:n_k}\right)$, which demonstrates a typical feature of the VBI.
\end{remark}

$\bullet$\quad For the measurement rate $\lambda_k^{\left(n\right)},n=1,\cdots,n_k$ in $\mathbf{\Lambda}_k^{1:n_k}$, its variational posterior density $q_{\lambda_k^{\left(n\right)}}\left(\lambda_k^{\left(n\right)}\right)$ can be computed as:
\begin{equation}
q_{\lambda_k^{\left(n\right)}}\left(\lambda_k^{\left(n\right)}\right)\sim\mathcal{G}\left(\lambda_k^{\left(n\right)};{\hat{\alpha}}_{k|k}^{\left(n\right)}, {\hat{\beta}}_{k|k}^{\left(n\right)}\right) \label{eq31}
\end{equation}

\noindent with
\begin{align}
&{\hat{\alpha}}_{k|k}^{\left(n\right)}={\hat{\alpha}}_{k|k-1}^{\left(n\right)}+\sum_{l=1}^{L_k^\theta}{q_{\boldsymbol{\theta}}\left(w_k^l\right)\phi_k^{l-\left(n\right)}} \tag{\ref{eq31}{a}}\\
&{\hat{\beta}}_{k|k}^{\left(n\right)}={\hat{\beta}}_{k|k-1}^{\left(n\right)}+\sum_{l=1}^{L_k^\theta}{q_{\boldsymbol{\theta}}\left(w_k^l\right)} \tag{\ref{eq31}{b}}
\end{align}

\noindent where $q_{\boldsymbol{\theta}}\left(w_k^l\right)$ denotes the marginal probability of the event indicator $w_k^l=1$ under $q_{\boldsymbol{\Theta}}\left(\boldsymbol{\Theta}_k^{1:L_k^\theta}\right)$, which corresponds to the JAE $\boldsymbol{\theta}_k^l$. The expectation of the measurement rate of the $n$-th target $\mathbb{E}_{\lambda_k^{\left(n\right)}}\left(\lambda_k^{\left(n\right)}\right)$ can be derived as:
\begin{equation}
\label{eq32}
\mathbb{E}_{\lambda_k^{\left(n\right)}}\left(\lambda_k^{\left(n\right)}\right)=\frac{{\hat{\alpha}}_{k|k}^{\left(n\right)}}{{\hat{\beta}}_{k|k}^{\left(n\right)}}
\end{equation}

$\bullet$\quad For the kinematic state of the target $\boldsymbol{x}_k^{\left(n\right)}, n=1,\cdots,n_k$ in $\mathbf{x}_k^{1:n_k}$ , its variational posterior density $q_{\boldsymbol{x}_k^{\left(n\right)}}\left(\boldsymbol{x}_k^{\left(n\right)}\right)$ can be calculated as:
\begin{align}
&q_{\boldsymbol{x}_k^{\left(n\right)}}\left(\boldsymbol{x}_k^{\left(n\right)}\right)\sim\mathcal{N}\left(\boldsymbol{x}_k^{\left(n\right)};{\hat{\boldsymbol{m}}}_{k|k}^{\left(n\right)},{\hat{\boldsymbol{P}}}_{k|k}^{\left(n\right)}\right) \label{eq33} \\
\text{with} \nonumber \\
&{\hat{\boldsymbol{m}}}_{k|k}^{\left(n\right)}={\hat{\boldsymbol{m}}}_{k|k-1}^{\left(n\right)}+\boldsymbol{K}^{\left(n\right)}\left(\frac{\sum_{l=1}^{L_k^\theta}{q_{\boldsymbol{\theta}}\left(w_k^l\right)\phi_k^{l-\left(n\right)}{\bar{\boldsymbol{y}}}_k^{l-\left(n\right)}}}{\sum_{l=1}^{L_k^\theta}{q_{\boldsymbol{\theta}}\left(w_k^l\right)\phi_k^{l-\left(n\right)}}}-\mathbf{H}_k{\hat{\boldsymbol{m}}}_{k|k-1}^{\left(n\right)}\right) \tag{\ref{eq33}{a}}\\
&{\hat{\boldsymbol{P}}}_{k|k}^{\left(n\right)}={\hat{\boldsymbol{P}}}_{k|k-1}^{\left(n\right)}-\boldsymbol{K}^{\left(n\right)}\mathbf{H}_k{\hat{\boldsymbol{P}}}_{k|k-1}^{\left(n\right)} \tag{\ref{eq33}{b}}\\
&\boldsymbol{K}^{\left(n\right)}={\hat{\boldsymbol{P}}}_{k|k-1}^{\left(n\right)}\mathbf{H}_k^{\mathrm{T}}\left(\mathbf{H}_k{\hat{\boldsymbol{P}}}_{k|k-1}^{\left(n\right)}\mathbf{H}_k^{\mathrm{T}}+\frac{\boldsymbol{D}_k^{\left(n\right)}\mathbb{E}_{\boldsymbol{X}_k^{\left(n\right)}}\left(\boldsymbol{X}_k^{\left(n\right)}\right)\left(\boldsymbol{D}_k^{\left(n\right)}\right)^{\mathrm{T}}}{\sum_{l=1}^{L_k^{{\theta}}}{q_{\boldsymbol{\theta}}\left(w_k^l\right)\phi_k^{l-\left(n\right)}}}\right)^{-1} \tag{\ref{eq33}{c}}
\end{align}

\noindent where $\mathbb{E}_{\boldsymbol{X}_k^{\left(n\right)}}\left(\boldsymbol{X}_k^{\left(n\right)}\right)$ is the expected value of the $n$-th target's extension matrix, which can be calculated according to Eq. (\ref{eq35}). 

$\bullet$\quad For the extension matrix $\boldsymbol{X}_k^{\left(n\right)}, n=1,2,...,n_k$ in $\mathbf{{X}}_k^{1:n_k}$, its variational posterior density $q_{\boldsymbol{X}_k^{\left(n\right)}}\left(\boldsymbol{X}_k^{\left(n\right)}\right)$ can be deduced as: 
\begin{equation}
q_{\boldsymbol{X}_k^{\left(n\right)}}\left(\boldsymbol{X}_k^{\left(n\right)}\right)\sim\mathcal{IW}\left(\boldsymbol{X}_k^{\left(n\right)};{\hat{v}}_{k|k}^{\left(n\right)},{\hat{\boldsymbol{V}}}_{k|k}^{\left(n\right)}\right) \label{eq34}
\end{equation}

\noindent with
\begin{align}
&{\hat{v}}_{k|k}^{\left(n\right)}={\hat{v}}_{k|k-1}^{\left(n\right)}+\sum_{l=1}^{L_k^\theta}{q_{\boldsymbol{\theta}}\left(w_k^l\right)\phi_k^{l-\left(n\right)}} \tag{\ref{eq34}{a}}\\
&{\hat{\boldsymbol{V}}}_{k|k}^{\left(n\right)}={\hat{\boldsymbol{V}}}_{k|k-1}^{\left(n\right)}+\sum_{l=1}^{L_k^\theta}q_{\boldsymbol{\theta}}\left(w_k^l\right)\boldsymbol{T}^{l-(n)} \tag{\ref{eq34}{b}}
\end{align}

\begin{figure*}[t!]
\begin{equation}
\label{eqaddadd1}
\boldsymbol{T}^{l-(n)}=\left({\boldsymbol{D}_k^{\left(n\right)}}^{-1}\left({\bar{\boldsymbol{Y}}}_k^{l-\left(n\right)}+\phi_k^{l-\left(n\right)}\left({\bar{\boldsymbol{y}}}_k^{l-\left(n\right)}-\mathbf{H}_k{\hat{\boldsymbol{m}}}_{k|k}^{\left(n\right)}\right)\left({\bar{\boldsymbol{y}}}_k^{l-\left(n\right)}-\mathbf{H}_k{\hat{\boldsymbol{m}}}_{k|k}^{\left(n\right)}\right)^{\mathrm{T}}+\phi_k^{l-\left(n\right)}\mathbf{H}_k{\hat{\boldsymbol{P}}}_{k|k}^{\left(n\right)}\mathbf{H}_k^{\mathrm{T}}\right)\left(\boldsymbol{D}_k^{\left(n\right)}\right)^{-\mathrm{T}}\right)
\end{equation}
\end{figure*}

The calculation of the auxiliary variable $\boldsymbol{T}^{l-(n)}$ is shown in Eq. (\ref{eqaddadd1}). The expectation $\mathbb{E}_{\boldsymbol{X}_k^{\left(n\right)}}\left(\boldsymbol{X}_k^{\left(n\right)}\right)$ of the extended matrix $\boldsymbol{X}_k^{\left(n\right)}$ is:
\begin{equation}
\label{eq35}
\mathbb{E}_{\boldsymbol{X}_k^{\left(n\right)}}\left(\boldsymbol{X}_k^{\left(n\right)}\right)=\frac{{\hat{\boldsymbol{V}}}_{k|k}^{\left(n\right)}}{{\hat{v}}_{k|k}^{\left(n\right)}-2n_d-2}
\end{equation}

The derivation of Eq. (\ref{eq31}), Eq. (\ref{eq33}), and Eq. (\ref{eq34}) are given in \cite[Appendix B, C, and D]{Cheng2025arvix}.

\begin{remark}
For the iterative process of the VBI, the termination of the iteration can usually be decided by monitoring the Evidence Lower BOund (ELBO). However, for multi-target tracking where real-time performance is required, this step can be practically ignored as the computation of ELBO consumes a lot of resources \cite{2024Gan}. As an alternative, we can check convergence by monitoring changes in the statistics of the variational posterior (e.g., the posterior mean of the kinematic state), or the maximum number of iterations $\mathtt{n}_{VB}$ can be specified directly \cite{2020Zhang}. 
\end{remark}

In this study, we terminate the variational iteration by specifying $\mathtt{n}_{VB}$. The pseudo-code of the proposed method in a Bayesian cycle is summarized as Algorithm \ref{Algorithm1}. Note that we use the superscript ``$-[t]$'' to specify the $t$-th variational iteration.

\begin{algorithm}[!t]
\small
\caption{One Cycle of the Proposed Method.}
\label{Algorithm1}
\KwData{Posterior estimate of state parameters $\left \{\boldsymbol{\zeta}_{{k-1}\mid {k-1}}^{\left(n \right) }  \right \} _{n=1}^{n_{k-1}} $ at time $k-1$.}
\KwResult{Posterior estimate of state parameters $\left \{\boldsymbol{\zeta}_{k\mid k}^{\left( n \right)}  \right \} _{n=1}^{n_{k} } $ at time $k$, $n_k=n_{k-1}$.}
\hrulefill \\
\textbf{Time Update:} \\
\For{$n=1,2,\cdots,n_k$}{$\hat{\boldsymbol{m}}_{k|k-1}^{\left(n\right)}$, $\hat{\boldsymbol{P}}_{k|k-1}^{\left(n\right)}$ $\gets$ Eq. (\ref{eq20}) , ${\hat{v}}_{k|k-1}^{\left(n\right)}$,${\hat{\boldsymbol{V}}}_{k|k-1}^{\left(n\right)}$ $\gets$ Eq.(\ref{eq21}) \\
${\hat{\alpha}}_{k|k-1}^{\left(n\right)}$, ${\hat{\beta}}_{k|k-1}^{\left(n\right)}$ $\gets$ Eq. (\ref{eq22}).}
\hrulefill \\
\textbf{Measurement Update:}\\
\textrm{Initialization:}\\
\For{$n=1,2,\cdots,n_k$}{
$\hat{\boldsymbol{m}}_{k|k}^{(n)-[0]} = \hat{\boldsymbol{m}}_{k|k-1}^{(n)}$ , $\hat{\boldsymbol{P}}_{k|k}^{(n)-[0]} = \hat{\boldsymbol{P}}_{k|k-1}^{(n)}$ , ${\hat{v}}_{k|k}^{(n)-[0]}={\hat{v}}_{k|k-1}^{(n)}$, \\
${\hat{\boldsymbol{V}}}_{k|k}^{(n)-[0]} = {\hat{\boldsymbol{V}}}_{k|k-1}^{(n)}$ , $\hat{\alpha}_{k|k}^{(n)-[0]} = \hat{\alpha}_{k|k-1}^{(n)}$ , $\hat{\beta}_{k|k}^{(n)-[0]} = \hat{\beta}_{k|k-1}^{(n)}$.
}
\textrm{Iterations:} \\
\While{$t\le \mathtt{n}_{VB}$}{
$q_{\boldsymbol{\Theta}}^{[t]}\left(\boldsymbol{\Theta}_k^{1:L_k^\theta}\right)$ $\gets$ Eq. (\ref{eq28})\\
\For{$n=1,2,\cdots,n_k$}{
 $\hat{\boldsymbol{m}}_{k|k}^{(n)-[t]}$,$\hat{\boldsymbol{P}}_{k|k}^{(n)-[t]}$ $\gets$ Eq.  (\ref{eq33}), ${\hat{v}}_{k|k}^{(n)-[t]}$, ${\hat{\boldsymbol{V}}}_{k|k}^{(n)-[t]}$ $\gets$ Eq. (\ref{eq34}), \\
 $\hat{\alpha}_{k|k}^{(n)-[t]}$, $\hat{\beta}_{k|k}^{(n)-[t]}$ $\gets$ Eq. (\ref{eq31}), \\
$\mathbb{E}_{\lambda_k^{\left(n\right)}}^{[t]}\left(\lambda_k^{\left(n\right)}\right)$ $\gets$ Eq. (\ref{eq32}).
}
}

\textrm{Posterior Estimate:}
\For{$n=1,2,\cdots,n_k$}{
$\hat{\boldsymbol{m}}_{k|k}^{(n)} = \hat{\boldsymbol{m}}_{k|k}^{(n)-[\mathtt{n}_{VB}]}$ , $\hat{\boldsymbol{P}}_{k|k}^{(n)} = \hat{\boldsymbol{P}}_{k|k}^{(n)-[\mathtt{n}_{VB}]}$ , ${\hat{v}}_{k|k}^{(n)}={\hat{v}}_{k|k}^{(n)-[\mathtt{n}_{VB}]}$, \\
${\hat{\boldsymbol{V}}}_{k|k}^{(n)} = {\hat{\boldsymbol{V}}}_{k|k}^{(n)-[\mathtt{n}_{VB}]}$ , $\hat{\alpha}_{k|k}^{(n)} = \hat{\alpha}_{k|k}^{(n)-[\mathtt{n}_{VB}]}$ , $\hat{\beta}_{k|k}^{(n)} = \hat{\beta}_{k|k}^{(n)-[\mathtt{n}_{VB}]}$.
}

\end{algorithm}

\section{LIGHTWEIGHT SCHEMES}
\label{sec5}
Notice that computing $q_{\boldsymbol{\theta}}\left(w_k^l\right)$ requires enumerating all feasible JAEs. In multi-target tracking with $m$ targets, a feasible JAE implies dividing the measurement set $\mathcal{Y}=\left\{\boldsymbol{y}^j\right\}_{j=1}^{m}$ containing $n$ measurements into $n+1$ subsets. The total number of partitions for a set with $m$ elements is the result of the $m$-th Bell number \cite{r1} that increases exponentially with increasing $m$. Therefore, enumerating all JAEs is infeasible for real‑time tracking, motivating computationally lightweight variants for practical use.

\subsection{Lightweight Scheme {\uppercase\expandafter{\romannumeral1}}: Gating and Clustering}
\label{sec5-1}
Based on the property that clutter is sparsely distributed in the surveillance area, the gating technique can be used to remove some of the clutter. By reducing the number of unknown measurements, the total number of feasible JAEs can be pruned. Since the kinematic state of the target is assumed to be a Gaussian distribution, an elliptic gating can be employed. The validation region of the $n$-th target is denoted as:
\begin{equation}
\mathcal{T}_k^{\left(n\right)}=\left\{{\dot{\boldsymbol{y}}}_k^{\left(n\right)};\left({\dot{\boldsymbol{y}}}_k^{\left(n\right)}-\mathbf{H}_k{\hat{\boldsymbol{m}}}_{k|k-1}^{\left(n\right)}\right)^{\mathrm{T}}{\boldsymbol{S}_k^{\left(n\right)}}^{-1}\left({\dot{\boldsymbol{y}}}_k^{\left(n\right)}-\mathbf{H}_k{\hat{\boldsymbol{m}}}_{k|k-1}^{\left(n\right)}\right)\le \mathtt{g}^2\right\}
\end{equation}

\noindent where the constant $\mathtt{g}$ is the threshold parameter, and the innovation covariance $\boldsymbol{S}_k^{\left(n\right)}$ for the $n$-th target is given by:
\begin{equation}
\label{eq37}
\boldsymbol{S}_k^{\left(n\right)}=\mathbf{H}_k\hat{\boldsymbol{P}}_{k|k-1}^{\left(n\right)}\mathbf{H}_k^{\mathrm{T}}+\boldsymbol{D}_k^{\left(n\right)}{\bar{\boldsymbol{X}}}_{k|k-1}^{\left(n\right)}\left(\boldsymbol{D}_k^{\left(n\right)}\right)^{\mathrm{T}}
\end{equation}

A measurement $\boldsymbol{y}_k^j$ is valid if it falls within the validation region of any target. Measurements in $\mathcal{Y}_k$ that are not valid are treated as clutter\footnote{For the RFS-based method, these invalid measurements are still retained and used to judge the newborn targets.}. The gating operation relies on a human-specified threshold which cannot exclude all the clutter. There is also a need to partition the remaining valid measurements to obtain multiple JAEs, which is still time-consuming. Fortunately, the measurements of an individual target are clustered in their spatial distribution. Based on this, the clustering methods can be used to further reduce the number of partitioning schemes.

The lightweight scheme combining gating and clustering is widely integrated in multi-target tracking applications \cite{2013Lundquist,2023Fowdur}. This lightweight processing reduces the original exponential complexity to quadratic complexity. When employing traditional clustering algorithms such as K-Means or density-based spatial clustering, the algorithm exhibits a computational complexity of $\mathcal{O}\left(n_k^2m_k^2\right)$ \cite{2024Xue}.

However, the performance of traditional clustering algorithms is heavily dependent on the prespecified algorithm parameters. This means that when two targets are close to each other, these methods cannot effectively divide the measurements that originate from them into two clusters. Of course, more clustering schemes can be obtained by specifying multiple different algorithm parameters, but it is still problematic as it results in a higher computational cost. Therefore, another feasible lightweight scheme is proposed. It is based on the approximation of the posterior density by the marginal association probability between the measurements and the targets without enumeration of measurement partitions.

\subsection{Lightweight Scheme {\uppercase\expandafter{\romannumeral2}}: Marginal Association Probability}
\label{sec5-2}
Consider a measurement $\boldsymbol{y}_k^j,j=1,...,m_k$, in the set $\mathcal{Y}_k$ at time $k$. 
Let $\epsilon_{\left(n\right)j}$ represent the marginal association probability that measurement $\boldsymbol{y}_k^j$ belongs to the $n$-th target. Then
Eq. (\ref{eq31}), Eq. (\ref{eq33}), and Eq. (\ref{eq34}) can be approximated as follows.

The parameters of Gaussian distribution:
\begin{subequations}
\label{eq38}
\begin{align}
&{\hat{\boldsymbol{m}}}_{k|k}^{\left(n\right)}\approx{\hat{\boldsymbol{m}}}_{k|k}^{\ast\left(n\right)}={\hat{\boldsymbol{m}}}_{k|k-1}^{\ast\left(n\right)}+\boldsymbol{K}^{\ast\left(n\right)}\left(\frac{\sum_{j=1}^{{m}_k}{\epsilon_{\left(n\right)j}\boldsymbol{y}_k^j}}{\sum_{j=1}^{m_k}\epsilon_{\left(n\right)j}}-\mathbf{H}_k{\hat{\boldsymbol{m}}}_{k|k-1}^{\ast\left(n\right)}\right) \\
&{\hat{\boldsymbol{P}}}_{k|k}^{\left(n\right)}\approx{\hat{\boldsymbol{P}}}_{k|k}^{\ast\left(n\right)}={\hat{\boldsymbol{P}}}_{k|k-1}^{\ast\left(n\right)}-\boldsymbol{K}^{\ast\left(n\right)}\mathbf{H}_k{\hat{\boldsymbol{P}}}_{k|k-1}^{\ast\left(n\right)} \\
& \boldsymbol{K}^{\ast\left(n\right)}={\hat{\boldsymbol{P}}}_{k|k-1}^{\ast\left(n\right)}\mathbf{H}_k^{\mathrm{T}}\left(\mathbf{H}_k{\hat{\boldsymbol{P}}}_{k|k-1}^{\ast\left(n\right)}\mathbf{H}_k^{\mathrm{T}}+\frac{\boldsymbol{D}_k^{\left(n\right)}{\hat{\boldsymbol{V}}}_{k|k}^{\ast\left(n\right)}\left(\boldsymbol{D}_k^{\left(n\right)}\right)^{\mathrm{T}}}{\sum_{j=1}^{m_k}\epsilon_{\left(n\right)j}\left({\hat{v}}_{k|k}^{\ast\left(n\right)}-2n_d-2\right)}\right)^{-1} 
\end{align}
\end{subequations}

The parameters of inverse Wishart distribution:
\begin{subequations}
\label{38-2}
\begin{align}
& {\hat{v}}_{k|k}^{\left(n\right)}\approx{\hat{v}}_{k|k}^{\ast\left(n\right)}={\hat{v}}_{k|k-1}^{\ast\left(n\right)}+\sum_{j=1}^{m_k}\epsilon_{\left(n\right)j} \\
& {\hat{\boldsymbol{V}}}_{k|k}^{\left(n\right)}\approx{\hat{\boldsymbol{V}}}_{k|k}^{\ast\left(n\right)}={\hat{\boldsymbol{V}}}_{k|k-1}^{\ast\left(n\right)}+\sum_{j=1}^{m_k}\epsilon_{\left(n\right)j}\boldsymbol{U}^{\left(n\right)j} \\
&
\boldsymbol{U}^{\left(n\right)j} \nonumber\\
& =\left({\boldsymbol{D}_k^{\left(n\right)}}^{-1}\left(\left(\boldsymbol{y}_k^j-\mathbf{H}_k{\hat{\boldsymbol{m}}}_{k|k}^{\left(n\right)}\right)\left(\boldsymbol{y}_k^j-\mathbf{H}_k{\hat{\boldsymbol{m}}}_{k|k}^{\left(n\right)}\right)^{\mathrm{T}}+\mathbf{H}_k{\hat{\boldsymbol{P}}}_{k|k}^{\left(n\right)}\mathbf{H}_k^{\mathrm{T}}\right)\left(\boldsymbol{D}_k^{\left(n\right)}\right)^{-\mathrm{T}}\right)
\end{align}
\end{subequations}

The parameters of gamma distribution:
\begin{subequations}
\begin{align}
&{\hat{\alpha}}_{k|k}^{\left(n\right)}\approx{\hat{\alpha}}_{k|k}^{\ast\left(n\right)}={\hat{\alpha}}_{k|k-1}^{\ast\left(n\right)}+\sum_{j=1}^{m_k}\epsilon_{\left(n\right)j}  \\
&{\hat{\beta}}_{k|k}^{\left(n\right)}\approx{\hat{\beta}}_{k|k}^{\ast\left(n\right)}={\hat{\beta}}_{k|k-1}^{\ast\left(n\right)}+1 
\end{align}
\end{subequations}

\noindent where the symbol `$\ast$' in the superscript represents an approximation. The marginal probability $\epsilon_{\left(n\right)j}$ can be significantly simplified in a similar way as the LMIPDA \cite{2015Baum}, i.e., 

\begin{equation}
\label{eq39}
\epsilon_{\left(n\right)j}\propto\frac{\mathbb{E}_{\lambda_k^{\left(n\right)}}\left(\lambda_k^{\left(n\right)}\right)\mathcal{L}_{\left(n\right)j}}{\left[\sum_{i=1}^{n_k}{\mathbb{E}_{\lambda_k^{\left(i\right)}}\left(\lambda_k^{\left(i\right)}\right)\mathcal{L}_{\left(i\right)j}}\right]+\rho\cdot\lambda_c}
\end{equation}

\noindent where $\mathcal{L}_{\left(n\right)j}=p\left(\boldsymbol{y}_k^j\mid\boldsymbol{\xi}_k^{\left(n\right)}\right)$ is the likelihood of the measurement $\boldsymbol{y}_k^j$ on the $n$-th target and $\rho$ is the clutter density presented in Eq. (\ref{new_eqadd5}). The calculation of $\mathbb{E}_{\lambda_k^{\left(n\right)}}\left(\lambda_k^{\left(n\right)}\right)$ can be referred to Eq. (\ref{eq32}). In \cite[Appendix E]{Cheng2025arvix}, we provide a brief explanation of the feasibility of this lightweight approach in Eq. (\ref{eq38}). 

It can be seen from Eq. (\ref{eq39}) that the calculation of the marginal association probability is linear in the number of objects and linear in the number of measurements. Therefore, after adopting this lightweight strategy, the overall complexity becomes $n_k\ast\mathcal{O}\left(n_km_k\right)\sim\mathcal{O}\left(n_k^2m_k\right)$. Compared to the $\mathcal{O}\left(n_k^2m_k^2\right)$ complexity of the clustering given in Section \ref{sec5-1}, it is significantly smaller. 

\begin{table*}
\centering
\caption{Time Complexity Comparison: Baseline vs Lightweight Scheme 1/2 ($\mathcal{O}$)}
\label{TimeComplex}
\begin{tabular}{c c cccc c}
\hline
\multirow{2}{*}{Methods} &  \multirow{2}{3cm}{\centering Time update} & \multicolumn{4}{c}{Measurement update} &  \multirow{2}{3cm}{\centering Overall complexity}  \\
\cline{3-6}
&  & Gating & Clustering & Data association &  Variational inference &  \\
\hline
RM-VB (non-lightweight)     & $\mathcal{O}(n_k)$ & -- & -- & $\mathcal{O}(e^{n_k}\cdot m_k^{n_k})$ & $\mathcal{O}(n_km_k)$ & $\mathcal{O}(e^{n_k}\cdot m_k^{n_k})$  \\
RM-VB (lightweight scheme 1) & $\mathcal{O}(n_k)$ & $\mathcal{O}(n_km_k)$  & $\mathcal{O}(m_k^2)$ & $\mathcal{O}(n_k^2m_k^2)$ & $\mathcal{O}(n_km_k)$ & $\mathcal{O}(n_k^2m_k^2)$ \\
RM-VB (lightweight scheme 2) & $\mathcal{O}(n_k)$ & -- & -- & $\mathcal{O}\left(n_k^2m_k\right)$ & $\mathcal{O}(n_km_k)$ & $\mathcal{O}\left(n_k^2m_k\right)$ \\
\hline
\end{tabular}
\end{table*}

Table \ref{TimeComplex} summarizes a detailed time-complexity analysis of our method across three versions: without a lightweight scheme, Lightweight Scheme 1, and Lightweight Scheme 2. For each version, we provide the time complexities of the time update and measurement update stages, and we conclude with a brief summary of the overall complexity of each version. It can be seen that both lightweight schemes reduce the method’s original exponential complexity to an acceptable level; however, the reduction in computational complexity achieved by Scheme 2 is significant. There is no doubt that such a lightweight operation loses information about the number of measurements $\phi$ assigned to each target under each JAE. The loss of information will lead to some errors in the shape estimation of the target \cite{2023LANnew}, but this lightweight scheme reduces the original complexity by an order of magnitude, so it is acceptable.

\subsection{Discussion}
\label{discuss}
In this section, we discuss our proposed method and analyze avenues for future improvement. Based on its characteristics, Table \ref{comparsionTable} presents a comparison between our method and existing data-association-based approaches and RFS-based approaches. The Table \ref{comparsionTable} shows that our method achieves accuracy close to that of data association–based methods at lower computational cost and with greater applicability, while avoiding the reliance on auxiliary trajectory generation required by RFS-based approaches. However, our method assumes a fixed number of targets and does not handle target birth and death. It also treats the detection probability as constant,  which may deviate from practice. Incorporating the detection and existence probabilities of targets as unknown parameters into our variational inference framework could further broaden its applicability. Relevant extension schemes are outlined in \cite{2023Yang}\cite{2025Gan}.

\begin{table*}[!b]
\centering
\caption{Comparative Overview: Our Method vs. Existing Tracking Methods}
\footnotesize
\setlength{\tabcolsep}{4pt}
\renewcommand{\arraystretch}{1.2}
\begin{tabular}{C{0.15\linewidth} C{0.15\linewidth} C{0.10\linewidth} C{0.15\linewidth} C{0.10\linewidth} C{0.08\linewidth}  C{0.08\linewidth} C{0.10\linewidth} C{0.10\linewidth}}
\hline
Method & Representative methods & Efficiency & Need auxiliary methods to generate trajectories & Accuracy & Target count & Detection probability  & Measurement rate  \\
\hline
Our method & RM-VB-Marginal & Medium & No & Medium & Fixed & Fixed & Variable \\
\hline
Data-association-based & RM-JPDA \cite{2015Schuster} & Low & No & High & Fixed & Fixed & Fixed \\
\hline
RFS-based & RM-PHD \cite{2012Granstrom} & Relatively high & Yes & Medium & Variable & Variable & Variable \\
\hline
\end{tabular}
\label{comparsionTable}
\end{table*}

In addition, the measurement model used in this paper (see Eq. (\ref{eqadd4})) imposes constraints on target-generated measurements. Specifically, the spatial extent of measurements is constrained to a Gaussian distribution centered at the target, and the measurement count is constrained to follow a Poisson distribution. In practice, however, spatial measurement patterns are often nonuniform—measurements may lie along the target’s boundary or concentrate on one side—and the count is often not purely random, typically depending on target size. To address such cases, existing methods take two approaches: on the one hand, they seek more suitable distributional models, such as using a skewed Gaussian \cite{2021Zhang} or a shape-based Gaussian mixture \cite{2025WenNew} for spatial patterns, and adopting a Gamma-like distribution \cite{2023LANnew} for the measurement count; on the other hand, they impose control or constraint conditions on measurements \cite{2025SBCNew}, as we do in this paper. While the latter yields simplicity and computational efficiency, it may limit applicability in more realistic scenarios. Therefore, combining our approach with the aforementioned distributional models is a promising direction for future research.

Note that this work addresses probabilistic estimation for extended and indistinguishable group targets. We do not synthesize a feedback controller, and no control inputs are considered; motion models are used as stochastic priors for the filtering recursion. A full closed‑loop stability analysis or controller design is beyond the scope of this estimation‑focused study and is left for future work on active sensing or joint estimation‑control.

\section{EXPERIMENTAL RESULTS}
\label{sec6}
\subsection{Methods for Comparison}
\label{sec6-1}
In this section, we compare our method with other state-of-the-art methods in simulation experiments to demonstrate the superior performance of our method. The methods involved in the comparison are:
\begin{itemize}
\item MEM-LJPDA: the MEM with Linear-time JPDA (LJPDA) for tracking proposed in \cite{2018Yang_LJPDA}.
\item RM-JPDA: the RMM with traditional JPDA for tracking proposed in \cite{2015Schuster}.
\item RM-T-PMB: the RMM with RFS-based PMBM and trajectory set \cite{2019Xia_T}, and we adopt the PMB \cite{2022Xia} approximation\footnote{MATLAB code can be found in https://github.com/yuhsuansia. Thanks to the authors for their contributions.}.
\item RM-VB-DBSCAN (proposed method): the RMM with the VBI for tracking, and use the lightweight scheme based on gating and DB-SCAN \cite{2020Deng} clustering described in Section \ref{sec5-1}.
\item RM-VB-Marginal (proposed method): the RMM with the VBI for tracking and use the lightweight scheme based on the marginal association probability given in Section \ref{sec5-2}.
\end{itemize}

\subsection{Metrics}
\label{sec6-2}
For performance evaluation of targets estimated with an ellipsoidal extent, \cite{2016Yang} has shown a reasonable metric named the Gaussian Wasserstein distance (GWD). Define the GWD between two ellipses $\boldsymbol{\varrho}_k^1=\left\{\boldsymbol{x}_k^1,\boldsymbol{X}_k^1\right\}$ and $\boldsymbol{\varrho}_k^2=\left\{\boldsymbol{x}_k^2,\boldsymbol{X}_k^2\right\}$ as:
\begin{equation}
\begin{split}
&d_{GWD}\left(\boldsymbol{\varrho}_k^1,\boldsymbol{\varrho}_k^2\right)\\ &=\sqrt{\left\Vert\mathbf{H}\left(\boldsymbol{x}_k^1-\boldsymbol{x}_k^2\right)\right\Vert^2+\mathrm{tr}\left(\boldsymbol{X}_k^1+\boldsymbol{X}_k^2-2\sqrt{\left(\sqrt{\boldsymbol{X}_k^1}\boldsymbol{X}_k^2\sqrt{\boldsymbol{X}_k^1}\right)}\right)}
\end{split}
\end{equation}

\noindent where $\boldsymbol{\varrho}_1$ is the actual value of the elliptic parameter and $\boldsymbol{\varrho}_2$ is the estimated value of the elliptic parameter obtained using the tracking method. Symbol `$\boldsymbol{x}$’ denotes the kinematic state of the center of the ellipse and symbol `$\boldsymbol{X}$’ represents the extended shape of the ellipse.

Furthermore, to evaluate the performance of the method on the estimation of the center position of the ellipse and the extended shape, we compute the Root-Mean-Square Error (RMSE) of them at the time $k$, i.e.,
\begin{equation}
{\rm RMSE}_{\mathrm{Pos}}(k) = \sqrt{\frac{1}{\mathtt{m}_{\mathrm{MC}}}\sum_{s=1}^{\mathtt{m}_{\mathrm{MC}}}{\left\Vert\mathbf{H}\left(\boldsymbol{x}_k^1-\boldsymbol{x}_k^{2,s}\right)\right\Vert^2}}
\end{equation}
\begin{equation}
{\rm RMSE}_\mathrm{Ext}(k)=\sqrt{\frac{1}{\mathtt{m}_{\mathrm{MC}}}\sum_{s=1}^{\mathtt{m}_{\mathrm{MC}}}\left({\mathrm{tr}\left[\boldsymbol{X}_k^1-\boldsymbol{X}_k^{2,s}\right]}\right)^2}
\end{equation}

\noindent where $\mathtt{m}_{\mathrm{MC}}$ denotes the number of Monte Carlo (MC) runs. $\boldsymbol{x}_k^{2,s}$ and $\boldsymbol{X}_k^{2,s}$ represent the estimated position and extended matrix of a single target at time $k$ of the $s$-th MC run, respectively. Furthermore, we will calculate the mean and variance of the above RMSE over the time. For the evaluated state `$\star$’, the mean and variance of its RMSE over time are defined as:
\begin{equation}
T_{Mean}\left({\rm RMSE}_\star\right)=\frac{1}{\mathtt{K}}\sum_{k=1}^{\mathtt{K}}{{\mathrm{RMSE}}_\star(k)}
\end{equation}
\begin{equation}
T_{Cov}\left({\rm RMSE}_\star\right)=\frac{1}{\mathtt{K}}\sum_{k=1}^{\mathtt{K}}\left({\rm RMSE}_\star(k)-T_{Mean}\left({\rm RMSE}_\star\right)\right)^2
\end{equation}

\noindent where $\mathtt{K}$ is the number of time steps.

\subsection{The Simulation Scenario}
\label{section6_3}
Consider a 2D simulation scene of size $[-50\mathrm{m}, 650\mathrm{m}]\times[-320\mathrm{m}, 320\mathrm{m}]$, where the simulation process consists of 60 time steps, with a time interval of $\mathtt{T}$=1s. At the initial time, there are two targets of different sizes. We refer to the two targets as Tar. 1 and Tar. 2, respectively. After starting the simulation, these two targets gradually approach with the same speed in the period of 1s$\sim$30s. Then they cross at the center of the scene at the 30s with the crossing process lasting for 3s, i.e., from 30s to 33s. Thereafter, these two targets gradually move away from each other from 33s to 60s until the end of the simulation. More details about these two targets are shown in Table \ref{table2}.
\begin{table}[ht]
    \centering
    \caption{Details of Tar. 1 and Tar. 2} 
    \label{table2}
    \begin{tabular}{|c|c|c|c|c|c|}
        \hline
         & Initial position (m) & Velocity (m/s) & Size (m) & Orientation (rad) \\
        \hline
        Tar. 1 & $[0, -300]$ & $[11, 7.7]$  & $[60, 30]$ & $-\pi/3$\\
        \hline
        Tar. 2 & $[0, 300]$ & $[11, -7.7]$ & $[40, 20]$ & $-\pi/4$\\
        \hline
    \end{tabular}
\end{table}

We show the Ground Truth (GT) of each target in the simulation scenario in Fig. \ref{fig3}:
\begin{figure}[!htp]
\centering                                         
\includegraphics[width=3.5in]{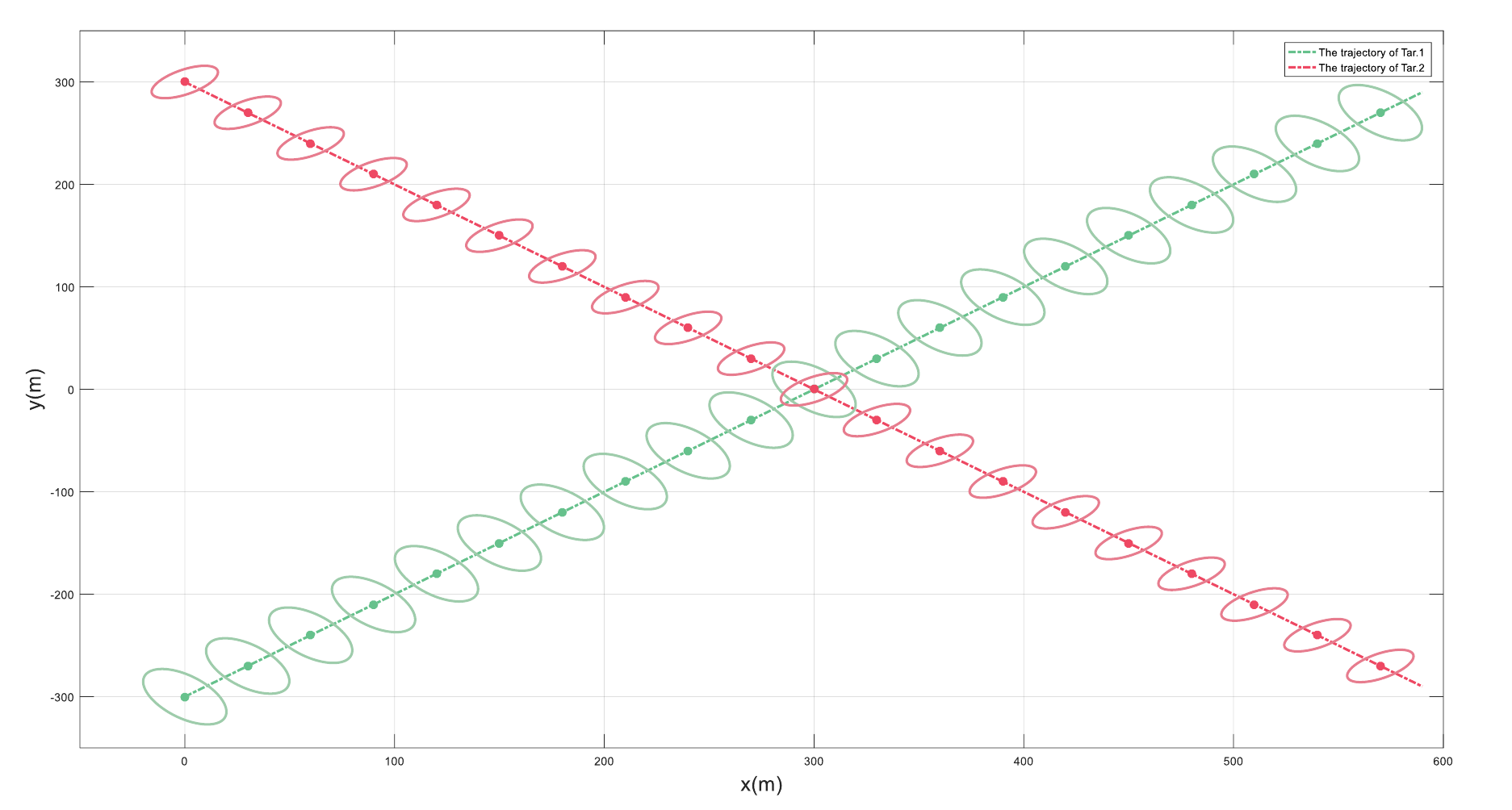}     
\caption{Ground truth of two targets in the simulation scenario. (\small{The circle shows the extended shape of the targets. We use the green dashed line to mark the trajectory of Tar. 1, and the red dashed line to mark the trajectory of Tar. 2. The points on the trajectory specify the position of the targets' center})} 
\label{fig3}                                     
\end{figure}

Other simulation parameters are set as follows:
\begin{itemize}
\item Assuming that both targets obey the most common uniform linear motion, the state transition matrix in Eq. (\ref{eq2}) is set as:
\begin{equation}
\mathbf{\Phi}_k=\left[\begin{matrix}1&\mathtt{T}\\0&1\\\end{matrix}\right]\otimes \mathbf{I}_2\ \forall k
\end{equation}
The measurement matrix in Eq. (\ref{eq4}) is set as:
\begin{equation}
\mathbf{H}_k=\left[\begin{matrix} 1&0&0&0\\0&0&1&0\end{matrix}\right]\ ,\forall k
\end{equation}
The process noise and measurement noise are assumed to be time-invariant and satisfied:
\begin{equation}
\mathbf{G}_k=\mathrm{diag}\left([1,1,0.1,0.1]\right); \mathbf{R}_k=\mathrm{diag}\left([0.01, 0.01]\right)
\end{equation}
\item 	Assuming that the clutter obeys a homogeneous PPP with Poisson parameters $\lambda_c=25$ and is uniformly distributed in the scene. The target generation measure obeys a non-homogeneous PPP, whose measurement rate is assumed to be $\lambda_t=20$. The detection probability is set as $P_d=0.98$ that the target is not completely occluded. Of course, the value of $\lambda_t$ is only used for generating the measurements in simulation, when initializing the measurement rate, we assume that it is unknown and artificially specified as another parameter. The two parameters $\lambda_t$ and $\lambda_c$ can be combined as a set of adjustable parameters to simulate different detection situations in the scenario. Table \ref{tab4new} summarizes the key parameters used in this study - process noise covariance $\mathbf{G}$, measurement noise covariance $\mathbf{R}$, clutter Poisson rate $\lambda_c$, and target measurement Poisson rate $\lambda_t$—along with their dependencies, empirical settings, and subsequent adjustment strategies.  The table provides a unified reference for experimental configuration and replication, facilitating quick comparison of the effects of different parameter choices.

\begin{table*}[htbp]
\centering
\caption{Summary of Key Parameters and Empirical Settings.}
\footnotesize
\setlength{\tabcolsep}{4pt}
\renewcommand{\arraystretch}{1.2}
\begin{tabular}{C{0.18\linewidth} C{0.26\linewidth} C{0.26\linewidth} C{0.26\linewidth}}
\hline
Parameter & Dependencies & Empirical settings & Adjustment strategy \\
\hline

Process noise covariance $\mathbf{G}$ & Dynamics model, sampling period $\mathtt{T}$, target maneuverability & Derive $\mathbf{G}$ from the spectral density of typical maneuver acceleration according to the model and $\mathtt{T}$ & Increase: rapid accelerations, sharp turns, or vibration cause response lag; large $\mathtt{T}$ or unmodeled disturbances due to wind and road irregularities. Decrease: track jitter and unstable velocity estimates; scene is stable, constant-velocity motion dominates, and short-term predictions are reliable. \\
\hline
Measurement noise covariance $\mathbf{R}$ & Sensor resolution, calibration error, Signal-to-Noise Ratio (SNR), measurement type & Diagonal matrix $\mathbf{R}$ built from the standard deviation of each sensor channel & Increase: longer range/lower SNR/harsher environment make measurements less accurate. Decrease: near range/high SNR/good calibration. \\
\hline
Clutter Poisson rate $\lambda_c$ (per frame) & Sensor type and detection threshold, environmental complexity, field of view (FOV), association gate volume ${\rm V}_{\rm gate}$ & Count clutter in target-free segments $n_c$ →  $\lambda_c = n_c / {\rm FOV} \cdot {\rm V}_{\rm gate}$ & Increase: rain/fog, multipath, strong reflections, or crowded backgrounds raise false detection density. Decrease: environment is clean or higher threshold markedly reduces false alarms. \\
\hline
Target measurement Poisson rate $\lambda_t$ (per target per frame) & Target size, resolution cell size, range and SNR & Use labeled data to estimate the per-target, per-frame mean number of valid measurements & Increase: larger/closer targets or higher SNR produce more measurements per target. Decrease: smaller/farther targets or lower SNR, fewer measurements per frame, or near point-like targets. \\
\hline
\end{tabular}
\label{tab4new}
\end{table*}

\item The parameter settings for each method under comparison can be summarized as follows:
\begin{enumerate}
\item[$\blacktriangleright$] For all methods, the initial kinematic states of these two targets are set to be consistent, and their measurement rates are assumed to be 80.
\item[$\blacktriangleright$] For MEM-LJPDA, The initial shape states of both targets are set to [orientation, semi-minor axis length, semi-major axis length]=[0rad,20m,20m].
\item[$\blacktriangleright$] For RM-JPDA and RM-T-PMB, we set the initial extended matrix of the two targets as ${\hat{v}}_{0|0}^{\left(1\right)}={\hat{v}}_{0|0}^{\left(2\right)}=7$, ${\hat{\boldsymbol{V}}}_{0|0}^{\left(1\right)}={\hat{\boldsymbol{V}}}_{0|0}^{\left(2\right)}=\mathrm{diag}([400,400])$, matching MEM‑LJPDA’s initial extent.
\item[$\blacktriangleright$] For our methods RM-VB-DBSCAN and RM-VB-Marginal. For the former, the DB-SCAN with different distance thresholds between 0.5 and 10 is applied. For the latter, we set the maximum number of VB iterations to $\mathtt{n}_{VB}=10$. And their parameters of the initialized extended matrix are consistent with RM-JPDA and RM-T-PMB.

\item[$\blacktriangleright$] We perform 100 MC runs on each tracking method\footnote{All methods are implemented in MATLAB R2016a with Intel (R) Core (TM) i9-10900 CPU and 16 GB RAM.}.
\end{enumerate}
\end{itemize}

\begin{figure}[!htp]
\centering                                         
\includegraphics[width=3in]{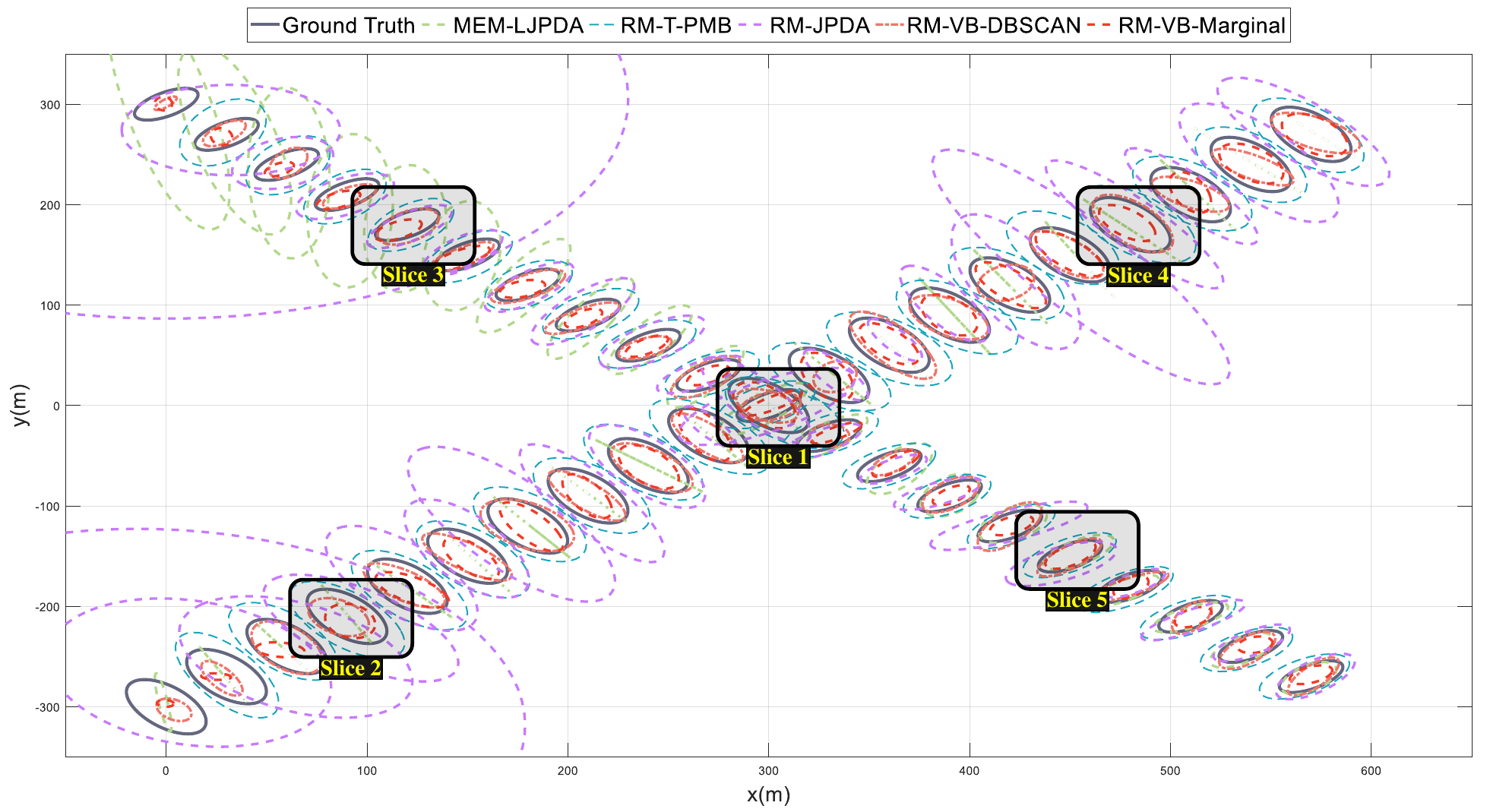}     
\caption{An example MC run of the simulation scenario.}
\label{figS_1}                                     
\end{figure}

\normalsize
Fig. \ref{figS_1} shows a typical MC run for the simulation scenario with the compared methods (plotted in every 3 time steps). The trajectories and contours of the targets are shown as thicker lines to facilitate comparison. In addition, several representative time-step slices are selected and shown in Fig. \ref{figS_2} in an enlarged view. It can be roughly seen from Fig. \ref{figS_2} that although each method can effectively track the two targets, there are differences in detail:
\begin{itemize}
\item For MEM-LJPDA, it suffers from an `estimation collapse' in the shape estimation of Tar. 1 (Slice-2 of Fig. \ref{figS_2}), which results in the collapse of the elliptical contour into a straight line, which is a defect of MEM. Although this phenomenon only occurs in a very limited number of MC runs, it must be emphasized that the effect of this phenomenon on the accuracy of shape estimation is fatal even if it occurs just once. Meanwhile, since MEM-LJPDA lacks the modeling of measurement rate, its convergence to the shape estimation is slower than other methods when the initialized measurement rate is different from the actual measurement rate (Slice-3 of Fig. \ref{figS_2}).
\item For RM-T-PMB and RM-JPDA, their shortcomings seem to be the same, that is, inaccurate estimation of the target shape, since they are based on the original RMM, which captures the change in the shape of the targets by the forgetting factor. Therefore, the accuracy of their shape estimation is highly dependent on the initialization of the extended matrix. As in the simulation, we initialized the shape matrix for RM-T-PMB and RM-JPDA with extended matrices larger and smaller than the real size, respectively. As a result, the shape estimation of RM-T-PMB is larger than the real shape while that of RM-JPDA is smaller.
\item Both of our proposed methods, RM-VB-DBSCAN and RM-VB-Marginal, have superior tracking performance. Among them, RM-VB-DBSCAN has the most accurate tracking results because it introduces a shape evolution model and uses the VBI to estimate the state parameters. RM-VB-Marginal is slightly less accurate in shape estimation than RM-VB-DBSCAN, as we analyzed in Section \ref{sec5-2}, which is due to neglecting of information about the number of measurements produced by the target.
\end{itemize}

\begin{figure}[!htp]
\centering                                         
\includegraphics[width=3in]{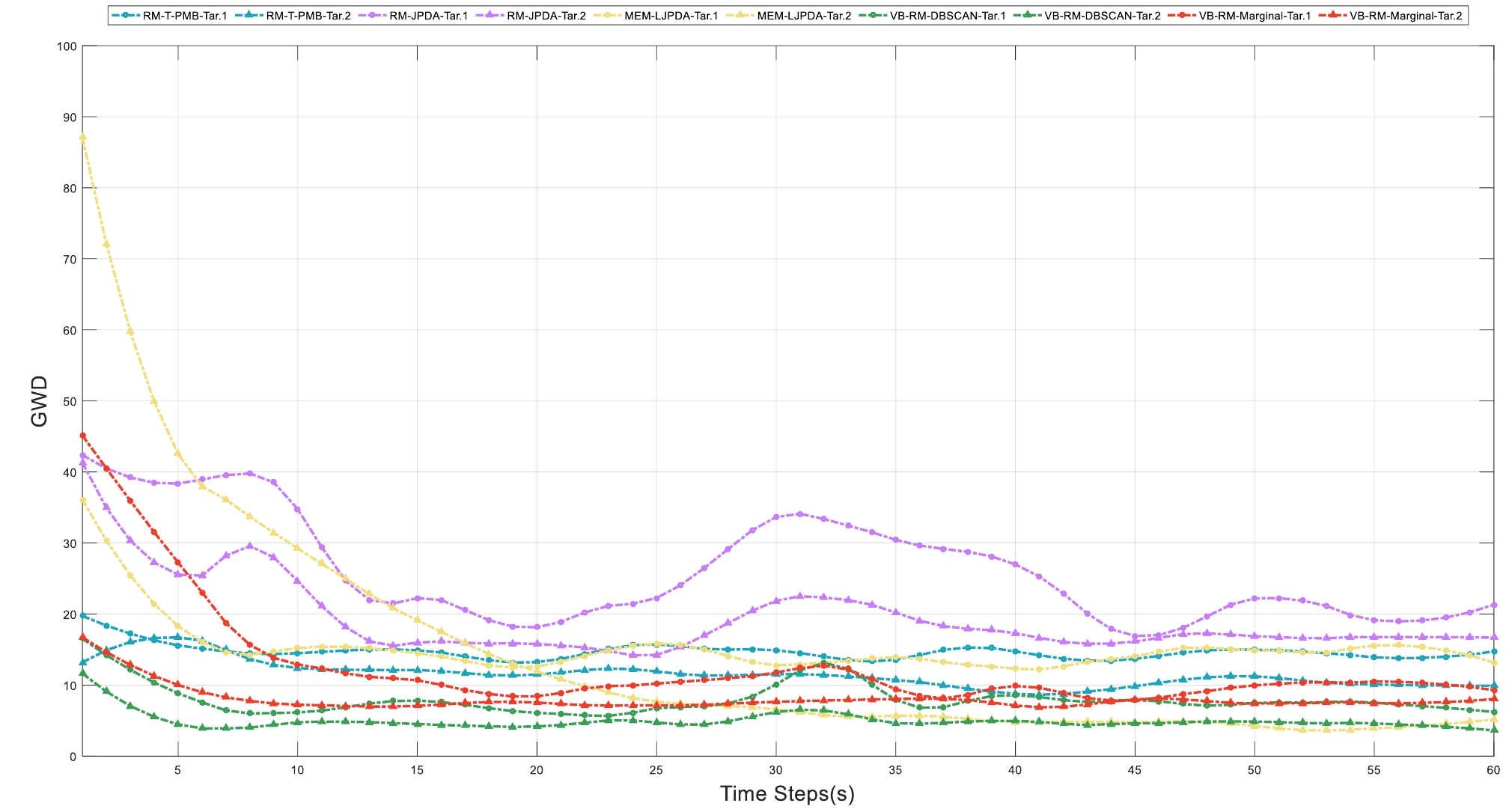}   

\caption{Simulation results of GWD. (\small{We use different colors to distinguish methods and different markers to distinguish targets. The 
 marker of Tar. 1 is `$\bullet$', and the marker of Tar. 2 is `$\blacktriangle$'}) }
\label{figS_3}                                     
\end{figure}

\begin{figure}[!htp]
\centering                                         
\includegraphics[width=3in]{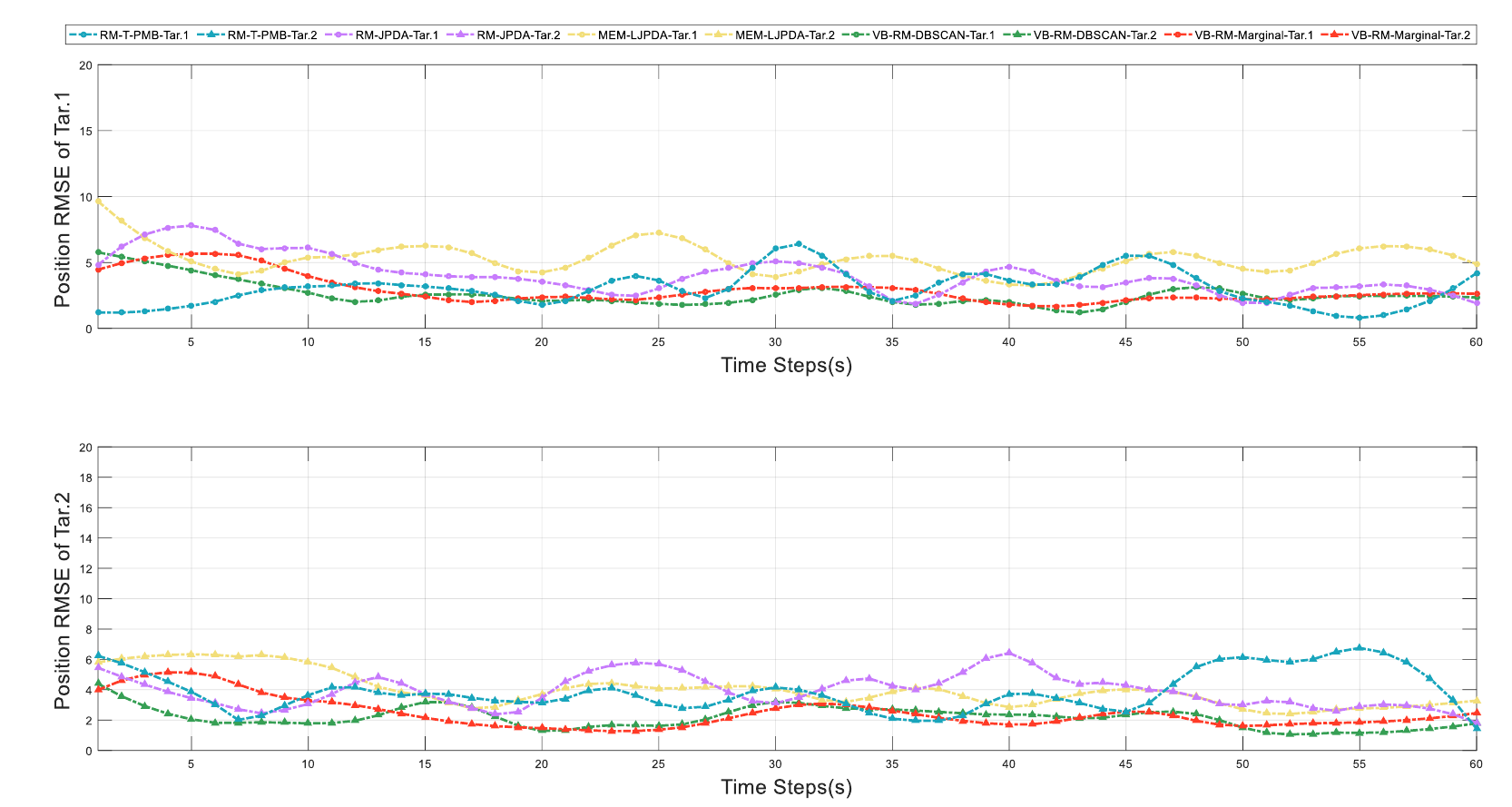}    
\caption{Simulation results of ${\rm RMSE}_{\mathrm{Pos}}$. (\small{The upper subfigure corresponds to Tar. 1, while the lower subfigure corresponds to Tar. 2})}
\label{figS_4}                                     
\end{figure}

We plot the acquired metrics, i.e., GWD (averaged over 100 MC runs), ${\rm RMSE}_{\mathrm{Pos}}$, and ${\rm RMSE}_{\mathrm{Ext}}$, in Fig. \ref{figS_3}, Fig. \ref{figS_4} and Fig. \ref{figS_5}, respectively.

The above metrics further support our conclusions that the MEM-LJPDA converges more slowly (Fig. \ref{figS_5}, yellow line). RM-T-PMB and RM-JPDA have large deviations in shape estimation (Fig. \ref{figS_5}, cyan line and purple line). RM-VB-DBSCAN has the best accuracy in both position estimation and shape estimation (Fig. \ref{figS_4} and Fig. \ref{figS_5}, green line). RM-VB-Marginal has slightly inferior tracking accuracy, this is mainly reflected in the larger bias in shape estimation it has than RM-VB-DBSCAN (Fig. \ref{figS_5}, red line), but there is no significant difference in the position estimation of the target between these two methods (Fig. \ref{figS_4}, red line, and green line).

\begin{figure}[!h]
\centering                                         
\includegraphics[width=3.2in]{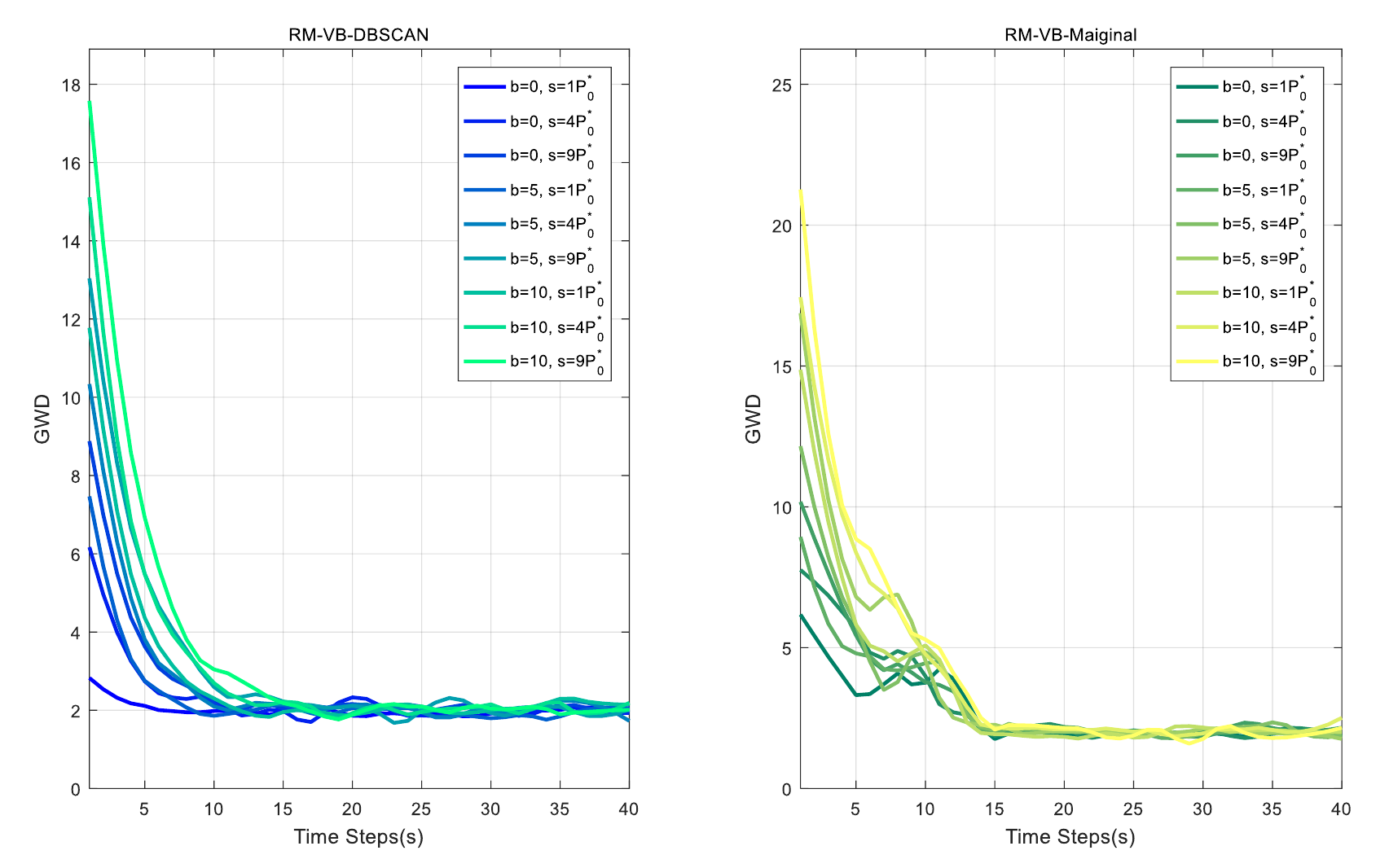}    
\caption{Time evolution for GWD of RM-VB with two lightweight schemes under varying initializations.}
\label{figN_1}                                     
\end{figure}

\begin{figure}[!htp]
\centering                                         
\includegraphics[width=3in]{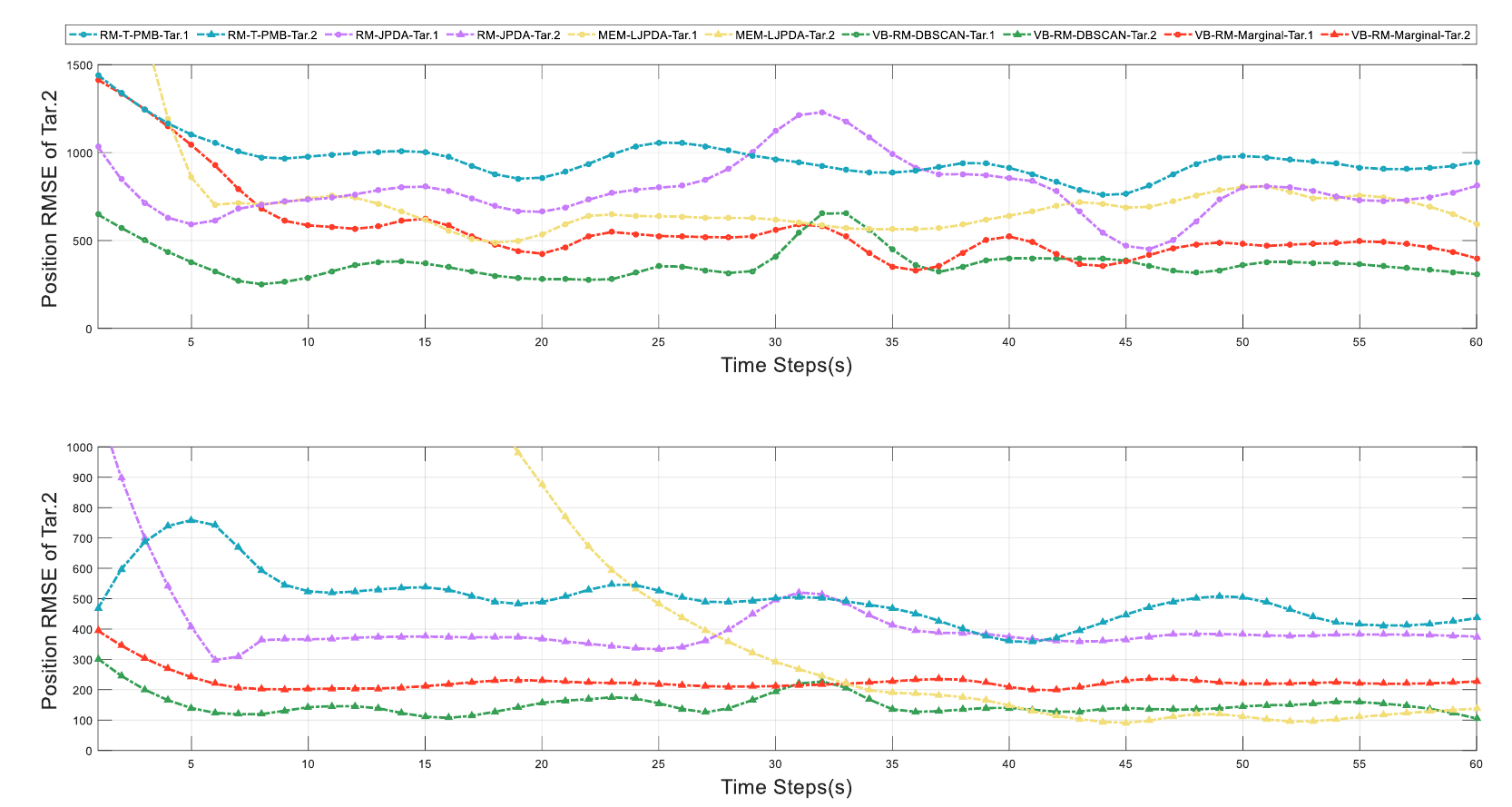}     
\caption{Simulation results of ${\rm RMSE}_{\mathrm{Ext}}$. (\small{The upper subfigure corresponds to Tar. 1, while the lower subfigure corresponds to Tar. 2})}
\label{figS_5}                                     
\end{figure}

\begin{figure*}[!htp]
\centering                                         
\includegraphics[width=7.3in]{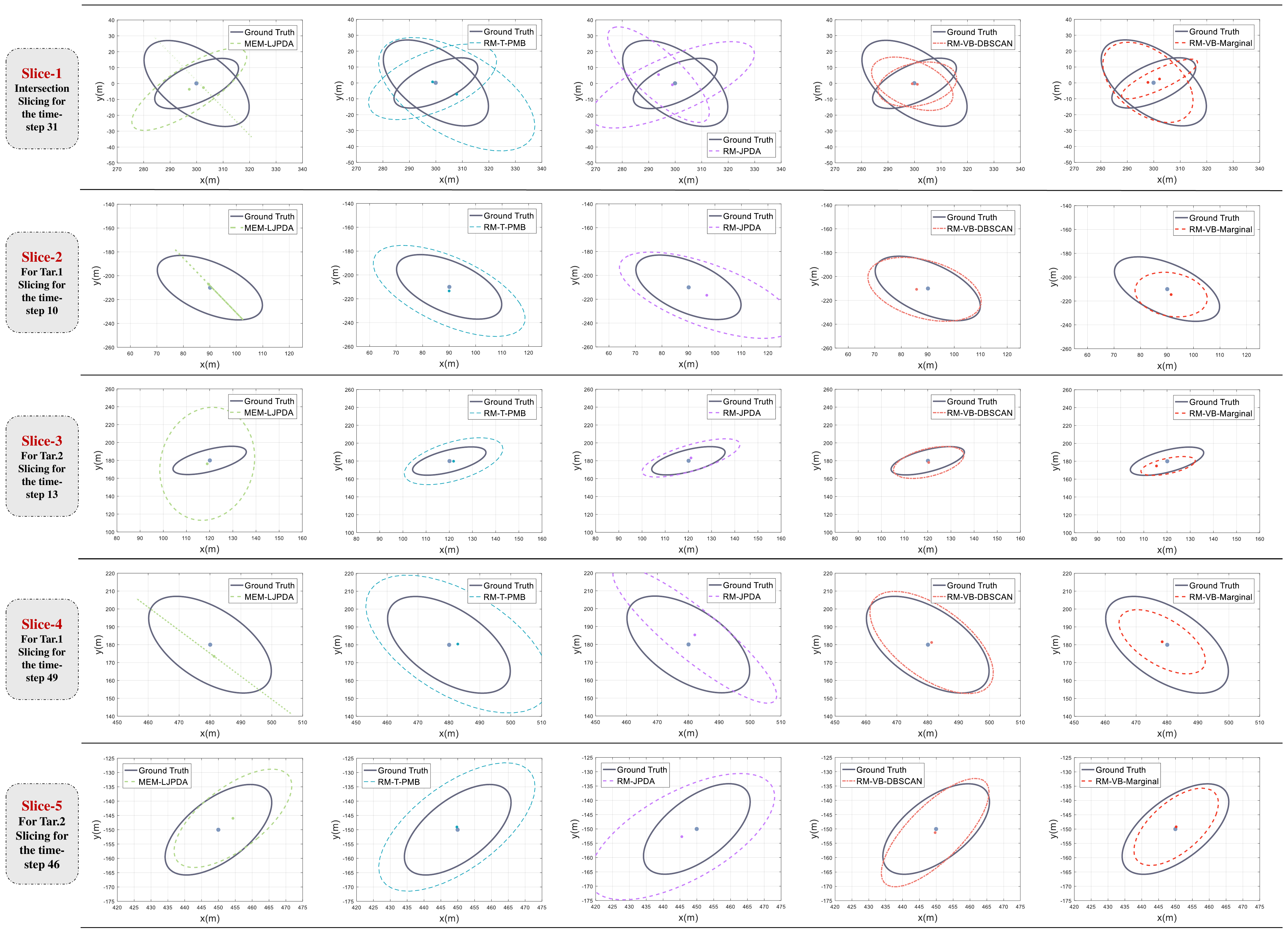}     
\caption{Enlarged display of the selected slice in Fig. \ref{figS_1}. 
	\small{(Showing slice 1, slice 2, slice 3, slice 4 and slice 5 from top to bottom of Fig.  \ref{figS_2}. And the shape estimation of MEM-LJPDA, RM-T-PMB, RM-JPDA, RM-VB-DBSCAN and RM-VB-Marginal are displayed from the left to the right in each row.)}} 
\label{figS_2}                                     
\end{figure*}

To assess how initialization affects RM-VB’s transient performance, we present a figure of the time evolution of the GWD between the estimated and ground-truth multi-target states under varying initial position biases and covariance scales. In Fig. \ref{figN_1}, differently colored lines show the mean over 100 MC runs. We vary the position bias $b\in\left\{0, 5, 10\right\}$ m and scale the initial covariance by factors $s\in\left\{1\times, 4\times, 9\times\right\}  \left(P_0 = s P^{\star}_{0}, \text{where} P^{\star}_{0} = \mathrm{diag}([1, 1, 0.1, 0.1])\right)$. Larger initialization errors increase the GWD at early time steps, but both lightweight schemes of RM-VB exhibit gradual convergence: all settings decrease and stabilize within about 5–10 time steps at similarly low GWD levels, indicating robustness to poor initialization.

In addition, to verify the performance of our proposed method under different measurement conditions, we vary the combinations of parameters $\left(\lambda_c,\lambda_t\right)$ to construct the following scenarios with different measurement conditions:
\begin{itemize}
\item $\left(\lambda_c,\lambda_t\right)=\left(25,10\right)$: There is dense clutter in the scene and the target is detected with few measurements. This combination of parameters is considered the baseline.
\item $\left(\lambda_c,\lambda_t\right)=\left(25,20\right)$: There is dense clutter in the scene, and the target is able to make more measurements. This combination of parameters is more in line with the actual measurement environment.
\item $\left(\lambda_c,\lambda_t\right)=\left(5,10\right)$: Sparse clutter in the scene, less measurement of target feedback.
\item $\left(\lambda_c,\lambda_t\right)=\left(5,20\right)$: Sparse clutter in the scene, and more measurements of the target are captured, this parameter combination meets the ideal detection conditions.
\end{itemize}

We obtained the statistics of GWD for each tracking method under the above four different combinations of parameters. In addition to $T_{Mean}\left({\rm RMSE}_{Pos}\right)$, $T_{Cov}\left({\rm RMSE}_{Ext}\right)$, and the corresponding covariances are listed in Table \ref{tab3}, respectively.

It can be seen from Table \ref{tab3}, in terms of both accuracy (Mean.) and stability of the estimate (Var.), the proposed method RM-VB-DBSCAN obtains the best performance in the scenarios with the first three parameter combinations (we also mark the "Best performance (Best PRF.)" in this table). 

\begin{table*}[t]
\caption{Statistical information of metrics with different combinations of $(\lambda_c, \lambda_t)$}
\label{tab3}
\centering
\scriptsize
\renewcommand{\arraystretch}{1.5}
\begin{tabular}{|cc|cc|cc|cc|}
\hline
\multicolumn{2}{|c|}{\multirow{2}{*}{Methods with parameters ($\lambda_c$, $\lambda_t$)}}                                             & \multicolumn{2}{c|}{GWD [Tar. 1, Tar. 2] $\downarrow$ }                                          & \multicolumn{2}{c|}{${\rm RMSE}_{\mathrm{Pos}}\ \downarrow$}                                     & \multicolumn{2}{c|}{${\rm RMSE}_{\mathrm{Ext}}\ \downarrow$}                                                     \\ \cline{3-8} 
\multicolumn{2}{|c|}{}                                                                                         & \multicolumn{1}{c|}{Mean.}                & Var.                  & 

\multicolumn{1}{c|}{Mean.}                 & Var.               & \multicolumn{1}{c|}{Mean.}                       & Var.                         \\ \hline
\multicolumn{1}{|c|}{\multirow{4}{*}{MEM-LJPDA}}      & (25,10)                                                & \multicolumn{1}{c|}{{[}15.218, 14.775{]}} & {[}17.367,309.156{]}  & \multicolumn{1}{c|}{{[}5.297, 3.997{]}}   & {[}2.175, 2.038{]} & \multicolumn{1}{c|}{{[}753.231, 1555.341{]}}    & {[}1.943e+05, 1.368e+07{]}   \\ \cline{2-8} 
\multicolumn{1}{|c|}{}                                & (25,20)                                                & \multicolumn{1}{c|}{{[}10.716, 7.929{]}}  & {[}61.779, 36.524{]}  & \multicolumn{1}{c|}{{[}4.7768, 3.0174{]}} & {[}1.375, 1.089{]} & 
\multicolumn{1}{c|}{{[}868.1, 970.2427{]}}  & {[}6.307e+04, 1.765e+03{]}   \\ \cline{2-8} 
\multicolumn{1}{|c|}{}                                & (5,10)                                                 & \multicolumn{1}{c|}{{[}16.505, 10.622{]}} & {[}192.918, 95.226{]} & \multicolumn{1}{c|}{{[}5.958, 3.560{]}}   & {[}5.504,1.339{]}  & \multicolumn{1}{c|}{{[}1.857e+03, 1.162e+03{]}} & {[}9.170e+06, 3.397e+06{]}   \\ \cline{2-8} 
\multicolumn{1}{|c|}{}                                & \begin{tabular}[c]{@{}c@{}}(5,20)\\ \textbf{Best PRF.}\end{tabular}  & \multicolumn{1}{c|}{{[}5.129, 3.114{]}}   & {[}9.141, 1.964{]}    & \multicolumn{1}{c|}{{[}3.712, 2.203{]}}   & {[}0.694, 0.325{]} & \multicolumn{1}{c|}{{[}343.908, 85.823{]}}      & {[}4.960e+05, 5.131e+03{]}   \\ \hline
\multicolumn{1}{|c|}{\multirow{4}{*}{RM-VB-DBSCAN}}   & \begin{tabular}[c]{@{}c@{}}(25,10)\\ \textbf{Best PRF.}\end{tabular} & \multicolumn{1}{c|}{{[}7.954, 4.932{]}}   & {[}7.3180, 2.4351{]}  & \multicolumn{1}{c|}{{[} 2.584, 2.156{]}}  & {[}1.281, 0.717{]} & \multicolumn{1}{c|}{{[}373.959, 150.218{]}}     & {[}2.0334e+04, 3.1934e+03{]} \\ \cline{2-8} 
\multicolumn{1}{|c|}{}                                & \begin{tabular}[c]{@{}c@{}}(25,20)\\ \textbf{Best PRF.}\end{tabular} & \multicolumn{1}{c|}{{[}5.965, 4.283{]}}   & {[}4.8079, 4.6313{]}  & \multicolumn{1}{c|}{{[}2.217, 1.644{]}}   & {[}0.962, 0.539{]} & \multicolumn{1}{c|}{{[}294.944, 145.695{]}}     & {[}1.294e+04, 1.394e+03{]}   \\ \cline{2-8} 
\multicolumn{1}{|c|}{}                                & \begin{tabular}[c]{@{}c@{}}(5,10)\\ \textbf{Best PRF.} \end{tabular}  & \multicolumn{1}{c|}{{[}7.602, 4.839{]}}   & {[}8.1840, 3.0073{]}  & \multicolumn{1}{c|}{{[}3.146, 2.077{]}}   & {[}1.841, 0.637{]} & \multicolumn{1}{c|}{{[}370.502, 159.689{]}}     & {[}2.577e+04, 1.667e+04{]}   \\ \cline{2-8} 
\multicolumn{1}{|c|}{}                                & (5,20)                                                 & \multicolumn{1}{c|}{{[}5.933, 3.793{]}}   & {[}3.6136, 1.6841{]}  & \multicolumn{1}{c|}{{[}2.189, 1.582{]}}   & {[}0.572, 0.248{]} & \multicolumn{1}{c|}{{[}393.357, 121.320{]}}     & {[}1.187e+04, 2.485e+03{]}   \\ \hline
\multicolumn{1}{|c|}{\multirow{4}{*}{RM-VB-Marginal}} & (25,10)                                                & \multicolumn{1}{c|}{{[}12.741, 7.998{]}}  & {[}63.235, 3.161{]}   & \multicolumn{1}{c|}{{[}2.913, 2.470{]}}   & {[}1.271, 1.190{]} & \multicolumn{1}{c|}{{[}593.958, 226.095{]}}     & {[}6.446e+04, 1.148e+03{]}   \\ \cline{2-8} 
\multicolumn{1}{|c|}{}                                & (25,20)                                                & \multicolumn{1}{c|}{{[}8.046, 7.782{]}}   & {[}32.838, 2.333{]}   & \multicolumn{1}{c|}{{[}2.996,1.815{]}}    & {[}2.230, 0.464{]} & \multicolumn{1}{c|}{{[}406.932, 229.025{]}}     & {[}1.566e+05, 1.206e+03{]}   \\ \cline{2-8} 
\multicolumn{1}{|c|}{}                                & (5,10)                                                 & \multicolumn{1}{c|}{{[}8.929, 5.426{]}}   & {[}24.919, 2.705{]}   & \multicolumn{1}{c|}{{[}3.232, 2.125{]}}   & {[}2.196, 0.338{]} & \multicolumn{1}{c|}{{[}437.979, 161.454{]}}     & {[}6.307e+04, 1.765e+03{]}   \\ \cline{2-8} 
\multicolumn{1}{|c|}{}                                & (5,20)                                                 & \multicolumn{1}{c|}{{[}6.567, 4.239{]}}   & {[}15.075, 3.369{]}   & \multicolumn{1}{c|}{{[}2.339, 1.646{]}}   & {[}0.698, 0.271{]} & \multicolumn{1}{c|}{{[}325.955, 127.456{]}}     & {[}3.698e+04, 2.253e+03{]}   \\ \hline
\multicolumn{1}{|c|}{\multirow{4}{*}{RM-T-PMB}}      & (25,10)                                                & \multicolumn{1}{c|}{{[}14.686, 11.530{]}} & {[}1.898, 4.169{]}    & \multicolumn{1}{c|}{{[}3.037, 2.962{]}}   & {[}3.882, 2.957{]} & \multicolumn{1}{c|}{{[}953.821, 499.523{]}}     & {[}1.885e+04, 1.004e+04{]}   \\ \cline{2-8} 
\multicolumn{1}{|c|}{}                                & (25,20)                                                & \multicolumn{1}{c|}{{[}15.134, 12.330{]}} & {[}9.593, 5.164{]}    & \multicolumn{1}{c|}{{[}3.130, 2.231{]}}   & {[}2.153, 2.498{]} & \multicolumn{1}{c|}{{[}1.013e+03, 566.231{]}}   & {[}7.816e+04, 2.054e+04{]}   \\ \cline{2-8} 
\multicolumn{1}{|c|}{}                                & (5,10)                                                 & \multicolumn{1}{c|}{{[}15.116, 13.528{]}} & {[}9.802, 12.094{]}   & \multicolumn{1}{c|}{{[}3.190, 2.492{]}}   & {[}2.447, 2.067{]} & \multicolumn{1}{c|}{{[}947.487, 653.663{]}}     & {[}8.017e+04, 4.504e+04{]}   \\ \cline{2-8} 
\multicolumn{1}{|c|}{}                                & (5,20)                                                 & \multicolumn{1}{c|}{{[}12.544, 10.874{]}} & {[}1.799, 3.431{]}    & \multicolumn{1}{c|}{{[}3.030, 2.051{]}}   & {[}2.603, 2.552{]} & \multicolumn{1}{c|}{{[}649.251, 465.895{]}}     & {[}1.127e+04, 1.077e+04{]}   \\ \hline
\end{tabular}
\end{table*}

The accuracy of MEM-LJPDA has improved a lot with the reduction of the clutter rate and the increase of the target measurement rate. Under the most ideal observation conditions, i.e., $(\lambda_c,\lambda_t)=(5,20)$, this method has the best performance. However, such ideal measurement conditions are often not available in practice.

The adaptability to changes in parameter combinations is also demonstrated by another proposed method, RM-VB-Marginal. Moreover, we find that the gap between its accuracy and that of RM-VB-DBSCAN narrows as the clutter decreases (see the metrics of the combinations of parameters $(\lambda_c,\lambda_t)=(5,20)$ and $(\lambda_c,\lambda_t)=(25,20))$. This pattern is also reflected in the case of an increase in the target measurement rate (see the metrics of the parameter combinations $(\lambda_c,\lambda_t)=(25,10)$ and $(\lambda_c,\lambda_t)=(25,20))$.

\begin{figure}[!htp]
\centering                                         
\includegraphics[width=3.5in]{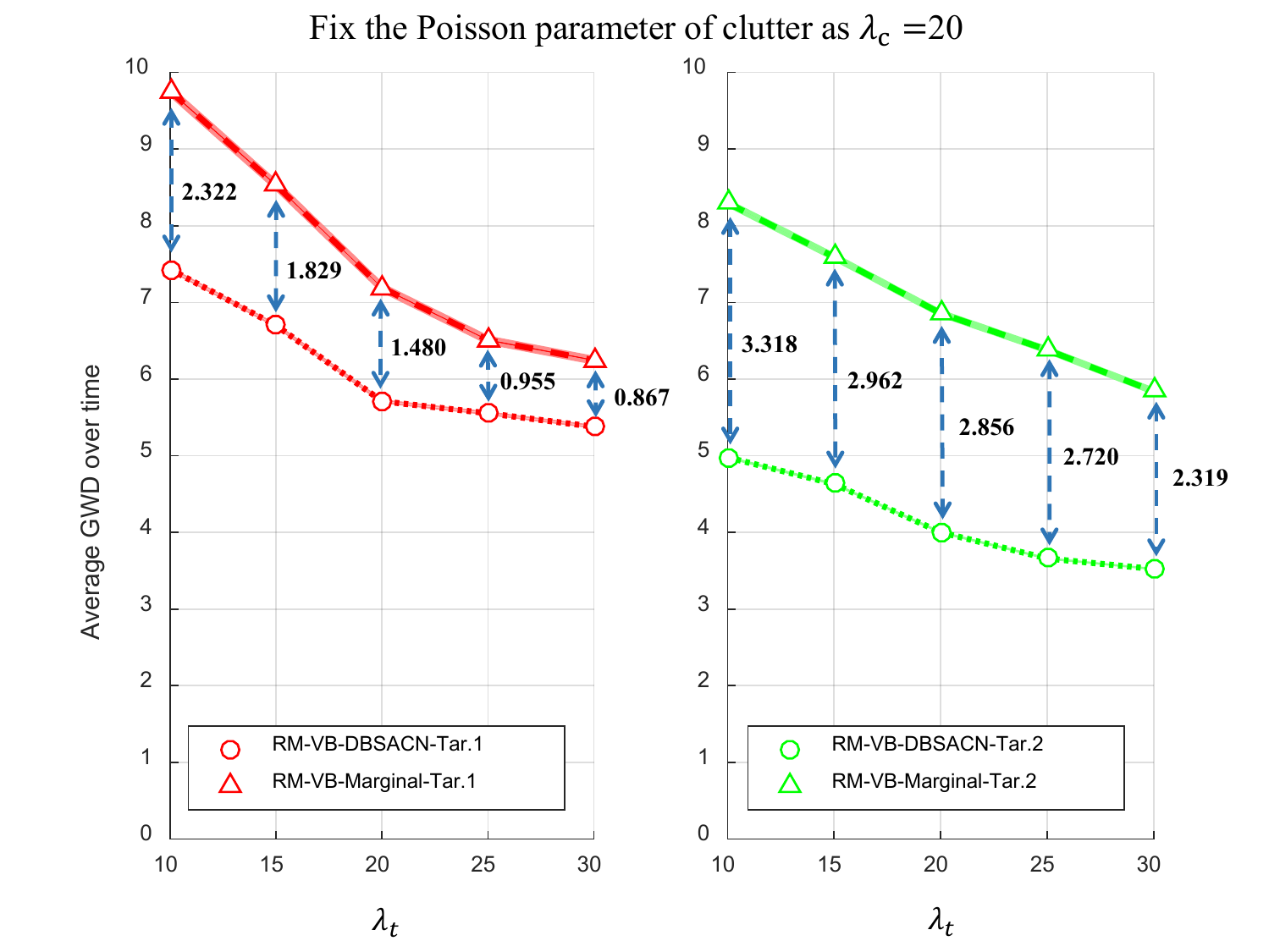}     
\caption{Average GWD over time with fixed $\lambda_c$ and variable $\lambda_t$ of two lightweight schemes.}
\label{figS_6}                                     
\end{figure}

\begin{figure}[!htp]
\centering                                         
\includegraphics[width=3.5in]{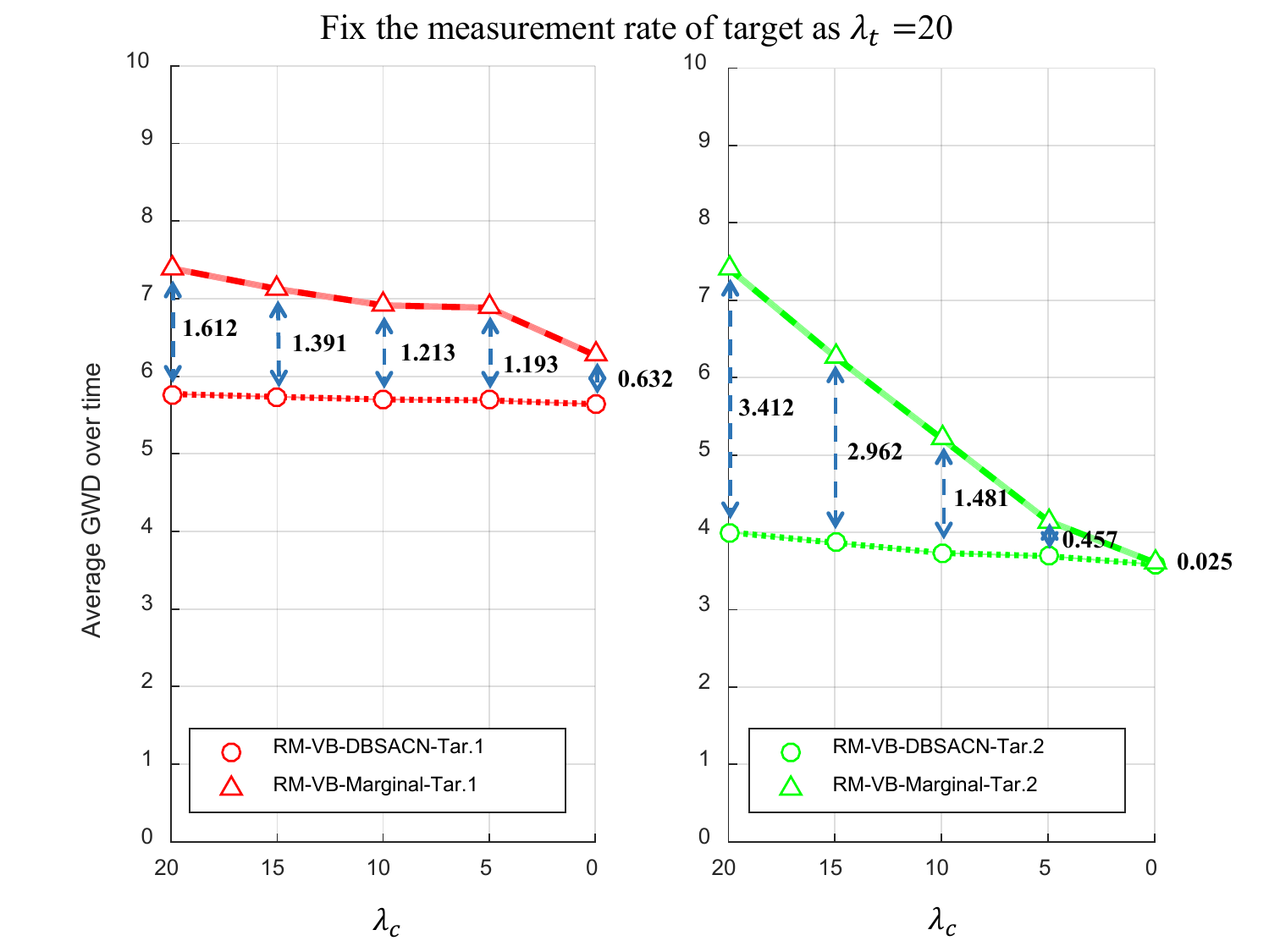}     
\caption{Average GWD over time with fixed $\lambda_t$ and variable $\lambda_c$ of two lightweight schemes.}
\label{figS_7}                                     
\end{figure}

We fixed the clutter rate to $\lambda_c=20$ and adjusted the target measurement rate $\lambda_t$ from 5 to 30 at intervals of 5. The average GWD of the two proposed methods are shown in Fig. \ref{figS_6}. Similarly, we fix the target's measurement rate to $\lambda_t=20$ and then reduce the Poisson parameter of the clutter $\lambda_c$ from 20 to 0 in intervals of 5. The variation of the average GWD for the two methods is shown in Fig. \ref{figS_7}.

Fig. \ref{figS_6} and Fig. \ref{figS_7} prove that increasing the target measurement rate can narrow the accuracy gap between RM-VB-DBSCAN and RM-VB-Marginal. Meanwhile, reducing the Poisson rate of clutter does not greatly improve the accuracy of RM-VB-DBSCAN, but the improvement of the accuracy of RM-VB-Marginal is significant. The lower clutter rate can quickly reduce the accuracy gap between the two methods. This means that if there is less clutter in the scene, RM-VB-Marginal will be more competitive.

The computational complexity of the tracking methods can be directly judged by the average computational time per MC run, as shown in Table \ref{table4}. In this table, we use the standard combination of parameters $(\lambda_c,\lambda_t)=(25,10)$. It can be clearly seen that RM-T-PMB has the highest time cost, because it uses stochastically optimized sampling and considers the case where the number of targets changes. Both RM-VB-DBSCAN and RM-JPDA use the DB-SCAN clustering for lightweight, but the computational time of RM-VB-DBSCAN is slightly higher than that of RM-JPDA because it performs several VBI iterations. Both MEM-LJPDA and RM-VB-Marginal are essentially based on marginal association probabilities for parameter estimation, and thus they have a significant reduction in time cost to linear complexity. However, since RM-VB-Marginal also performs multiple VBI iterations, its complexity is slightly higher than that of MEM-LJPDA.

\begin{table}[!h]
\caption{Average computational time per MC run of the tracking methods}
\centering
\label{table4}
\renewcommand{\arraystretch}{1.5}
\begin{tabular}{|c|c|}
\hline
Tracking Methods & Average computational times per MC run (s) \\ \hline
MEM-LJPDA        & 6.68                                     \\ \hline
RM-T-PMB           & 187.27                                   \\ \hline
RM-JPDA          & 142.121                                  \\ \hline
RM-VB-DBSCAN     & 150.116                                  \\ \hline
RM-VB-Marginal   & 11.203                                   \\ \hline
\end{tabular}
\end{table}

\subsection{The Real Data Scenario}
\label{section6_4}
In this section, we further illustrate the advantages of the proposed method using real data. The test data are collected on an urban road in Guangdong \cite{bilibili}. The real scenario involves multiple vehicles traveling at an approximately constant speed on the road while using a drone fixed in mid-air to capture images in the surveillance area.

\begin{figure}
    \centering
    \includegraphics[width=0.98\linewidth]{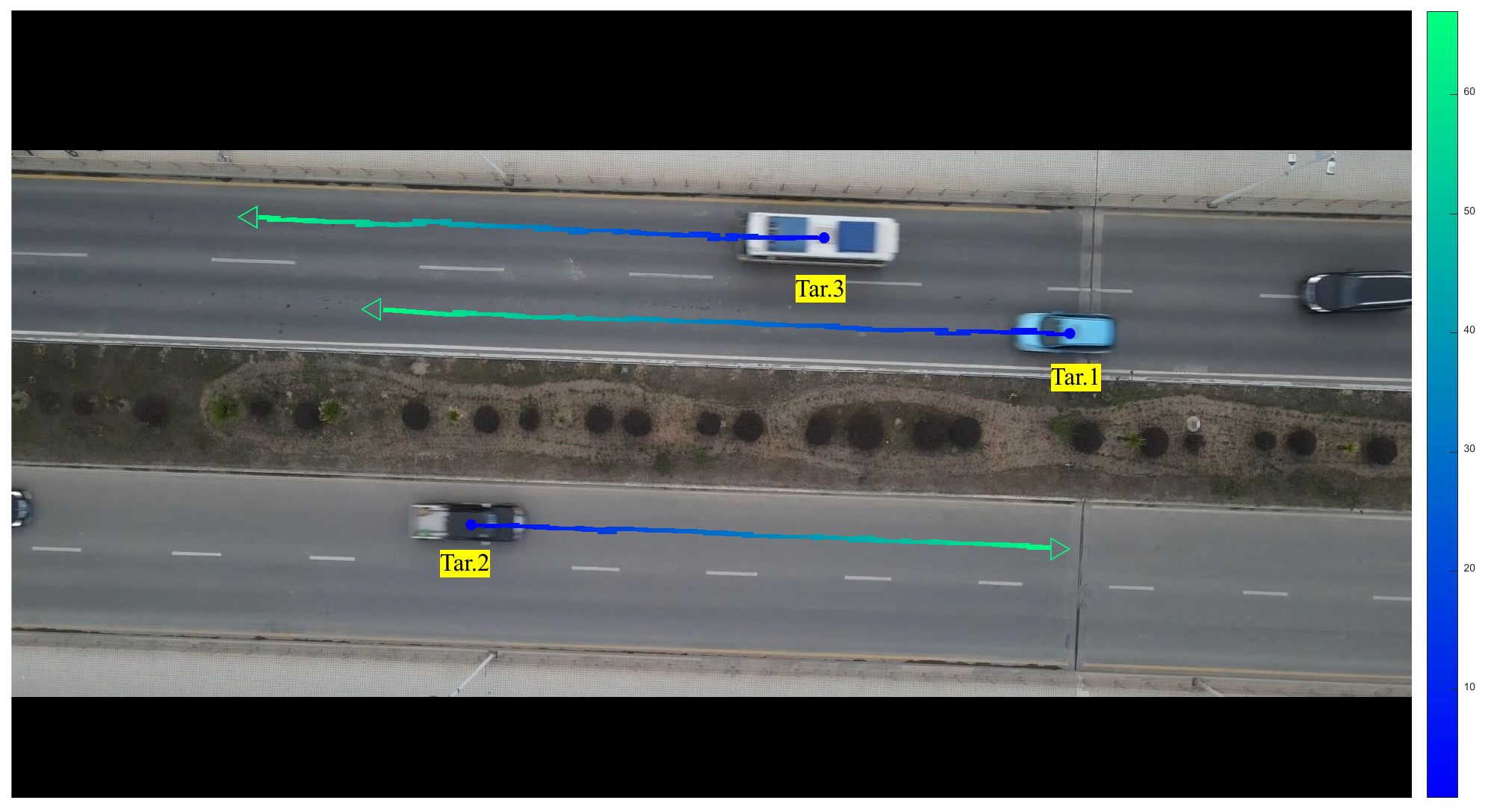}
    \caption{The trajectories of the targets in real data scenario. \\
    \footnotesize{In the figure, an initial frame is shown. The color bar shows the time delay of the targets' movement.}}
    \label{fig_R1_GT}
\end{figure}

In this scene, we have selected three vehicles of interest for tracking, including a royal blue saloon car (Tar. 1), a black saloon car (Tar. 2), and a white bus (Tar. 3). Their actual trajectories are shown in Fig. {\ref{fig_R1_GT}}. We use solid lines with gradient colors to indicate the center trajectories. At the beginning of the scene, targets are located at the positions indicated by the corresponding dark blue dots, and then they drive along the paths marked by the lines until surveillance ended at the locations marked by the respective green triangles. The sequence contains 67 frames at a sampling interval $\mathrm{T}=0.0625s$.

Throughout the scenario, images captured by the drone are processed for measurement acquisition. We first use the image processing algorithm--SURF \cite{2006SURF} to separate objects of interest from the background based on gradient features, and then a median filter is employed to reduce the number of clutter that comes from the background\footnote{We retain a small amount of clutter to ensure the authenticity of the scenario.}. Finally, the target pixels are uniformly sampled to obtain measurements generated by the target surface. This measurement acquisition scheme is consistent with \cite{2021Tuncer}. In addition, since the scene and the size of the target are different from the previous simulation experiments, the parameters of various methods used have been modified as follows:
\begin{itemize}
    \item For RM-T-PMB, RM-JPDA and our proposed methods, the initial shape parameters are adjusted to be the same and smaller than the actual shape size of the targets, i.e.: ${\hat{v}}_{0|0}^{\left(1\right)}={\hat{v}}_{0|0}^{\left(2\right)}={\hat{v}}_{0|0}^{\left(3\right)}=14$ and ${\hat{V}}_{0|0}^{\left(1\right)}={\hat{V}}_{0|0}^{\left(2\right)}={\hat{V}}_{0|0}^{\left(3\right)}=5\mathbf{I}_2$.
    \item The initial position of each target is set to the initial frame based on their position in the initial frame and plus a zero-mean Gaussian perturbation $\mathcal{N}\!\left(\boldsymbol{0},\mathrm{diag}([5,5])\right)$, and the initial kinematic state covariance is reduced by a factor of ten.
\end{itemize}

\begin{figure*}
    \centering
    \includegraphics[width=0.9\linewidth]{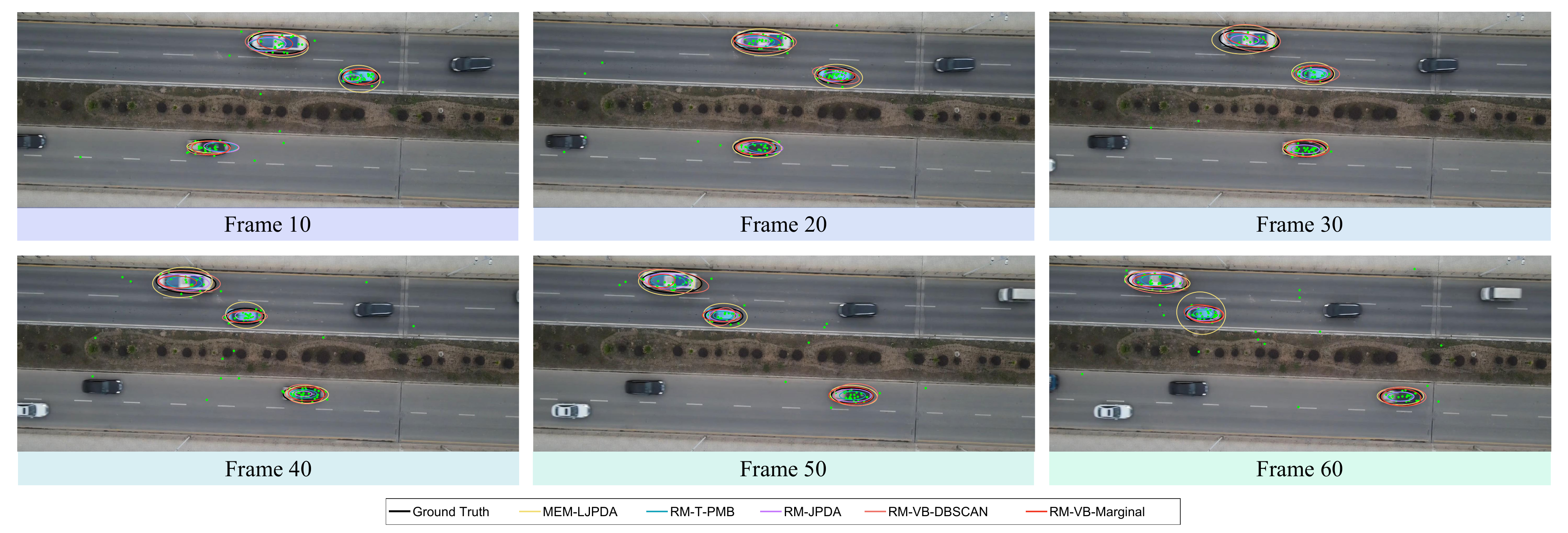}
    \caption{A representative run on the real data sequence (\footnotesize{Contours denote estimated extents; green dots denote measurements}).}
    \label{fig_R1_example}
\end{figure*}

\begin{figure}[!h]
    \centering
    \includegraphics[width=3in]{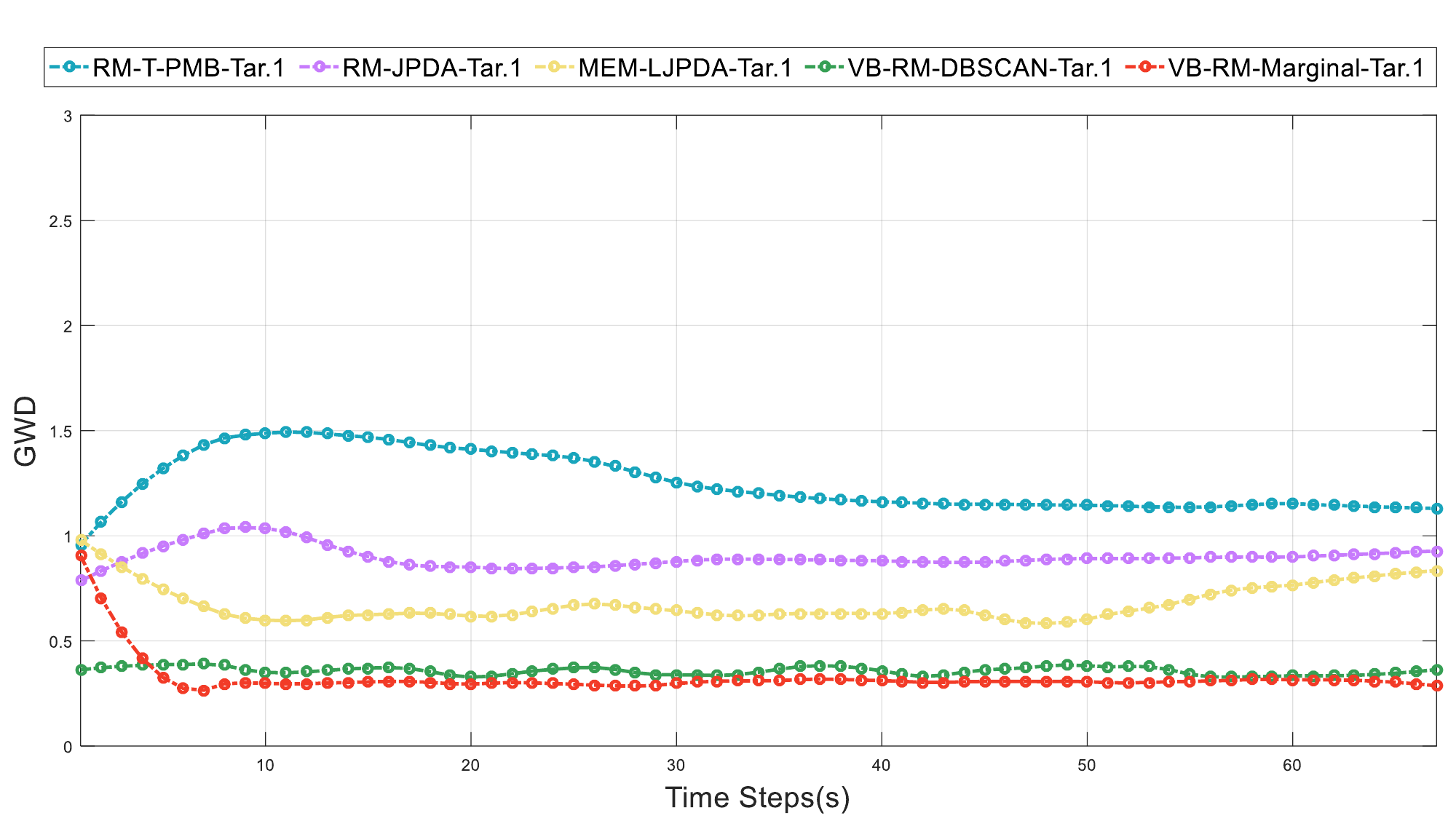}
    \caption{The GWD of Tar. 1 in real data scenario. (Averaged over 100 MC runs)}
    \label{fig_R1_GWD_Tar1}
\end{figure}

\begin{figure}[!h]
    \centering
    \includegraphics[width=3in]{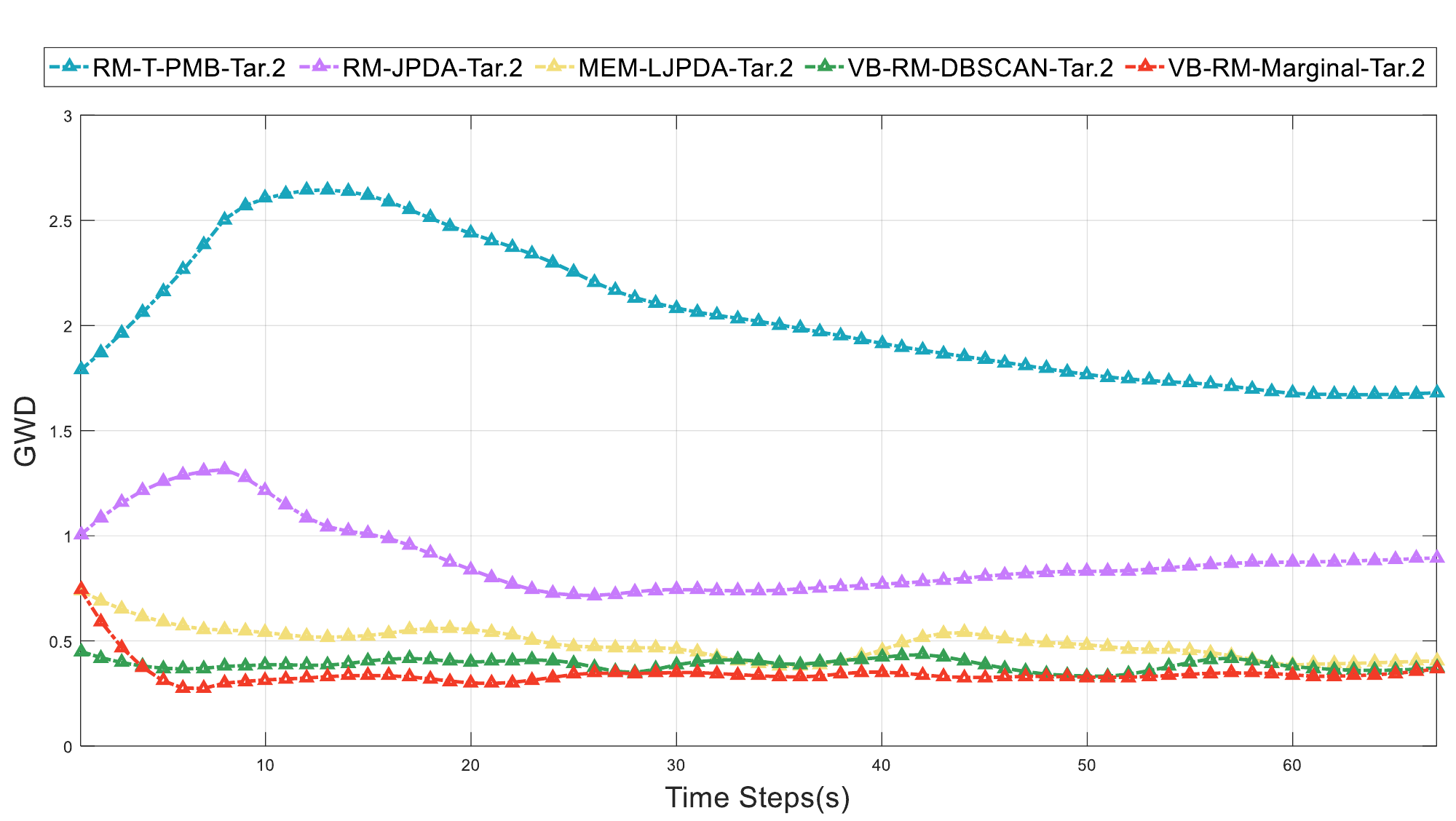}
    \caption{The GWD of Tar. 2 in real data scenario.}
    \label{fig_R1_GWD_Tar2}
\end{figure}

\begin{figure}[!h]
    \centering
    \includegraphics[width=3in]{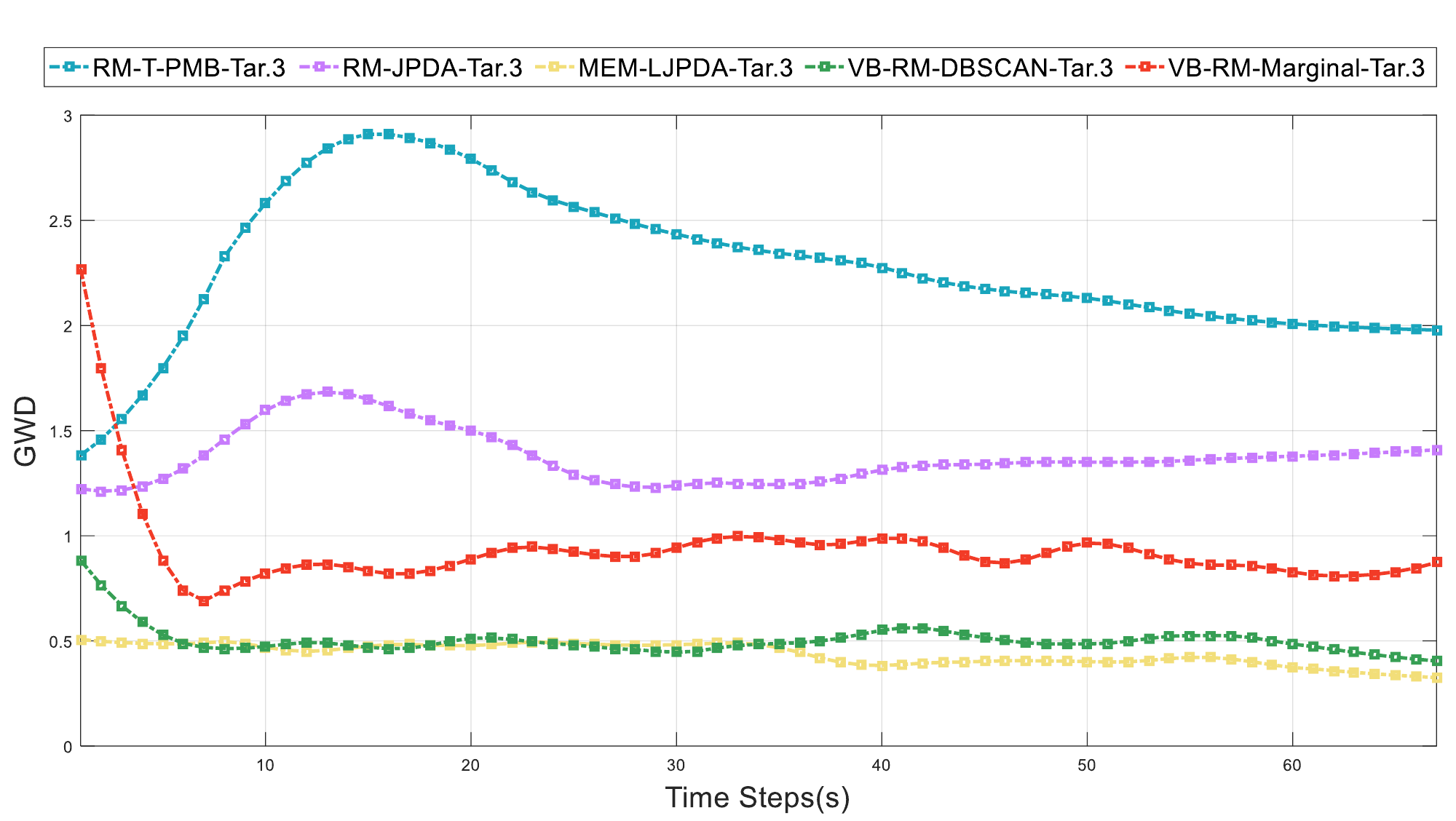}
    \caption{The GWD of Tar. 3 in real data scenario.}
    \label{fig_R1_GWD_Tar3}
\end{figure}

Fig. {\ref{fig_R1_example}} shows the estimate of the extent of the target corresponding to frames ${10, 20, 30, 40, 50, 60}$. These slices are chosen to illustrate the performance differences between methods more clearly, and the GWD of the three targets is counted and presented in Fig. {\ref{fig_R1_GWD_Tar1}}, Fig. {\ref{fig_R1_GWD_Tar2}} and Fig. {\ref{fig_R1_GWD_Tar3}} respectively.

The conclusions drawn from the simulation experiment in Section \ref{section6_3} are further confirmed by the results shown in the figures. The two traditional RMM-based methods, RM-T-PMB and RM-JPDA, rely on a prior shape parameter setting and cannot accurately estimate the shape of the target. Since MEM-LJPDA does not have parameter modeling for the target measurement rate, its shape estimation converges more slowly and is not accurate enough in the shape estimation of Tar. 1 (Fig. {\ref{fig_R1_GWD_Tar1}}, yellow line). Interestingly, due to less clutter in the scene, RM-VB-Marginal achieved an accuracy that is very close to or even higher than RM-VB-DBSCAN, especially for Tar. 1 and Tar. 2 (Fig. {\ref{fig_R1_GWD_Tar1}} and Fig. {\ref{fig_R1_GWD_Tar2}}, red line), while ensuring the efficiency of calculation.

\section{CONCLUSION}
\label{sec7}
In this work, we propose a novel VBI-based method for multi-target tracking in a cluttered environment. We use the RMM with shape evolution to model the targets and obtain approximate posterior distributions for the targets' joint states based using VBI. Moreover, two lightweight schemes—clustering-based and marginal-association-based—are presented as potential solutions to improve the practicality of our method. The efficacy of the proposed algorithm is validated through both simulation and real data experiments. The experimental results show that, in comparison to the existing multi-target tracking methods, our proposed approach yields superior estimation precision and shows robust performance across different measurement environments. Additional discussion and directions for future improvement are outlined in Section \ref{discuss}.

Our method offers satisfactory scalability and ease of integration owing to the use of the RMM for target modeling. Consequently, the proposed VBI-based multi-target parameter estimator can be integrated with various existing VBI-based single-target state-estimation models (e.g., \cite{2020Zhang,2021Tuncer,2021Zhang}) to extend them to multi-target tracking.

\vspace{1\baselineskip}
\noindent \textbf{AUTHOR CONTRIBUTIONS:} Project administration, visualization, writing - original draft and writing - review and editing, Yuanhao Cheng; funding acquisition, Yunhe Cao; validation, Tat-Soon Yeo; investigation, Jie Fu and Yulin Zhang.

\vspace{1\baselineskip}
\noindent \textbf{FUNDING:} 
This research was funded by the National Natural Science Foundation of China under 728 Grant number 61771367.

\vspace{1\baselineskip}
\noindent \textbf{CONFLICT OF INTEREST:} 
The authors declare that they have no conflicts of interest.

\vspace{1\baselineskip}
\noindent \textbf{DATA AVAILABILITY STATEMENT:} 
The data that support the findings of this study are available in video (BV1RA41157dN) at \url{https://www.bilibili.com}. These data were derived from the following resources available in the public domain: \url{https://www.bilibili.com/video/BV1RA41157dN}.

\vspace{1\baselineskip}
\noindent \textbf{ORCID:}
Yuanhao Cheng \\
\url{https://orcid.org/0009-0003-6595-3005}.

\bibliography{ref}

\newpage
\onecolumn
\begin{center}
\Large\MakeUppercase{Supporting Materials}
\end{center}

\appendices
\section{The Derivation of Eq.(\ref{eq28})}
\label{appendixA1}
Using the rule of Eq.(\ref{eq26}) we have:
\begin{equation}
\begin{split}
\label{eqappenda1}
q_{\boldsymbol{\Theta}}\left(\boldsymbol{\Theta}_k^{1:L_k^\theta}\right)\propto\mathrm{exp}\left\{\int{\int{\int{\left[\mathrm{ln}\ p\left(\mathbf{x}_k^{1:n_k},\mathbf{X}_k^{1:n_k},\mathbf{\Lambda}_k^{1:n_k},\boldsymbol{\Theta}_k^{1:L_k^\theta},\mathcal{Y}_k\mid\mathcal{Y}^{k-1}\right)q_\mathbf{x}\left(\mathbf{x}_k^{1:n_k}\right)q_\mathbf{X}\left(\mathbf{X}_k^{1:n_k}\right)q_\mathbf{\Lambda}\left(\mathbf{\Lambda}_k^{1:n_k}\right)\right]d\mathbf{x}_k^{1:n_k}d\mathbf{X}_k^{1:n_k}d\mathbf{\Lambda}_k^{1:n_k}}\ }}\right\}
\end{split}
\end{equation}

In Eq.(\ref{eqappenda1}), we have:
\begin{equation}
\label{eqappenda2}
\begin{split}
&p\left(\mathbf{x}_k^{1:n_k},\mathbf{X}_k^{1:n_k},\mathbf{\Lambda}_k^{1:n_k},\boldsymbol{\Theta}_k^{1:L_k^\theta},\mathcal{Y}_k\mid\mathcal{Y}^{k-1}\right)=p\left(\boldsymbol{\Xi}_k^{1:n_k},\boldsymbol{\Theta}_k^{1:L_k^\theta},\mathcal{Y}_k\mid\mathcal{Y}^{k-1}\right) \\
&=p\left(\mathcal{Y}_k\mid\boldsymbol{\Xi}_k^{1:n_k},\boldsymbol{\Theta}_k^{1:L_k^\theta}\right)p\left(\boldsymbol{\Theta}_k^{1:L_k^\theta}\mid\boldsymbol{\Xi}_k^{1:n_k},\mathcal{Y}^{k-1}\right)p\left(\boldsymbol{\Xi}_k^{1:n_k}\mid\mathcal{Y}^{k-1}\right) \\
&=p\left(\boldsymbol{\Theta}_k^{1:L_k^\theta}\mid\mathcal{Y}_k,\boldsymbol{\Xi}_k^{1:n_k}\right)p\left(\boldsymbol{\Xi}_k^{1:n_k}\mid\mathcal{Y}^{k-1}\right)
\end{split}
\end{equation}

Notice that Eq.(\ref{eqappenda2}) holds because:
\begin{equation}
\label{eqappenda3}
p\left(\boldsymbol{\Theta}_k^{1:L_k^\theta}\mid\mathcal{Y}_k,\boldsymbol{\Xi}_k^{1:n_k}\right)=p\left(\mathcal{Y}_k\mid\boldsymbol{\Theta}_k^{1:L_k^\theta},\boldsymbol{\Xi}_k^{1:n_k}\right)p\left(\boldsymbol{\Theta}_k^{1:L_k^\theta}\mid\boldsymbol{\Xi}_k^{1:n_k}\right)
\end{equation}

The Eq.(\ref{eqappenda3}) follows from the Bayesian expansion shown in Eq.(\ref{eq15}), so the term $p\left(\mathbf{x}_k^{1:n_k},\mathbf{X}_k^{1:n_k},\mathbf{\Lambda}_k^{1:n_k},\boldsymbol{\Theta}_k^{1:L_k^\theta},\mathcal{Y}_k\mid\mathcal{Y}^{k-1}\right)$ can be further expanded as:
\begin{equation}
\label{eqappenda4}
\begin{split}
&p\left(\mathbf{x}_k^{1:n_k},\mathbf{X}_k^{1:n_k},\mathbf{\Lambda}_k^{1:n_k},\boldsymbol{\Theta}_k^{1:L_k^\theta},\mathcal{Y}_k\mid\mathcal{Y}^{k-1}\right)\propto p\left(\boldsymbol{\Theta}_k^{1:L_k^\theta}\mid\mathcal{Y}_k,\boldsymbol{\Xi}_k^{1:n_k}\right)p\left(\boldsymbol{\Xi}_k^{1:n_k}\mid\mathcal{Y}^{k-1}\right) \\
&=\prod_{l=1}^{L_k^\theta}\left[\left(\rho\right)^{\phi_k^{l-\left(0\right)}}\left(\lambda_c\right)^{\phi_k^{l-\left(0\right)}}\prod_{n=1}^{n_k}\left(e^{-\lambda_k^{\left(n\right)}}\left(\lambda_k^{\left(n\right)}\right)^{\phi_k^{l-\left(n\right)}}\prod_{\boldsymbol{y}_k^{\widetilde{j}}\in{\widetilde{\mathcal{Y}}}_k^{l-\left(n\right)}}{p\left(\boldsymbol{y}_k^{\widetilde{j}}\mid\boldsymbol{\xi}_k^{\left(n\right)}\right)}\right)\right]^{w_k^l}\times\prod_{n=1}^{n_k}\mathcal{GGIW}\left(\boldsymbol{\xi}_k^{\left(n\right)};{\hat{\boldsymbol{\zeta}}}_{k|k-1}^{\left(n\right)}\right) \\
&=\prod_{l=1}^{L_k^\theta}\left[\left(\rho\right)^{\phi_k^{l-\left(0\right)}}\left(\lambda_c\right)^{\phi_k^{l-\left(0\right)}}\prod_{n=1}^{n_k}\left(\mathcal{GGIW}\left(\boldsymbol{\xi}_k^{\left(n\right)}; {\hat{\boldsymbol{\zeta}}}_{k|k-1}^{\left(n\right)}\right)e^{-\lambda_k^{\left(n\right)}}\left(\lambda_k^{\left(n\right)}\right)^{\phi_k^{l-\left(n\right)}}\prod_{\boldsymbol{y}_k^{\widetilde{j}}\in{\widetilde{\mathcal{Y}}}_k^{l-\left(n\right)}}{p\left(\boldsymbol{y}_k^{\widetilde{j}}\mid\boldsymbol{\xi}_k^{\left(n\right)}\right)}\right)\right]^{w_k^l}
\end{split}
\end{equation}

\noindent where
${\hat{\boldsymbol{\zeta}}}_{k|k-1}^{\left(n\right)}=\left\{{\hat{\boldsymbol{m}}}_{k|k-1}^{\left(n\right)},{\hat{\boldsymbol{P}}}_{k|k-1}^{\left(n\right)},{\hat{v}}_{k|k-1}^{\left(n\right)},{\hat{\boldsymbol{V}}}_{k|k-1}^{\left(n\right)},{\hat{\alpha}}_{k|k-1}^{\left(n\right)},{\hat{\beta}}_{k|k-1}^{\left(n\right)}\right\}$, $n=1,2,...,n_k$.

Taking logarithms on both sides of Eq.(\ref{eqappenda4}), we have:
\begin{equation}
\label{eqappenda5}
\begin{split}
\mathrm{ln}\ &p\left(\mathbf{x}_k^{1:n_k},\mathbf{X}_k^{1:n_k},\mathbf{\Lambda}_k^{1:n_k},\boldsymbol{\Theta}_k^{1:L_k^\theta},\mathcal{Y}_k\mid\mathcal{Y}^{k-1}\right) \\
&={\underbrace{w_k^l\sum_{l=1}^{L_k^\theta}{\mathrm{ln}\ \left[\left(\rho\right)^{\phi_k^{l-\left(0\right)}}\left(\lambda_c\right)^{\phi_k^{l-\left(0\right)}}\prod_{n=1}^{n_k}\left(e^{-\lambda_k^{\left(n\right)}}\left(\lambda_k^{\left(n\right)}\right)^{\phi_k^{l-\left(n\right)}}\prod_{\boldsymbol{y}_k^{\widetilde{j}}\in{\widetilde{\mathcal{Y}}}_k^{l-\left(n\right)}}{p\left(\boldsymbol{y}_k^{\widetilde{j}}\mid\xi_k^{\left(n\right)}\right)}\right)\right]} +\sum_{n=1}^{n_k}{\mathrm{ln}\ \left[\begin{matrix}\mathcal{N}\left(\boldsymbol{x}_{k}^{\left(n\right)};{\hat{\boldsymbol{m}}}_{k|k-1}^{\left(n\right)},{\hat{\boldsymbol{P}}}_{k|k-1}^{\left(n\right)}\right)\cdot\mathcal{IW}\left(\boldsymbol{X}_{k}^{\left(n\right)};{\hat{v}}_{k|k-1}^{\left(n\right)},{\hat{\boldsymbol{V}}}_{k|k-1}^{\left(n\right)}\right)\\\cdot\mathcal{G}\left(\lambda_{k}^{\left(n\right)};{\hat{\alpha}}_{k|k-1}^{\left(n\right)},{\hat{\beta}}_{k|k-1}^{\left(n\right)}\right)\\\end{matrix}\right]}}_{\Diamond}}
\end{split}
\end{equation}

Then the Eq.(\ref{eqappenda1}) can then be rewritten as:
\begin{equation}
\label{eqappenda6}
\begin{split}
&\mathrm{ln}\ q_{\boldsymbol{\Theta}}\left(\boldsymbol{\Theta}_k^{1:L_k^\theta}\right) \\ &\propto\int{\int{\int{\left\{\Diamond\cdot q_\mathbf{x}\left(\mathbf{x}_k^{1:n_k}\right)q_\mathbf{X}\left(\mathbf{X}_k^{1:n_k}\right)q_\mathbf{\Lambda}\left(\mathbf{\Lambda}_k^{1:n_k}\right)\right\}d\mathbf{x}_k^{1:n_k}d\mathbf{X}_k^{1:n_k}d\mathbf{\Lambda}_k^{1:n_k}}}} \\
& \propto w_k^l\sum_{l=1}^{L_k^\theta}{\left[\mathrm{ln}\ \left(\rho\right)^{\phi_k^{l-\left(0\right)}}+ \mathrm{ln}\ \left(\lambda_c\right)^{\phi_k^{l-\left(0\right)}}+\sum_{n=1}^{n_k}\mathbb{E}_{\lambda_k^{\left(n\right)}}\left[\mathrm{ln}\ \left(\lambda_k^{\left(n\right)}\right)^{\phi_k^{l-\left(n\right)}}\right]\sum_{\boldsymbol{y}_k^{\widetilde{j}}\in{\widetilde{\mathcal{Y}}}_k^{l-\left(n\right)}}\left[\mathrm{ln}\ \mathcal{N}\left(\boldsymbol{y}_k^{\widetilde{j}};\mathbf{H}_k\hat{\boldsymbol{m}}_{k|k-1}^{\left(n\right)}, \boldsymbol{S}_k^{\left(n\right)}\right)\right]\right]}+\mathtt{C}_{\boldsymbol{\Theta}}
\end{split}
\end{equation}

\noindent where $\mathtt{C}_{\boldsymbol{\Theta}}$ is a constant term formed by absorbing all the terms in Eq.(\ref{eqappenda4}) that are unrelated to $\boldsymbol{\Theta}$.

Exponentiating the last line of Eq.(\ref{eqappenda6}) at the same time, we can obtain that:
\begin{equation}
\begin{split}
&q_{\boldsymbol{\Theta}}\left(\boldsymbol{\Theta}_k^{1:L_k^\theta}\right)\propto\prod_{l=1}^{L_k^\theta}\left[\left(\rho\right)^{\phi_k^{l-\left(0\right)}}\left(\lambda_c\right)^{\phi_k^{l-\left(0\right)}}\prod_{n=1}^{n_k}{\bar{\mathcal{U}}}_{\lambda_k^{\left(n\right)}}\left(\phi_k^{l-\left(n\right)}\right)\prod_{\boldsymbol{y}_k^{\widetilde{j}}\in{\widetilde{\mathcal{Y}}}_k^{l-\left(n\right)}}{\mathcal{N}\left(\boldsymbol{y}_k^{\widetilde{j}};\mathbf{H}_k\hat{\boldsymbol{m}}_{k|k-1}^{\left(n\right)},\boldsymbol{S}_k^{\left(n\right)}\right)}\right]^{w_k^l}
\end{split}
\end{equation}

\noindent where ${\bar{\mathcal{U}}}_{\lambda_k^{\left(n\right)}}\left(\phi_k^{l-\left(n\right)}\right)=\left(\mathbb{E}_{\lambda_k^{\left(n\right)}}\left(\lambda_k^{\left(n\right)}\right)\right)^{\phi_k^{l-\left(n\right)}}$. In addition, the term $\prod_{n=1}^{n_k}\prod_{\boldsymbol{y}_k^{\widetilde{j}}\in{\widetilde{\mathcal{Y}}}_k^{\left(n\right)}}\mathcal{N}\left(\boldsymbol{y}_k^{\widetilde{j}};\mathbf{H}_k\hat{\boldsymbol{m}}_{k|k-1}^{\left(n\right)},\boldsymbol{S}_k^{\left(n\right)}\right)$ is the general likelihood of the measurements generated by the $n$-th target under the $l$-th JAE $\boldsymbol{\theta}_k^l$.

\section{The Derivation of Eq.(\ref{eq33})}
\label{appendixA2}
Consistent with the unfolding rules used in Eq.(\ref{eqappenda1}), we have:	
\begin{equation}
\label{eqappendb1}
\begin{split}
q_\mathbf{x}\left(\mathbf{x}_k^{1:n_k}\right)&\propto\mathrm{exp}\left\{\int{\int{\int{\left[\mathrm{ln}\ p\left(\mathbf{x}_k^{1:n_k},\mathbf{X}_k^{1:n_k},\mathbf{\Lambda}_k^{1:n_k},\boldsymbol{\Theta}_k^{1:L_k^\theta},\mathcal{Y}_k\mid\mathcal{Y}^{k-1}\right)q_\mathbf{X}\left(\mathbf{X}_k^{1:n_k}\right)q_\mathbf{\Lambda}\left(\mathbf{\Lambda}_k^{1:n_k}\right)q_{\boldsymbol{\Theta}}\left(\boldsymbol{\Theta}_k^{1:L_k^\theta}\right)\right]d\mathbf{X}_k^{1:n_k}d\mathbf{\Lambda}_k^{1:n_k}}d\boldsymbol{\Theta}_k^{1:L_k^\theta}}\ }\right\} \\
&\propto\mathrm{exp}\left\{\sum_{w_k^l,l=1,2...,L_k^\theta}{q_{\boldsymbol{\Theta}}\left(\boldsymbol{\Theta}_k^{1:L_k^\theta}\right)\left[\int\int{\left[\mathrm{ln}\ p\left(\mathbf{x}_k^{1:n_k},\mathbf{X}_k^{1:n_k},\mathbf{\Lambda}_k^{1:n_k},\boldsymbol{\Theta}_k^{1:L_k^\theta},\mathcal{Y}_k\mid\mathcal{Y}^{k-1}\right)q_\mathbf{X}\left(\mathbf{X}_k^{1:n_k}\right)q_\mathbf{\Lambda}\left(\mathbf{\Lambda}_k^{1:n_k}\right)\right]d\mathbf{X}_k^{1:n_k}d\mathbf{\Lambda}_k^{1:n_k}}\right]}\right\}
\end{split}
\end{equation}

The last step of Eq.(\ref{eqappendb1}) holds because the JAEs are discrete and the rule can be referred to Eq. (58) in \cite{2023Yang}. Notice that the idea of an update process is consistent with traditional JPDA since we have modeled the JAE. 

Given an individual JAE, JPDA assumes the targets’ states are mutually independent. Therefore, for each component $q_{\boldsymbol{x}_k^{\left(n\right)}}\left(\boldsymbol{x}_k^{\left(n\right)}\right),n=1,2,...,n_k$ in $q_\mathbf{x}\left(\mathbf{x}_k^{1:n_k}\right)$, we obtain:
\begin{equation}
\small
\begin{split}
\label{eqappendb2}
&q_{x_k^{\left(n\right)}}\left(x_k^{\left(n\right)}\right)\\
&\propto \mathrm{exp}\left\{\sum_{w_k^l,l=1,2...,L_k^\theta}{q_\mathbf{\Theta}\left(\mathbf{\Theta}_k^{1:L_k^\theta}\right)\left[\int{{\underbrace{\cdots}_{{\left(3n_k-1\right)\times\int}}}\int{\left[\mathrm{ln}\ p\left(\mathbf{x}_k^{1:n_k},\mathbf{X}_k^{1:n_k},\mathbf{\Lambda}_k^{1:n_k},\mathbf{\Theta}_k^{1:L_k^\theta},\mathcal{Y}_k\mid\mathcal{Y}^{k-1}\right)\prod_{\widetilde{n}=1,n\neq\widetilde{n}}^{n_k}{q_{\boldsymbol{x}_k^{\left(\widetilde{n}\right)}}\left(\boldsymbol{x}_k^{\left(\widetilde{n}\right)}\right)}\prod_{n=1}^{n_k}{q_{\boldsymbol{X}_k^{\left(n\right)}}\left(\boldsymbol{X}_k^{\left(n\right)}\right)q_{\lambda_k^{\left(n\right)}}\left(\lambda_k^{\left(n\right)}\right)}\right] {\underbrace{d\boldsymbol{x}_k^{(1:n_k)}}_{{\widetilde{n}\neq n}}}d\boldsymbol{X}_k^{(1:n_k)}d\lambda_k^{(1:n_k)}}}\right]}\right\}
\end{split}
\end{equation}

Then, the joint posterior $p\left(\mathbf{x}_k^{1:n_k},\mathbf{X}_k^{1:n_k},\mathbf{\Lambda}_k^{1:n_k},\mathbf{\Theta}_k^{1:L_k^\theta},\mathcal{Y}_k\mid\mathcal{Y}^{k-1}\right)$ in Eq.(\ref{eqappendb2}) can be split into two parts of the relevant and irrelevant parameters with respect to the $n$-th target's state, i.e.,
\begin{equation}
\label{eqappendb3}
\normalsize
\begin{split}
&p\left(\mathbf{x}_k^{1:n_k},\mathbf{X}_k^{1:n_k},\mathbf{\Lambda}_k^{1:n_k},\mathbf{\Theta}_k^{1:L_k^\theta},\mathcal{Y}_k\mid\mathcal{Y}^{k-1}\right) \\
&=\prod_{l=1}^{L_k^\theta}\left[\mathcal{GGIW}\left(\boldsymbol{\xi}_k^{\left(n\right)};{\hat{\boldsymbol{\zeta}}}_{k|k-1}^{\left(n\right)}\right)e^{-\lambda_k^{\left(n\right)}}\left(\lambda_k^{\left(n\right)}\right)^{\phi_k^{l-\left(n\right)}}\prod_{\boldsymbol{y}_k^{\widetilde{j}}\in{\widetilde{\mathcal{Y}}}_k^{l-\left(n\right)}}{\mathcal{N}\left(\boldsymbol{y}_k^{\widetilde{j}};\mathbf{H}_k\boldsymbol{x}_{k|k-1}^{\left(n\right)},\boldsymbol{S}_k^{\left(n\right)}\right)}\right]^{w_k^l} \\
&\times{
	\prod_{l=1}^{L_k^\theta}\left[	\underbrace{\left(\rho\right)^{\phi_k^{l-\left(0\right)}}\left(\lambda_c\right)^{\phi_k^{l-\left(0\right)}}\prod_{\widetilde{n}=1,n\neq\widetilde{n}}^{n_k}\left(\mathcal{GGIW}\left(\boldsymbol{\xi}_k^{\left(\widetilde{n}\right)};{\hat{\zeta}}_{k|k-1}^{\left(\widetilde{n}\right)}\right)e^{-\lambda_k^{\left(\widetilde{n}\right)}}\left(\lambda_k^{\left(n\right)}\right)^{\phi_k^{l-\left(\widetilde{n}\right)}}\prod_{\boldsymbol{y}_k^{\widetilde{j}}\in{\widetilde{\mathcal{Y}}}_k^{l-\left(\widetilde{n}\right)}}{\mathcal{N}\left(y_k^{\widetilde{j}};\mathbf{H}_k\boldsymbol{x}_{k|k-1}^{\left(\widetilde{n}\right)},\boldsymbol{S}_k^{\left(\widetilde{n}\right)}\right)}\right)}_{\divideontimes}\right]^{w_k^l}
}
\end{split}
\end{equation}

As term `$\divideontimes$' is independent of $n$-th target’s state, it is absorbed into constant $\mathtt{C}_{\boldsymbol{x}_k^{\left(n\right)}}$ in subsequent integrals, allowing Eq.(\ref{eqappendb2}) to be reduced to:
\begin{equation}
\small
\label{eqappendb4}
\begin{split}
&\mathrm{ln}\ q_{\boldsymbol{x}_k^{\left(n\right)}}\left(\boldsymbol{x}_k^{\left(n\right)}\right) \\
&\propto\sum_{w_k^l,l=1,2...,L_k^\theta}{q_\mathbf{\Theta}\left(\mathbf{\Theta}_k^{1:L_k^\theta}\right)}\left\{\int{\int{\mathrm{ln}\ \prod_{l=1}^{L_k^\theta}\left[\mathcal{GGIW}\left(\boldsymbol{\xi}_k^{\left(n\right)};{\hat{\boldsymbol{\zeta}}}_{k|k-1}^{\left(n\right)}\right)e^{-\lambda_k^{\left(n\right)}}\left(\lambda_k^{\left(n\right)}\right)^{\phi_k^{l-\left(n\right)}}\prod_{\boldsymbol{y}_k^{\widetilde{j}}\in{\widetilde{\mathcal{Y}}}_k^{l-\left(n\right)}}{\mathcal{N}\left(\boldsymbol{y}_k^{\widetilde{j}};\mathrm{H}_k\boldsymbol{x}_{k|k-1}^{\left(n\right)},\boldsymbol{S}_k^{\left(n\right)}\right)}\right]^{w_k^l}q_{\boldsymbol{X}_k^{\left(n\right)}}\left(\boldsymbol{X}_k^{\left(n\right)}\right)q_{\lambda_k^{\left(n\right)}}\left(\lambda_k^{\left(n\right)}\right)d\boldsymbol{X}_k^{\left(n\right)}d\lambda_k^{\left(n\right)}}\ }\right\}+\mathtt{C}_{\boldsymbol{x}_k^{\left(n\right)}}\\
&\propto\sum_{w_k^l,l=1,2...,L_k^\theta}{q_\mathbf{\Theta}\left(\mathbf{\Theta}_k^{1:L_k^\theta}\right)}\left\{\int{\int{\mathrm{ln}\ 		\prod_{l=1}^{L_k^\theta}\left[\bigstar\right]^{w_k^l}q_{\boldsymbol{X}_k^{\left(n\right)}}\left(\boldsymbol{X}_k^{\left(n\right)}\right)q_{\lambda_k^{\left(n\right)}}\left(\lambda_k^{\left(n\right)}\right)d\boldsymbol{X}_k^{\left(n\right)}d\lambda_k^{\left(n\right)}}\ }\right\}+\mathtt{C}_{\boldsymbol{x}_k^{\left(n\right)}}
\end{split}
\end{equation}

Eq.(\ref{eqappendb4}) is important because we will use the conclusion of the formula in the subsequent proof of Eq.(\ref{eqappendc1}) and Eq.(\ref{eqappendd1}), and the symbol `$\bigstar$' in Eq.(\ref{eqappendb4}) is the product of the following six terms:
\normalsize
\begin{equation}
\label{eqappendb5}
\begin{split}
\bigstar=&\mathcal{N}\left(\boldsymbol{x}_k^{\left(n\right)};{\hat{\boldsymbol{m}}}_{k|k-1}^{\left(n\right)},{\hat{\boldsymbol{P}}}_{k|k-1}^{\left(n\right)}\right)\cdot\mathcal{IW}\left(\boldsymbol{X}_k^{\left(n\right)};{\hat{v}}_{k|k-1}^{\left(n\right)},{\hat{\boldsymbol{V}}}_{k|k-1}^{\left(n\right)}\right)\cdot\mathcal{G}\left(\lambda_k^{\left(n\right)};{\hat{\alpha}}_{k|k-1}^{\left(n\right)},{\hat{\beta}}_{k|k-1}^{\left(n\right)}\right)\cdot \hookleftarrow \\
&\hookrightarrow\left(e^{-\lambda_k^{\left(n\right)}}\left(\lambda_k^{\left(n\right)}\right)^{\phi_k^{l-\left(n\right)}}\right)\cdot\mathcal{N}\left({\bar{\boldsymbol{y}}}_k^{\left(n\right)};\mathbf{H}_k\boldsymbol{x}_k^{\left(n\right)},\frac{\boldsymbol{D}_k^{\left(n\right)}\boldsymbol{X}_k^{\left(n\right)}\left(\boldsymbol{D}_k^{\left(n\right)}\right)^{\mathrm{T}}}{\phi_k^{l-\left(n\right)}}\right)\cdot\mathcal{W}\left({\bar{\boldsymbol{Y}}}_k^{\left(n\right)};\phi_k^{l-\left(n\right)}-1,\boldsymbol{D}_k^{\left(n\right)}\boldsymbol{X}_k^{\left(n\right)}\left(\boldsymbol{D}_k^{\left(n\right)}\right)^{\mathrm{T}}\right)
\end{split}
\end{equation}

So the term ``$\mathrm{ln}\ \prod_{l}^{L_k^\theta}\left[\bigstar\right]^{w_k^l}$'' in Eq.(\ref{eqappendb4}) can be expanded as:
\begin{equation}
\label{eqappendb6}
\begin{split}
&\mathrm{ln}\ \prod_{l}^{L_k^\theta}\left[\bigstar\right]^{w_k^l} = \\ &w_k^l\sum_{l=1}^{L_k^\theta}{\left[\begin{matrix}-\frac{1}{2}\left({\bar{\boldsymbol{y}}}_k^{(n)}-\mathbf{H}_k\boldsymbol{x}_k^{\left(n\right)}\right)^{\mathrm{T}}\left(\frac{\boldsymbol{D}_k^{\left(n\right)}\boldsymbol{X}_k^{\left(n\right)}\left(\boldsymbol{D}_k^{\left(n\right)}\right)^{\mathrm{T}}}{\phi_k^{l-\left(n\right)}}\right)^{-1}\left({\bar{\boldsymbol{y}}}_k^{(n)}-\mathbf{H}_k\boldsymbol{x}_k^{\left(n\right)}\right)-\frac{1}{2}\left(\boldsymbol{x}_k^{\left(n\right)}-{\hat{\boldsymbol{m}}}_{k|k-1}^{\left(n\right)}\right)^{\mathrm{T}}\left({\hat{\boldsymbol{P}}}_{k|k-1}^{\left(n\right)}\right)^{-1}\left(\boldsymbol{x}_k^{\left(n\right)}-{\hat{\boldsymbol{m}}}_{k|k-1}^{\left(n\right)}\right)-\frac{1}{2}\mathrm{tr}\left(\left(\boldsymbol{D}_k^{\left(n\right)}\boldsymbol{X}_k^{\left(n\right)}\left(\boldsymbol{D}_k^{\left(n\right)}\right)^{\mathrm{T}}\right)^{-1}{\bar{\boldsymbol{Y}}}_k^{\left(n\right)}\right)\\-\frac{1}{2}\mathrm{tr}\left({{\hat{\boldsymbol{V}}}_{k|k-1}^{\left(n\right)}\left(\boldsymbol{X}_k^{\left(n\right)}\right)}^{-1}\right)-\frac{1}{2}\left({\hat{v}}_{k|k-1}^{\left(n\right)}{+\phi}_k^{l-\left(n\right)}\right)\mathrm{ln}\left|\boldsymbol{X}_k^{\left(n\right)}\right|+\left({\hat{\alpha}}_{k|k-1}^{\left(n\right)}+\phi_k^{l-\left(n\right)}-1\right)\mathrm{ln}\ \lambda_k^{\left(n\right)}-\left({\hat{\beta}}_{k|k-1}^{\left(n\right)}+1\right)\lambda_k^{\left(n\right)}\\\end{matrix}\right]\ }
\end{split}
\end{equation}

To facilitate the presentation of the derivation process of Eq.(\ref{eqappendb6}), we list the specific expanded form of each term in Eq.(\ref{eqappendb6}) as follows:
\begin{subequations}
\label{eqappendb7}
\begin{align}
&\mathrm{ln}\ \left[\mathcal{N}\left(\boldsymbol{x}_k^{\left(n\right)};{\hat{\boldsymbol{m}}}_{k|k-1}^{\left(n\right)},{\hat{\boldsymbol{P}}}_{k|k-1}^{\left(n\right)}\right)\right]={\underbrace{\mathrm{ln}\left[\left(2\pi\right)^{-\frac{n_d}{2}}\right]-\frac{1}{2}\mathrm{ln}\left[\left|{\hat{\boldsymbol{P}}}_{k|k-1}^{\left(n\right)}\right|\right]}_{\text{discard}}}-\frac{1}{2}\left(\boldsymbol{x}_k^{\left(n\right)}-{\hat{\boldsymbol{m}}}_{k|k-1}^{\left(n\right)}\right)^{\mathrm{T}}\left({\hat{\boldsymbol{P}}}_{k|k-1}^{\left(n\right)}\right)^{-1}\left(\boldsymbol{x}_k^{\left(n\right)}-{\hat{\boldsymbol{m}}}_{k|k-1}^{\left(n\right)}\right) \\
& \mathrm{ln}\ \left[\mathcal{IW}\left(\boldsymbol{X}_k^{\left(n\right)};{\hat{v}}_{k|k-1}^{\left(n\right)},{\hat{\boldsymbol{V}}}_{k|k-1}^{\left(n\right)}\right)\right]={\underbrace{-\mathrm{ln}\left[2^\frac{{\hat{v}_{k|k-1}n_d}}{2}\right]-\frac{{\hat{v}}_{k|k-1}^{\left(n\right)}}{2}\mathrm{ln}\ \left|{\hat{\boldsymbol{V}}}_{k|k-1}^{\left(n\right)}\right|-\mathrm{ln}\left[\Gamma_{n_d}\left(\frac{{\hat{v}}_{k|k-1}^{\left(n\right)}}{2}\right)\right]}_{\text{discard}}}-\frac{{\hat{v}}_{k|k-1}^{\left(n\right)}}{2}\mathrm{ln}\left|\boldsymbol{X}_k^{\left(n\right)}\right|-\frac{\mathrm{tr}\left({\hat{\boldsymbol{V}}}_{k|k-1}^{\left(n\right)}\left(\boldsymbol{X}_k^{\left(n\right)}\right)^{-1}\right)}{2} \\
& \mathrm{ln}\ \left[\mathcal{G}\left(\lambda_k^{\left(n\right)};{\hat{\alpha}}_{k|k-1}^{\left(n\right)},{\hat{\beta}}_{k|k-1}^{\left(n\right)}\right)\right]={\underbrace{{\hat{\alpha}}_{k|k-1}^{\left(n\right)}\mathrm{ln}\left({\hat{\beta}}_{k|k-1}^{\left(n\right)}\right)}_{\text{discard}}}+\left({\hat{\alpha}}_{k|k-1}^{\left(n\right)}-1\right)\mathrm{ln}\lambda_k^{\left(n\right)}-{\hat{\beta}}_{k|k-1}^{\left(n\right)}\lambda_k^{\left(n\right)} \\
& \mathrm{ln}\ \left[e^{-\lambda_k^{\left(n\right)}}\left(\lambda_k^{\left(n\right)}\right)^{\phi_k^{l-\left(n\right)}}\right]=-\lambda_k^{\left(n\right)}+\phi_k^{l-\left(n\right)}\mathrm{ln}\lambda_k^{\left(n\right)} \\
& \mathrm{ln}\ \left[\mathcal{N}\left({\bar{\boldsymbol{y}}}_k^{\left(n\right)};\mathbf{H}_k\boldsymbol{x}_k^{\left(n\right)},\frac{\boldsymbol{D}_k^{\left(n\right)}\boldsymbol{X}_k^{\left(n\right)}\left(\boldsymbol{D}_k^{\left(n\right)}\right)^{\mathrm{T}}}{\phi_k^{l-\left(n\right)}}\right)\mathcal{W}\left({\bar{\boldsymbol{Y}}}_k^{\left(n\right)};\phi_k^{l-\left(n\right)}-1,\boldsymbol{D}_k^{\left(n\right)}\boldsymbol{X}_k^{\left(n\right)}\left(\boldsymbol{D}_k^{\left(n\right)}\right)^{\mathrm{T}}\right)\right] \nonumber \\
&=-\frac{1}{2}\left({\bar{\boldsymbol{y}}}_k^{(n)}-\mathbf{H}_k\boldsymbol{x}_k^{\left(n\right)}\right)^{\mathrm{T}}\left(\frac{\boldsymbol{D}_k^{\left(n\right)}\boldsymbol{X}_k^{\left(n\right)}\left(\boldsymbol{D}_k^{\left(n\right)}\right)^{\mathrm{T}}}{\phi_k^{l-\left(n\right)}}\right)^{-1}\left({\bar{\boldsymbol{y}}}_k^{(n)}-\mathbf{H}_k\boldsymbol{x}_k^{\left(n\right)}\right)-\frac{1}{2}\left(\phi_k^{l-\left(n\right)}\right)\mathrm{ln}\left(\left|\boldsymbol{D}_k^{\left(n\right)}\boldsymbol{X}_k^{\left(n\right)}\left(\boldsymbol{D}_k^{\left(n\right)}\right)^{\mathrm{T}}\right|\right) \hookleftarrow \nonumber \\
&\qquad \qquad\hookrightarrow+{\underbrace{\frac{1}{2}\left(\phi_k^{l-\left(n\right)}-4\right)\mathrm{ln} \left(\left|{\bar{\boldsymbol{Y}}}_k^{\left(n\right)}\right|\right)}_{\text{discard}}} -\frac{1}{2}\mathrm{tr}\left[\left(\boldsymbol{D}_k^{\left(n\right)}\boldsymbol{X}_k^{\left(n\right)}\left(\boldsymbol{D}_k^{\left(n\right)}\right)^{\mathrm{T}}\right)^{-1}{\bar{\boldsymbol{Y}}}_k^{\left(n\right)}\right]
\end{align}
\end{subequations}

Substituting Eq.(\ref{eqappendb6}) into Eq.(\ref{eqappendb4}) and integrating out the terms independent of the state $\boldsymbol{x}_k^{\left(n\right)}$, we can derive that:
\begin{equation}
\label{eqappendb8}
\begin{split}
&\mathrm{ln}\ q_{x_k^{\left(n\right)}}\left(x_k^{\left(n\right)}\right) \\
&\propto \sum_{w_k^l,l=1,2...,L_k^\theta} q\left(w_k^l\right)\left(-\frac{1}{2}\left({\bar{\boldsymbol{y}}}_k^{(n)}-\mathbf{H}_k\boldsymbol{x}_k^{\left(n\right)}\right)^T\left(\frac{\boldsymbol{D}_k^{\left(n\right)}\mathbb{E}_{\boldsymbol{X}_k^{\left(n\right)}}\left(\boldsymbol{X}_k^{\left(n\right)}\right)\left(\boldsymbol{D}_k^{\left(n\right)}\right)^{\mathrm{T}}}{\phi_k^{l-\left(n\right)}}\right)^{-1}\left({\bar{\boldsymbol{y}}}_k^{(n)}-\mathbf{H}_k\boldsymbol{x}_k^{\left(n\right)}\right)-\frac{1}{2}\left(\boldsymbol{x}_k^{\left(n\right)}-{\hat{\boldsymbol{m}}}_{k|k-1}^{\left(n\right)}\right)^{\mathrm{T}}\left({\hat{\boldsymbol{P}}}_{k|k-1}^{\left(n\right)}\right)^{-1}\left(\boldsymbol{x}_k^{\left(n\right)}-{\hat{\boldsymbol{m}}}_{k|k-1}^{\left(n\right)}\right)\right)+\mathtt{C}_{\boldsymbol{x}_k^{\left(n\right)}} \\
&= -\frac{1}{2}\left(\boldsymbol{x}_k^{\left(n\right)}-{\hat{\boldsymbol{m}}}_{k|k-1}^{\left(n\right)}\right)^{\mathrm{T}}\left({\hat{\boldsymbol{P}}}_{k|k-1}^{\left(n\right)}\right)^{-1}\left(\boldsymbol{x}_k^{\left(n\right)}-{\hat{\boldsymbol{m}}}_{k|k-1}^{\left(n\right)}\right)\cdot{\underbrace{\sum_{w_k^l,l=1,2...,L_k^\theta} q\left(w_k^l\right)}_{1}}-\frac{1}{2}\left(\frac{\boldsymbol{A}_1^{\left(n\right)}}{\boldsymbol{A}_2^{\left(n\right)}}-\mathbf{H}_k\boldsymbol{x}_k^{\left(n\right)}\right)^{\mathrm{T}}\left(\frac{\boldsymbol{D}_k^{\left(n\right)}\mathbb{E}_{\boldsymbol{X}_k^{\left(n\right)}}\left(\boldsymbol{X}_k^{\left(n\right)}\right)\left(\boldsymbol{D}_k^{\left(n\right)}\right)^{\mathrm{T}}}{\boldsymbol{A}_2^{\left(n\right)}}\right)^{-1}\left(\frac{\boldsymbol{A}_1^{\left(n\right)}}{\boldsymbol{A}_2^{\left(n\right)}}-\mathbf{H}_k\boldsymbol{x}_k^{\left(n\right)}\right)+\mathtt{C}_{\boldsymbol{x}_k^{\left(n\right)}}
\end{split}
\end{equation}

\noindent where $\mathbb{E}_{\boldsymbol{X}_k^{\left(n\right)}}\left(\boldsymbol{X}_k^{\left(n\right)}\right)=\frac{\boldsymbol{V}_{k|k}^{\left(n\right)}}{\left(v_{k|k}^{\left(n\right)}-n_d-1\right)}$ is the number of measures assigned to the first target in the $n$-th JAE. $q\left(w_k^l\right)$ satisfies that $q\left(w_k^l\right)=q_{\boldsymbol{\Theta}}\left(\boldsymbol{\Theta}_k^{1:L_k^\theta}\right)$ when $w_k^l=1$. And we have $\sum_{w_k^l,l=1,2...,L_k^\theta} q\left(w_k^l\right)=1$.  $\boldsymbol{A}_1^{\left(n\right)}$ and $\boldsymbol{A}_2^{\left(n\right)}$ in Eq.(\ref{eqappendb8}) are two auxiliary parameters that can be computed as:
\begin{equation}
\normalsize
\label{eqappendb9}
\boldsymbol{A}_1^{\left(n\right)}=\sum_{l=1}^{L_k^\theta}{q\left(w_k^l\right)\phi_k^{l-\left(n\right)}{\bar{\boldsymbol{y}}}_k^n}
\end{equation}
\begin{equation}
\label{eqappendb10}
\boldsymbol{A}_2^{\left(n\right)}=\sum_{l=1}^{L_k^\theta}{q\left(w_k^l\right)\phi_k^{l-\left(n\right)}}
\end{equation}

Ignoring the constant term and rewrite Eq.(\ref{eqappendb8}) as:
\begin{equation}
\label{eqappendb11}
\mathrm{ln}\ q_{\boldsymbol{x}_k^{\left(n\right)}}\left(\boldsymbol{x}_k^{\left(n\right)}\right)\propto \mathrm{ln}\ \mathcal{N}\left(\boldsymbol{x}_k^{\left(n\right)};{\hat{\boldsymbol{m}}}_{k|k-1}^{\left(n\right)},{\hat{\boldsymbol{P}}}_{k|k-1}^{\left(n\right)}\right)+ \mathrm{ln}\ \mathcal{N}\left(\frac{\boldsymbol{A}_1^{\left(n\right)}}{\boldsymbol{A}_2^{\left(n\right)}};\mathbf{H}_k\boldsymbol{x}_k^{\left(n\right)},\frac{\boldsymbol{D}_k^{\left(n\right)}\mathbb{E}_{\boldsymbol{X}_k^{\left(n\right)}}\left(\boldsymbol{X}_k^{\left(n\right)}\right)\left(\boldsymbol{D}_k^{\left(n\right)}\right)^{\mathrm{T}}}{\boldsymbol{A}_2^{\left(n\right)}}\right)
\end{equation}

Exponentiating both sides of Eq.(\ref{eqappendb11}) shows that it can be expressed in the form of the product of two Gaussian distributions. It is natural to express it as an individual Gaussian distribution according to Eq.(28) and Eq.(31) in \cite{2023Yang}, with mean ${\hat{\boldsymbol{m}}}_{k|k}^{\left(n\right)}$ and covariance ${\hat{\boldsymbol{P}}}_{k|k}^{\left(n\right)}$, i.e:
\begin{equation}
\label{eqappendb12}
q_{\boldsymbol{x}_k^{\left(n\right)}}\left(\boldsymbol{x}_k^{\left(n\right)}\right)\sim\mathcal{N}\left(\boldsymbol{x}_k^{\left(n\right)};{\hat{\boldsymbol{m}}}_{k|k}^{\left(n\right)},{\hat{\boldsymbol{P}}}_{k|k}^{\left(n\right)}\right)
\end{equation}

\noindent where the values of ${\hat{\boldsymbol{m}}}_{k|k}^{\left(n\right)}$ and ${\hat{\boldsymbol{P}}}_{k|k}^{\left(n\right)}$ have been given by Eq.(\ref{eq33}).

\section{The Derivation of Eq.(\ref{eq31})}
\label{appendixA3}
Similarly, we inherit the conclusions in Eq.(\ref{eqappendb4}) and obtain that:
\begin{equation}
\label{eqappendc1}
\mathrm{ln}\ q_{\lambda_k^{\left(n\right)}}\left(\lambda_k^{\left(n\right)}\right)\propto\sum_{w_k^l,l=1,2...,L_k^\theta}{q_\mathbf{\Theta}\left(\mathbf{\Theta}_k^{1:L_k^\theta}\right)\left\{\int{\int{\mathrm{ln}\ \prod_{{l^{'}}}^{L_k^\theta}\left[\bigstar\right]^{w_k^{l^{'}}}q_{\boldsymbol{x}_k^{\left(n\right)}}\left(\boldsymbol{x}_k^{\left(n\right)}\right)q_{\boldsymbol{X}_k^{\left(n\right)}}\left(\boldsymbol{X}_k^{\left(n\right)}\right)d\boldsymbol{x}_k^{\left(n\right)}d\boldsymbol{X}_k^{\left(n\right)}}\ }\right\}}+\mathtt{C}_{\lambda_k^{\left(n\right)}}
\end{equation}

Expanding the integral term in Eq.(\ref{eqappendc1}) and absorbing the term independent of $\lambda_k^{\left(n\right)}$ into the constant term, we can derive that:
\begin{equation}
\label{eqappendc2}
\begin{split}
\mathrm{ln}\ q_{\lambda_k^{\left(n\right)}}\left(\lambda_k^{\left(n\right)}\right)\propto\sum_{w_k^l,l=1,2...,L_k^\theta}{q_\mathbf{\Theta}\left(\mathbf{\Theta}_k^{1:L_k^\theta}\right)\left[\left({\hat{\alpha}}_{k|k-1}^{\left(n\right)}+\phi_k^{l-\left(n\right)}-1\right)\mathrm{ln}\ \lambda_k^{\left(n\right)}-\left({\hat{\beta}}_{k|k-1}^{\left(n\right)}+1\right)\lambda_k^{\left(n\right)}\right]}+\mathtt{C}_{\lambda_k^{\left(n\right)}}
\end{split}
\end{equation}

Ignoring the constant term $\mathtt{C}_{\lambda_k^{\left(n\right)}}$ and simultaneously taking exponents and normalizing both sides of Eq.(\ref{eqappendc2}), we have:
\begin{equation}
q_{\lambda_k^{\left(n\right)}}\left(\lambda_k^{\left(n\right)}\right)\sim\mathcal{G}\left(\lambda_k^{\left(n\right)};{\hat{\alpha}}_{k|k}^{\left(n\right)},{\hat{\beta}}_{k|k}^{\left(n\right)}\right)
\end{equation}

\noindent where the values of ${\hat{\alpha}}_{k|k}^{\left(n\right)}$ and ${\hat{\beta}}_{k|k}^{\left(n\right)}$ are indicated by Eq.(\ref{eq31}).

\section{The Derivation of Eq.(\ref{eq34})}
\label{appendixA4}
According to Eq.(\ref{eqappendb4}), we have:
\begin{equation}
\label{eqappendd1}
\mathrm{ln}\ q_{\boldsymbol{X}_k^{\left(n\right)}}\left(\boldsymbol{X}_k^{\left(n\right)}\right)\propto\sum_{w_k^l,l=1,2...,L_k^\theta}{q_\mathbf{\Theta}\left(\mathbf{\Theta}_k^{1:L_k^\theta}\right)\left\{\int{\int{\mathrm{ln}\ 		\prod_{l^{'}}^{L_k^\theta}\left[\bigstar\right]^{w_k^{l^{'}}}q_{\boldsymbol{x}_k^{\left(n\right)}}\left(\boldsymbol{x}_k^{\left(n\right)}\right)q_{\lambda_k^{\left(n\right)}}\left(\lambda_k^{\left(n\right)}\right)d\boldsymbol{x}_k^{\left(n\right)}d\lambda_k^{\left(n\right)}}\ }\right\}}+\mathtt{C}_{\boldsymbol{X}_k^{\left(n\right)}}
\end{equation}

The expanded form of Eq.(\ref{eqappendd1}) can be written as:
\begin{equation}
\label{eqappendd2}
\begin{split}
&\mathrm{ln}\ q_{\boldsymbol{X}_k^{\left(n\right)}}\left(\boldsymbol{X}_k^{\left(n\right)}\right) \\
&\propto\sum_{w_k^l,l=1,2...,L_k^\theta}{q_\mathbf{\Theta}\left(\mathbf{\Theta}_k^{1:L_k^\theta}\right)\left[\begin{matrix}-\frac{1}{2}\underbrace{\mathbb{E}_{\boldsymbol{x}_k^{\left(n\right)}}\left(\left({\bar{\boldsymbol{y}}}_k^{(n)}-\mathbf{H}_k\boldsymbol{x}_k^{\left(n\right)}\right)^{\mathrm{T}}\left(\frac{\boldsymbol{D}_k^{\left(n\right)}\boldsymbol{X}_k^{\left(n\right)}\left(\boldsymbol{D}_k^{\left(n\right)}\right)^T}{\phi_k^{l-\left(n\right)}}\right)^{-1}\left({\bar{\boldsymbol{y}}}_k^{(n)}-\mathbf{H}_k\boldsymbol{x}_k^{\left(n\right)}\right)\right)}_{\blacktriangle}\hookleftarrow\\\hookrightarrow-\frac{1}{2}\mathrm{tr}\left(\left(\boldsymbol{D}_k^{\left(n\right)}\boldsymbol{X}_k^{\left(n\right)}\left(\boldsymbol{D}_k^{\left(n\right)}\right)^{\mathrm{T}}\right)^{-1}{\bar{\boldsymbol{Y}}}_k^{\left(n\right)}\right)-\frac{1}{2}\mathrm{tr}\left({{\hat{\boldsymbol{V}}}_{k|k-1}^{\left(n\right)}\left(\boldsymbol{X}_k^{\left(n\right)}\right)}^{-1}\right)-\frac{1}{2}\left({\hat{v}}_{k|k-1}^{\left(n\right)}+\phi_k^{l-\left(n\right)}\right)\mathrm{ln}\left|\boldsymbol{X}_k^{\left(n\right)}\right|\\\end{matrix}\right]}+\mathtt{C}_{\boldsymbol{X}_k^{\left(n\right)}}
\end{split}
\end{equation}

The term `$\blacktriangle$' in Eq.(\ref{eqappendd2}) can be expanded as:
\begin{equation}
\label{eqappendd3}
\begin{split}
&\mathbb{E}_{\boldsymbol{x}_k^{\left(n\right)}}\left(\left({\bar{\boldsymbol{y}}}_k^{(n)}-\mathbf{H}_k\boldsymbol{x}_k^{\left(n\right)}\right)^{\mathrm{T}}\left(\frac{\boldsymbol{D}_k^{\left(n\right)}\boldsymbol{X}_k^{\left(n\right)}\left(\boldsymbol{D}_k^{\left(n\right)}\right)^{\mathrm{T}}}{\phi_k^{l-\left(n\right)}}\right)^{-1}\left({\bar{\boldsymbol{y}}}_k^{(n)}-\mathbf{H}_k\boldsymbol{x}_k^{\left(n\right)}\right)\right) \\
&=\mathbb{E}_{x_k^{\left(n\right)}}\left(\mathrm{tr}\left[\left({\bar{\boldsymbol{y}}}_k^{(n)}-\mathbf{H}_k\boldsymbol{x}_k^{\left(n\right)}\right)\left({\bar{\boldsymbol{y}}}_k^{(n)}-\mathbf{H}_k\boldsymbol{x}_k^{\left(n\right)}\right)^{\mathrm{T}}\left(\frac{\boldsymbol{D}_k^{\left(n\right)}\boldsymbol{X}_k^{\left(n\right)}\left(\boldsymbol{D}_k^{\left(n\right)}\right)^{\mathrm{T}}}{\phi_k^{l-\left(n\right)}}\right)^{-1}\right]\right) \\
&=\phi_k^{l-\left(n\right)}\mathrm{tr}\left[{\mathbb{E}_{\boldsymbol{x}_k^{\left(n\right)}}\left(\left({\bar{\boldsymbol{y}}}_k^{(n)}-\mathbf{H}_k\boldsymbol{x}_k^{\left(n\right)}\right)\left({\bar{\boldsymbol{y}}}_k^{(n)}-\mathbf{H}_k\boldsymbol{x}_k^{\left(n\right)}\right)^{\mathrm{T}}\right)\left(\boldsymbol{D}_k^{\left(n\right)}\boldsymbol{X}_k^{\left(n\right)}\left(\boldsymbol{D}_k^{\left(n\right)}\right)^{\mathrm{T}}\right)}^{-1}\right] \\
&=\phi_k^{l-\left(n\right)}\mathrm{tr}\left[{\left(\left({\bar{\boldsymbol{y}}}_k^{(n)}-\mathbf{H}_k{\hat{\boldsymbol{m}}}_{k|k}^{\left(n\right)}\right)\left({\bar{\boldsymbol{y}}}_k^{(n)}-\mathbf{H}_k{\hat{\boldsymbol{m}}}_{k|k}^{\left(n\right)}\right)^{\mathrm{T}}+\mathbf{H}_k{\hat{\boldsymbol{P}}}_{k|k}^{\left(n\right)}\mathbf{H}_k^{\mathrm{T}}\right)\left(\boldsymbol{D}_k^{\left(n\right)}\boldsymbol{X}_k^{\left(n\right)}\left(\boldsymbol{D}_k^{\left(n\right)}\right)^{\mathrm{T}}\right)}^{-1}\right]
\end{split}
\end{equation}

Substituting Eq.(\ref{eqappendd3}) into Eq.(\ref{eqappendd2}) we have:
\begin{equation}
\label{eqappendd4}
\footnotesize
\begin{split}
&\mathrm{ln}\ q_{\boldsymbol{X}_k^{\left(n\right)}}\left(\boldsymbol{X}_k^{\left(n\right)}\right) \propto  {\underbrace{\sum_{w_k^l,l=1,2...,L_k^\theta}{q_{\boldsymbol{\Theta}}\left(\boldsymbol{\Theta}_k^{1:L_k^\theta}\right)\left\{-\frac{1}{2}\left({\hat{v}}_{k|k-1}^{\left(n\right)}+\phi_k^{l-\left(n\right)}\right)\mathrm{ln}\left|\boldsymbol{X}_k^{\left(n\right)}\right|\right\}}}_{\text{Part\ A}}}- \hookleftarrow\\
&\hookrightarrow {\underbrace{\sum_{w_k^l,l=1,2...,L_k^\theta}{q_\mathbf{\Theta}\left(\mathbf{\Theta}_k^{1:L_k^\theta}\right)\left\{-\frac{1}{2}\left[\phi_k^{l-\left(n\right)}\mathrm{tr}\left[\left(\left({\bar{\boldsymbol{y}}}_k^n-\mathbf{H}_k{\hat{\boldsymbol{m}}}_{k|k}^{\left(n\right)}\right)\left({\bar{\boldsymbol{y}}}_k^n-\mathbf{H}_k{\hat{\boldsymbol{m}}}_{k|k}^{\left(n\right)}\right)^{\mathrm{T}}+\mathbf{H}_k{\hat{\boldsymbol{P}}}_{k|k}^{\left(n\right)}\mathbf{H}_k^{\mathrm{T}}\right)\left(\boldsymbol{D}_k^{\left(n\right)}\boldsymbol{X}_k^{\left(n\right)}\left(\boldsymbol{D}_k^{\left(n\right)}\right)^{\mathrm{T}}\right)^{-1}\right]\right]-\frac{1}{2}\mathrm{tr}\left(\left(\boldsymbol{D}_k^{\left(n\right)}\boldsymbol{X}_k^{\left(n\right)}\left(\boldsymbol{D}_k^{\left(n\right)}\right)^{\mathrm{T}}\right)^{-1}{\bar{\boldsymbol{Y}}}_k^{\left(n\right)}\right)-\frac{1}{2}\mathrm{tr}\left({{\hat{\boldsymbol{V}}}_{k|k-1}^{\left(n\right)}\left(\boldsymbol{X}_k^{\left(n\right)}\right)}^{-1}\right)\right\}}}_{\text{Part\ B}}} +\mathtt{C}_{\boldsymbol{X}_k^{\left(n\right)}}
\end{split}
\end{equation}

For Part A in Eq.(\ref{eqappendd4}) we have:
\begin{equation}
\label{eqappendd5}
\begin{split}
\sum_{w_k^l,l=1,2...,L_k^\theta}{q_\mathbf{\Theta}\left(\mathbf{\Theta}_k^{1:L_k^\theta}\right)\left\{-\frac{1}{2}\left({\hat{v}}_{k|k-1}^{\left(n\right)}+\phi_k^{l-\left(n\right)}\right)\mathrm{ln}\left|\boldsymbol{X}_k^{\left(n\right)}\right|\right\}}&=-\frac{1}{2}\left({\hat{v}}_{k|k-1}^{\left(n\right)}+\sum_{l=1}^{L_k^\theta}{q\left(w_k^l\right)\phi_k^{l-\left(n\right)}}\right)\mathrm{ln}\left|\boldsymbol{X}_k^{\left(n\right)}\right|\\
&=-\frac{1}{2}\left({\hat{v}}_{k|k-1}^{\left(n\right)}+\boldsymbol{A}_2^{\left(n\right)}\right)\mathrm{ln}\left|\boldsymbol{X}_k^{\left(n\right)}\right|
\end{split}
\end{equation}

And we can derive from Part B in Eq.(\ref{eqappendd4}) that:
\begin{equation}
\footnotesize
\label{eqappendd6}
\begin{split}
\sum_{w_k^l,l=1,2...,L_k^\theta}&{q_\mathbf{\Theta}\left(\mathbf{\Theta}_k^{1:L_k^\theta}\right)\left\{-\frac{1}{2}\left[\phi_k^{l-\left(n\right)}\mathrm{tr}\left[\left[\left({\bar{\boldsymbol{y}}}_k^{l-(n)}-\mathbf{H}_k{\hat{\boldsymbol{m}}}_{k|k}^{\left(n\right)}\right)\left({\bar{\boldsymbol{y}}}_k^{l-(n)}-\mathbf{H}_k{\hat{\boldsymbol{m}}}_{k|k}^{\left(n\right)}\right)^{\mathrm{T}}+\mathbf{H}_k{\hat{\boldsymbol{P}}}_{k|k}^{\left(n\right)}\mathbf{H}_k^{\mathrm{T}}\right]\left(\boldsymbol{D}_k^{\left(n\right)}\boldsymbol{X}_k^{\left(n\right)}\left(\boldsymbol{D}_k^{\left(n\right)}\right)^{\mathrm{T}}\right)^{-1}\right]\right]-\frac{1}{2}\mathrm{tr}\left(\left(\boldsymbol{D}_k^{\left(n\right)}\boldsymbol{X}_k^{\left(n\right)}\left(\boldsymbol{D}_k^{\left(n\right)}\right)^{\mathrm{T}}\right)^{-1}{\bar{\boldsymbol{Y}}}_k^{l-\left(n\right)}\right)-\frac{1}{2}\mathrm{tr}\left({{\hat{\boldsymbol{V}}}_{k|k-1}^{\left(n\right)}\left(\boldsymbol{X}_k^{\left(n\right)}\right)}^{-1}\right)\right\}}\\
&=\sum_{w_k^l,l=1,2...,L_k^\theta}{q_\mathbf{\Theta}\left(\mathbf{\Theta}_k^{1:L_k^\theta}\right)}\left\{\frac{1}{2}\mathrm{tr}\left({\left[{\hat{\boldsymbol{V}}}_{k|k-1}^{\left(n\right)}+\left(\left(\boldsymbol{D}_k^{\left(n\right)}\right)^{-1}\left({\bar{\boldsymbol{Y}}}_k^{\left(n\right)}+\phi_k^{l-\left(n\right)}\left({\bar{\boldsymbol{y}}}_k^n-\mathbf{H}_k{\hat{\boldsymbol{m}}}_{k|k}^{\left(n\right)}\right)\left({\bar{\boldsymbol{y}}}_k^n-\mathbf{H}_k{\hat{\boldsymbol{m}}}_{k|k}^{\left(n\right)}\right)^{\mathrm{T}}+\phi_k^{l-\left(n\right)}\mathbf{H}_k{\hat{\boldsymbol{P}}}_{k|k}^{\left(n\right)}\boldsymbol{H}_k^T\right)\left(\boldsymbol{D}_k^{\left(n\right)}\right)^{-\mathrm{T}}\right)\right]\left(\boldsymbol{X}_k^{\left(n\right)}\right)}^{-1}\right)\right\}
\end{split}
\end{equation}

Notice that in Eq.(\ref{eqappendd6}) we used two simple matrix equivalence relations:
\begin{subequations}
\begin{align}
&\left(\mathtt{UOU}^{\mathrm{T}}\right)^{-1}=\mathtt{U}^{-\mathrm{T}}\mathtt{O}^{-1}\mathtt{U}^{-1} \\
\mathrm{tr}\left(\mathtt{U}\mathtt{O}\right)+\mathrm{tr}\left(\mathtt{U}\mathtt{G}\right)&= \mathrm{tr}\left(\mathtt{U}(\mathtt{O}+\mathtt{G})\right) \quad \text{when $\mathtt{U,O,G}$ are all PSDMs}
\end{align}
\end{subequations}

Let $\boldsymbol{T}_k^{l-(n)}=\left({\boldsymbol{D}_k^{\left(n\right)}}^{-1}\left({\bar{\boldsymbol{Y}}}_k^{l-\left(n\right)}+\phi_k^{l-\left(n\right)}\left({\bar{\boldsymbol{y}}}_k^{l-(n)}-\mathbf{H}_k{\hat{\boldsymbol{m}}}_{k|k}^{\left(n\right)}\right)\left({\bar{\boldsymbol{y}}}_k^{l-(n)}-\mathbf{H}_k{\hat{\boldsymbol{m}}}_{k|k}^{\left(n\right)}\right)^{\mathrm{T}}+\phi_k^{l-\left(n\right)}\mathbf{H}_k{\hat{\boldsymbol{P}}}_{k|k}^{\left(n\right)}\mathbf{H}_k^{\mathrm{T}}\right)\left(\boldsymbol{D}_k^{\left(n\right)}\right)^{\mathrm{-T}}\right)$, together with Eq.(\ref{eqappendd3}) and Eq.(\ref{eqappendd5}), then Eq.(\ref{eqappendd6}) can be rewritten as:
\begin{equation}
\label{eqappendd7}
\normalsize
\mathrm{ln}\ q_{\boldsymbol{X}_k^{\left(n\right)}}\left(\boldsymbol{X}_k^{\left(n\right)}\right)\propto-\frac{1}{2}\left({\hat{v}}_{k|k-1}^{\left(n\right)}+\boldsymbol{A}_2^{\left(n\right)}\right)\mathrm{ln}\left|\boldsymbol{X}_k^{\left(n\right)}\right|-\frac{1}{2}\mathrm{tr}\left(\left[{\hat{\boldsymbol{V}}}_{k|k-1}^{\left(n\right)}+\sum_{l=1}^{L_k^\theta}{q\left(w_k^l\right)\boldsymbol{T}_k^{l-\left(n\right)}}\right]\left(\boldsymbol{X}_k^{\left(n\right)}\right)^{-1}\right)+\mathtt{C}_{\boldsymbol{X}_k^{\left(n\right)}}
\end{equation}

Ignoring the constant term $\mathtt{C}_{\boldsymbol{X}_k^{\left(n\right)}}$ and exponentiating both sides of Eq.(\ref{eqappendd7}), after normalizing we can obtain that:
\begin{equation}
q_{\boldsymbol{X}_k^{\left(n\right)}}\left(\boldsymbol{X}_k^{\left(n\right)}\right)\sim\mathcal{IW}\left(\boldsymbol{X}_k^{\left(n\right)};{\hat{v}}_{k|k}^{\left(n\right)},{\hat{\boldsymbol{V}}}_{k|k}^{\left(n\right)}\right)
\end{equation}

\noindent where ${\hat{v}}_{k|k}^{\left(n\right)}$ and ${\hat{\boldsymbol{V}}}_{k|k}^{\left(n\right)}$ are given by Eq.(\ref{eq34}).

\section{The Derivation of Eq.(\ref{eq38})}
\label{appendixA5}
Take the kinematic state of the $n$-th target at time $k$ as an example. According to Eq. (\ref{eq33}), the two parameters of its variational posterior $q_{\boldsymbol{x}_k^{\left(n\right)}}\left(\boldsymbol{x}_k^{\left(n\right)}\right)\sim\mathcal{N}\left(\boldsymbol{x}_k^{\left(n\right)};{\hat{\boldsymbol{m}}}_{k|k}^{\left(n\right)},{\hat{\boldsymbol{P}}}_{k|k}^{\left(n\right)}\right)$ can be calculated as:
\begin{subequations}
\label{eqappende1}
\begin{align}
&{\hat{\boldsymbol{m}}}_{k|k}^{\left(n\right)}={\hat{\boldsymbol{m}}}_{k|k-1}^{\left(n\right)}+\boldsymbol{K}^{\left(n\right)}\left(\frac{\sum_{l=1}^{L_k^\theta}{q_\theta\left(w_k^l\right)\phi_k^{l-\left(n\right)}{\bar{\boldsymbol{y}}}_k^{l-\left(n\right)}}}{\sum_{l=1}^{L_k^\theta}{q_\theta\left(w_k^l\right)\phi_k^{l-\left(n\right)}}}-\mathbf{H}_k{\hat{\boldsymbol{m}}}_{k|k-1}^{\left(n\right)}\right) \\
&{\hat{\boldsymbol{P}}}_{k|k}^{\left(n\right)}={\hat{\boldsymbol{P}}}_{k|k-1}^{\left(n\right)}-\boldsymbol{K}^{\left(n\right)}\mathbf{H}_k{\hat{\boldsymbol{P}}}_{k|k-1}^{\left(n\right)}\\
&\boldsymbol{K}^{\left(n\right)}={\hat{\boldsymbol{P}}}_{k|k-1}^{\left(n\right)}\mathbf{H}_k^{\mathrm{T}}\left(\mathbf{H}_k{\hat{\boldsymbol{P}}}_{k|k-1}^{\left(n\right)}\mathbf{H}_k^{\mathrm{T}}+\frac{\boldsymbol{D}_k^{\left(n\right)}\mathbb{E}_{\boldsymbol{X}_k^{\left(n\right)}}\left(\boldsymbol{X}_k^{\left(n\right)}\right)\left(\boldsymbol{D}_k^{\left(n\right)}\right)^{\mathrm{T}}}{\sum_{l=1}^{L_k^\theta}{q_\theta\left(w_k^l\right)\phi_k^{l-\left(n\right)}}}\right)^{-1}
\end{align}
\end{subequations}

Considering time $k$, the marginal probability $\epsilon_k^{\left(n\right)j}$ of an individual measurement $\boldsymbol{y}_k^j$ assigned to the $n$-th target can be expressed as a summation over JAEs, namely:
\begin{equation}
\epsilon_k^{\left(n\right)j}=p\left(\vartheta_j=n\mid\mathcal{Y}_k,\boldsymbol{\Xi}_k^{1:n_k}\right)
=\sum_{\boldsymbol{\theta}_k^l\ with\ \vartheta_j=n}{p\left(\boldsymbol{\theta}_k^l\mid\mathcal{Y}_k,\boldsymbol{\Xi}_k^{1:n_k}\right)}
\end{equation}

It is straightforward to see that the calculation of $\epsilon_k^{\left(n\right)j}$ also includes the traversal of all JAEs, which is similar to operation  $\sum_{l=1}^{L_k^\theta}{q_\theta\left(w_k^l\right)\phi_k^{l-\left(n\right)}}$. Using marginal association probabilities, the resulting JAE can be viewed as relating each measurement $\boldsymbol{y}_k^j, j=1,2,\cdots m_k$ to the $n$-th target with the possibility of marginal probabilities. In this case, for each JAE, there is $\phi_k^{l-\left(n\right)}=1$, and the number of JAEs at time $k$ is reduced to the number of measurements $m_k$. Therefore, the term ``$\sum_{l=1}^{L_k^\theta}{q_\theta\left(w_k^l\right)\phi_k^{l-\left(n\right)}}$" in Eq.(\ref{eqappende1}) naturally degenerates into ``$\sum_{j=1}^{m_k}\epsilon_k^{\left(n\right)j}$''. Since only an independent measurement is considered, the equivalent measurement ${\bar{\boldsymbol{y}}}_k^{l-\left(n\right)}$ decays to $\boldsymbol{y}_k^j$, and the term ``$\sum_{l=1}^{L_k^\theta}{q_\theta\left(w_k^l\right)\phi_k^{l-\left(n\right)}{\bar{\boldsymbol{y}}}_k^{l-\left(n\right)}}$'' in the formula reduces to ``$\sum_{j=1}^{m_k}{\epsilon_k^{\left(n\right)j}\boldsymbol{y}_k^j}$''. Substituting the reduced terms back into Eq.(\ref{eqappende1}), then we can obtain the approximate estimates of the parameters as:
\begin{subequations}
\begin{align}
&{\hat{\boldsymbol{m}}}_{k|k}^{\left(n\right)}\approx{\hat{\boldsymbol{m}}}_{k|k-1}^{\left(n\right)}+\boldsymbol{K}^{\left(n\right)}\left(\frac{\sum_{j=1}^{m_k}{\epsilon_k^{\left(n\right)j}\boldsymbol{y}_k^j}}{\sum_{j=1}^{m_k}\epsilon_k^{\left(n\right)j}}-\mathbf{H}_k{\hat{\boldsymbol{m}}}_{k|k-1}^{\left(n\right)}\right)\\
& {\hat{\boldsymbol{P}}}_{k|k}^{\left(n\right)}={\hat{\boldsymbol{P}}}_{k|k-1}^{\left(n\right)}-\boldsymbol{K}^{\left(n\right)}\mathbf{H}_k{\hat{\boldsymbol{P}}}_{k|k-1}^{\left(n\right)} \\
& \boldsymbol{K}^{\left(n\right)}\approx{\hat{\boldsymbol{P}}}_{k|k-1}^{\left(n\right)}\mathbf{H}_k^{\mathrm{T}}\left(\mathbf{H}_k{\hat{\boldsymbol{P}}}_{k|k-1}^{\left(n\right)}\mathbf{H}_k^{\mathrm{T}}+\frac{\boldsymbol{D}_k^{\left(n\right)}\mathbb{E}_{\boldsymbol{X}_k^{\left(n\right)}}\left(\boldsymbol{X}_k^{\left(n\right)}\right)\left(\boldsymbol{D}_k^{\left(n\right)}\right)^T}{\sum_{j=1}^{m_k}\epsilon_k^{\left(n\right)j}}\right)^{-1}
\end{align}
\end{subequations}

We use ``$\approx$'' is because of the information loss caused by attenuation, mainly because we ignore the number ``$\phi_k^{l-\left(n\right)}$'' of the measurement associated with the target. In numerical experiments, we found that this mainly affects the target’s shape estimation, while the kinematic-state estimates are essentially unaffected. The same approximation can also be used to derive the other equations in Eq.(\ref{eq38}), we omit them here because their derivations are similar.

\end{document}